\documentclass[prd,aps,twocolumn,amsmath,amssymb,showpacs,floats,floatfix,superscriptaddress,nofootinbib]{revtex4-2}
\usepackage{mathrsfs,color,array}
\usepackage[usenames,dvipsnames]{xcolor} 
\usepackage{amssymb,amsmath,mathtools,mathrsfs,slashed} 
\usepackage{epsfig,subfigure,placeins,float} 
\usepackage{booktabs,longtable,ctable,multirow} 
\usepackage{exscale,relsize} 
\usepackage[normalem]{ulem} 
\usepackage{enumerate}
\usepackage{tabularx}
\usepackage{cancel}
\usepackage{tabulary}
\usepackage{verbatim}
\usepackage{times,mathptmx} 
\usepackage{xspace}
\usepackage{epstopdf}
\usepackage{color}
\usepackage{graphicx,graphics} 
\usepackage{bm} 
\usepackage{color}

\newcommand{\camb}{\textsc{CAMB}\xspace}
\newcommand{\eftcamb}{\textsc{EFTCAMB}\xspace}
\newcommand{\cosmomc}{\textsc{CosmoMC}\xspace}
\newcommand{\cosmolss}{\textsc{CosmoLSS}\xspace}
\newcommand{\hmcode}{\textsc{hmcode}\xspace}
\newcommand{\halofit}{\textsc{Halofit}\xspace}

\newcommand{\bpz}{\textsc{bpz}\xspace}

\newcommand{\be}{\begin{equation}}
\newcommand{\ee}{\end{equation}}
\newcommand{\bea}{\begin{eqnarray}}
\newcommand{\eea}{\end{eqnarray}}

\usepackage{color}

\def\tt{\Uptheta}

\def\dkmu2{\delta K_{\mu \nu}\delta K^{\mu \nu}}
\def\pmu2{  \phi_{\mu \nu}\phi^{\mu \nu}}

\newcommand{\mnras}{MNRAS}
\newcommand{\aap}{A\&A}

\newcommand{\apjs}{ApJS}
\newcommand{\jcap}{JCAP}

\newcommand{\apjl}{ApJ}

\usepackage{newtxtext}
\usepackage{newtxmath}
\usepackage{amsfonts}
\usepackage{amsbsy}
\usepackage[export]{adjustbox}
\usepackage{anyfontsize}
\usepackage{gensymb}
\usepackage{soul}
\usepackage{wrapfig}
\usepackage[utf8]{inputenc}
\usepackage{braket}
\usepackage{rotating}
\usepackage{needspace,enumitem,etoolbox}
\definecolor{linkcolor}{rgb}{0.0,0.3,0.5} 
\usepackage[unicode, colorlinks=true, linkcolor=linkcolor, citecolor=linkcolor, filecolor=linkcolor,urlcolor=linkcolor, pdfusetitle]{hyperref}
\definecolor{romared}{RGB}{142,0,28}
\definecolor{reddy}{rgb}{0,0,0.8}
\definecolor{greeny}{rgb}{0,0.3,0}
\definecolor{bluey}{rgb}{0.1,0.1,1}
\definecolor{magentay}{rgb}{0.79,0.08,0.48}
\definecolor{linkcolorish}{rgb}{0.6,0,0}
\definecolor{citecolorish}{rgb}{0,0,0.75}
\definecolor{urlcolorish}{rgb}{0.12,0.46,0.7}
\hypersetup{colorlinks=true,
            citecolor=blue,
            linkcolor=magenta,
            urlcolor=romared}

\begin{document}

\title{Testing Gravity on Cosmic Scales: A Case Study of Jordan-Brans-Dicke Theory}

\author{Shahab Joudaki}
\email{shahab.joudaki@physics.ox.ac.uk}
\affiliation{Astrophysics, University of Oxford, Denys Wilkinson Building, Keble Road, Oxford OX1 3RH, UK}
\author{Pedro G.~Ferreira}
\email{pedro.ferreira@physics.ox.ac.uk}
\affiliation{Astrophysics, University of Oxford, Denys Wilkinson Building, Keble Road, Oxford OX1 3RH, UK}
\author{Nelson A.~Lima}
\email{nelson.lima15@gmail.com}
\affiliation{Institute for Theoretical Physics, University of Heidelberg,
  	Philosophenweg 16, D-69120 Heidelberg, Germany}
\author{Hans A. Winther}
\email{h.a.winther@astro.uio.no}
\affiliation{Institute of Cosmology \& Gravitation, Dennis Sciama Building, University of Portsmouth, Portsmouth PO1 3FX, UK}
\affiliation{Institute of Theoretical Astrophysics, University of Oslo, 0315 Oslo, Norway}
\date{\today}

\begin{abstract}
We provide an end-to-end exploration of a distinct modified gravitational theory in Jordan-Brans-Dicke (JBD) gravity, from an analytical and numerical description of the background expansion and linear perturbations, to the nonlinear regime captured with a hybrid suite of $N$-body simulations, to the cosmological constraints from existing probes of the expansion history, the large-scale structure, and the cosmic microwave background~(CMB). We have focused on JBD gravity as it both approximates a wider class of Horndeski scalar-tensor theories on cosmological scales and allows us to adequately model the nonlinear corrections to the matter power spectrum. In a combined analysis of the Planck 2018 CMB temperature, polarization, and lensing reconstruction, together with Pantheon supernova distances and the Baryon Oscillation Spectroscopic Survey (BOSS) measurements of baryon acoustic oscillation distances, the Alcock-Paczynski effect, and the growth rate, we constrain the JBD coupling constant to $\omega_{\rm BD} > 970$ ($95\%$~confidence level;~CL) in agreement with the General Relativistic expectation given by $\omega_{\rm BD} \rightarrow \infty$. In the unrestricted JBD model, where the effective gravitational constant at present, $G_{\rm matter}/G$, is additionally varied, increased dataset concordance $($e.g.~within $1\sigma$ agreement in $\smash{S_8 = \sigma_8 \sqrt{\Omega_{\rm m}/0.3})}$ enables us to further include the combined (``$3\times2{\rm pt}$'') dataset of cosmic shear, galaxy-galaxy lensing, and overlapping redshift-space galaxy clustering from the Kilo Degree Survey and the 2-degree Field Lensing Survey (KiDS$\times$2dFLenS). In analyzing the weak lensing measurements, the nonlinear corrections due to baryons, massive neutrinos, and modified gravity are simultaneously modeled and propagated in the cosmological analysis for the first time. In the joint analysis of all datasets, we constrain $\omega_{\rm BD} > 1540$~($95\%$~CL), $G_{\rm matter}/G = 0.997 \pm 0.029$, the sum of neutrino masses, $\sum m_{\nu} < 0.12~{\rm eV}$~($95\%$~CL), and the baryonic feedback amplitude, $B < 2.8$~($95\%$~CL), all in agreement with the standard model expectation. In fixing the sum of neutrino masses, the lower bound on the coupling constant strengthens to $\omega_{\rm BD} > 1460$ and $\omega_{\rm BD} > 2230$ (both at $95\%$~CL) in the restricted and unrestricted JBD models, respectively. We explore the impact of the JBD modeling choices, and show that a more restrictive parameterization of the coupling constant degrades the neutrino mass bound by up to a factor of three. In addition to the improved concordance between KiDS$\times$2dFLenS and Planck, the tension in the Hubble constant between Planck and the direct measurement of Riess et al.~(2019) is reduced to $\sim3\sigma$;~however, we find no substantial model selection preference for JBD gravity relative to $\Lambda$CDM. We further show that a positive shift in the effective gravitational constant suppresses the CMB damping tail, which might complicate future inferences of small-scale physics, given its degeneracy with the primordial helium abundance, the effective number of neutrinos, and the running of the spectral index.
\end{abstract}


\maketitle

\defcitealias{riess2019}{R19}

\section{Introduction}
\label{introsec}
\setcounter{footnote}{0}
\renewcommand{\thefootnote}{\arabic{footnote}}

As the quantity and quality of data from ground and space-based telescopes increase, cosmological tests of Einstein's theory of General Relativity (GR) have become increasingly robust (e.g.~\cite{simpson13,mueller18,planckmg15,planck2018,Joudaki:2017zdt,desy1ext,blake20}). These tests can take a ``model-independent'' form, for instance through the measurement of possible deviations of the gravitational potentials $\Psi$ and $\Phi$ (denoting temporal and spatial perturbations to the spacetime metric, respectively, e.g.~\cite{bt10, mb95, tsujikawa07, jz08, ps16}), the index $\gamma_{\rm G}$ (parameterizing the linear growth rate \cite{aqg05,linder05}), the $E_{\rm G}$ parameter (encapsulating the ratio of galaxy-galaxy lensing and galaxy-velocity cross-correlations~\cite{zhang07,reyes10}), or the Bellini-Sawicki $\alpha_i$ parameters (encompassing a subset of effective field theories given by stable scalar-tensor theories universally coupled to gravity with at most second order equations of motion~\cite{Horndeski:1974wa,cheung08,gubitosi13,Bellini:2014fua}). These model-independent approaches are immensely useful for tests of GR on cosmic scales, where large classes of modified gravity (MG) models are simultaneously constrained (e.g.~\cite{simpson13,mueller18,Joudaki:2017zdt, desy1ext, daniel10, bt10, joudaki17, planckmg15, planck2018, reyes10, blake16eg, amon18, caldwell07, blake11, samushia12, johnson15, Okumura16, mancini19, ferte19, noller19, traykova19, zucca19, blake20}).

A common limitation of model-independent approaches, however, is the need to avoid or suppress nonlinear scales in the matter density and galaxy density fields (e.g.~\cite{planckmg15, planck2018, Joudaki:2017zdt, desy1ext}). This is due to the inability to adequately simulate the model-independent parameterizations. The screening mechanism responsible for the suppression of power is highly model dependent, and models that have similar signatures on linear scales can differ substantially in the nonlinear regime \cite{Clifton:2011jh}. However, these nonlinear scales are necessary to include to fully utilize the expected constraining power of probes of the large-scale structure with next-generation telescopes such as Euclid~\cite{euclid11}, the Vera C.~Rubin Observatory~\cite{lsst09}, the Dark Energy Spectroscopic Instrument (DESI)~\cite{desi16}, and the Nancy Grace Roman Space Telescope~\cite{wfirst15}. 

A further limitation of model-independent approaches is the common assumption of a $\Lambda$CDM expansion history (e.g.~\cite{planckmg15, planck2018, Joudaki:2017zdt, desy1ext}). While a more general expansion rate can be considered (e.g.~by allowing for a change in the dark energy equation of state), it will require additional free parameters and will generally not correspond to a given physical theory. In considering a distinct modified gravity theory, however, the background expansion is naturally determined. This implies that the distinct theory in principle has the ability to self-consistently resolve not only discordances between datasets that measure the growth of structure but also the expansion rate.

As a result, in this paper, we will consider a specific extension of GR in the form of Jordan-Brans-Dicke (JBD) gravity, where Newton's constant is promoted to a dynamical field~\cite{Brans:1961sx}. We take a comprehensive approach by providing the underlying theory, nonlinear description with numerical simulations, and constraints using the latest cosmological data. We choose this specific gravitational theory as it is the most extensively studied extension of GR and a fertile sandbox to explore the power of different observations for constraining gravity. It allows for a comparison of highly different regimes, the astrophysical and the cosmological, and an assessment of how adequately different types of data can be used to determine non-standard parameters. Moreover, it is one of the scalar-tensor theories that has survived the recent observation of a binary neutron star merger that places extremely tight constraints (at the level of one part in $10^{15}$) on the speed of gravitational waves~\cite{ligovirgo,Baker:2017hug,Creminelli:2017sry,Ezquiaga:2017ekz}.

To understand JBD theory (see e.g.~\cite{Brans:1961sx, Clifton:2011jh, Lima:2015xia}) it consists of a metric, $g_{\alpha\beta}$ (with determinant $g$), and a real scalar field, $\phi$, that satisfy the following non-minimally coupled action,
\begin{eqnarray}
S_{\rm BD}=
\int d^4 x\sqrt{-g}\left[-\frac{\alpha}{12}\varphi^2R
+\frac{1}{2}g^{\mu\nu}\partial_\mu\varphi\partial_\nu\varphi - 2\mathcal{V}
+\mathcal{L}_{\rm m}\right]. \nonumber \\
\label{JBD1}
\end{eqnarray}
Here, $R$ is the Ricci scalar, $\smash{\mathcal{L}_{\rm m}}$ is the Lagrangian density of matter minimally coupled to the metric (which can include a cosmological constant, $\Lambda$), $\mathcal{V}$ is the potential, and $\alpha$ is the single free coupling constant which vanishes in GR. It is customary to re-express this theory in terms of the scalar field $\frac{M^2_{\rm Pl}}{2}\phi \equiv -\frac{\alpha}{12}\varphi^2$ and potential $\frac{M_{\rm Pl}^2}{2} V \equiv \mathcal{V}$, where we have identified the reduced Planck mass $M^2_{\rm Pl}=(8\pi G)^{-1}$ (which includes the ``bare'' gravitational constant, $G$). The action can then be rewritten as
\begin{eqnarray}
S_{\rm BD}=
\int d^4 x\sqrt{-g}\left[\frac{M^2_{\rm Pl}}{2}\left({\phi}R
-\frac{\omega_{\rm BD}}{\phi}g^{\mu\nu}\partial_\mu\phi\partial_\nu\phi - 2V\right)
+\mathcal{L}_{\rm m}\right], \nonumber \\
\label{JBD2}
\end{eqnarray}
where the single free coupling constant in the theory is now $\omega_{\rm BD}=-\frac{3}{2\alpha}$ and GR is recovered in the limit $\omega_{\rm BD}\rightarrow \infty$. We restrict our analysis to the simplest case of a constant potential (which is {\it not} equivalent to replacing $R$ by $R-2\Lambda$ in Eq.~\ref{JBD2}). We note that it is possible to consider extensions of this model by, for instance, incorporating a self-interaction potential $V(\phi)$ in the action above (see e.g.~Lima~\&~Ferreira~2016~\cite{Lima:2015xia}).\footnote{The generalized JBD action, where $\omega_{\rm BD} \rightarrow \omega_{\rm BD}(\phi)$ and $V \rightarrow V(\phi)$, encapsulates other distinct theories such as coupled quintessence~\cite{amendola99} and $f(R)$ gravity~\cite{felice10}.}

The motivation behind JBD gravity was originally to implement Mach's principle in GR~\cite{Brans:1961sx}, but its presence has become ubiquitous, arising as the scalar-tensor component of unified field theories, as the low energy phenomenology of higher-dimensional theories, and as the decoupling limit of extensions of GR with higher spin fields (e.g.~\cite{Clifton:2011jh}). Over the years, extensions and generalizations of JBD gravity have been proposed, culminating in a cluster of results on general scalar-tensor theories~\cite{Deffayet:2009wt, deffayet11, Gleyzes:2014dya, Gleyzes:2014qga, Crisostomi:2016czh, BenAchour:2016fzp, Zumalacarregui:2013pma, Langlois:2017mxy}. Yet, even in this extended realm, JBD gravity still encapsulates, to some extent, the main long wavelength features of generalized scalar-tensor theories such~as Horndeski gravity~\cite{Horndeski:1974wa}. Indeed, for many scalar-tensor theories of gravity, both the kinetic and potential terms can be expanded as polynomials in derivatives of the scalar field which become subdominant on cosmological scales -- a form of gradient expansion which naturally reverts to JBD gravity on cosmological scales~\cite{Avilez:2013dxa}.

There are a number of rich phenomenological properties of JBD gravity, from cosmological tracker solutions, to black hole no-hair theorems, and a host of observational effects that can be measured with astrophysical observations (e.g.~\cite{Berti:2015itd}). One of the most stringent constraints on JBD gravity has been obtained from Shapiro time delay measurements by the Cassini satellite, where the parametrized post-Newtonian (PPN) parameter $\eta \equiv \Phi/\Psi = 1 + (2.1 \pm 2.3) \times 10^{-5}$ \cite{Bertotti:2003rm}. This translates into a bound on $\omega_{\rm BD}>4.0 \times 10^4$ at $95\%$ confidence level (CL; discarding the negative $\omega_{\rm BD}$ solution due to ghost instability; see Sec.~\ref{theorysec} for the relation between $\eta$ and $\omega_{\rm BD}$). 

A complementary strong bound is obtained from the analysis of the pulsar--white dwarf binary PSR J1738+0333, where $\omega_{\rm BD}>1.2 \times 10^4$~($95\%$~CL)~\cite{2012MNRAS.423.3328F}. The stellar triple system PSR J0337+1715, where the inner pulsar--white dwarf binary is in orbit with another white dwarf, was shown to provide an even stronger bound of $\omega_{\rm BD}>7.3 \times 10^4$~($95\%$~CL)~\cite{archibald18}. A reanalysis of this same stellar triple system with an independent observational dataset and updated analysis methodology (e.g.~pertaining to the timing model, the determination of the masses and orbital parameters, and treatment of systematic uncertainties) subsequently resulted in the strongest bound to date, $\omega_{\rm BD}>1.4 \times 10^5$~($95\%$~CL)~\cite{voisin20}.

Recent attempts at constraining JBD gravity with cosmological observations are promising, but not competitive with the astrophysical constraints (e.g.~\cite{ck99,nagata04,Acquaviva:2004ti,Wu:2009zb,Avilez:2013dxa,umilta15,Ballardini:2016cvy,ooba16,Ooba:2017gyn,Peracaula19,ballardini20,sola20}). For instance, in Avilez \& Skordis (2014)~\cite{Avilez:2013dxa}, a lower bound of $\omega_{\rm BD} > 1.9 \times 10^3$ ($95\%$ CL) was obtained using cosmic microwave background (CMB) temperature and lensing measurements from Planck 2013 \cite{planck2013}. In Ballardini et al.~(2016)~\cite{Ballardini:2016cvy}, an extended JBD model with a potential in Eq.~(\ref{JBD1}) of the form $V(\phi) \propto \phi^n$ was considered. For the case of a quadratic potential, the authors constrained $w_{\rm BD} > 330$ ($95\%$~CL) using CMB temperature, polarization, and lensing data from Planck 2015~\cite{planck2015} combined with baryon acoustic oscillation (BAO) distance measurements from the 6dF Galaxy Survey~\cite{beutler11}, SDSS Main Galaxy Sample~\cite{ross15}, and BOSS LOWZ/CMASS samples~\cite{anderson14}. In Ballardini et al.~(2020)~\cite{ballardini20}, this analysis was then updated to include CMB and BAO distance measurements from Planck 2018 \cite{planck2018} and BOSS DR12 \cite{alam17}, along with the Riess et al.~(2019) \cite{riess2019} direct measurement of the Hubble constant, such that $\omega_{\rm BD} > 450$ ($95\%$~CL).

In Ooba et al.~(2017)~\cite{Ooba:2017gyn}, a modified JBD model with a field-dependent $\omega_{\rm BD}(\phi)$ was considered instead. Using the same dataset combination as in Ballardini et al.~(2016)~\cite{Ballardini:2016cvy}, the authors obtained a current lower bound on $\omega_{{\rm BD}}$ between $2.0 \times 10^3$ and $3.3 \times 10^3$ ($95\%$~CL) depending on the shape of their prior on the JBD parameter. Another setup consisting of a JBD model with a constant potential, a scalar field that is unrestricted at early times (redshift of $z = 10^{14}$), and an inverse coupling constant that is allowed to be negative was considered in Sol\`{a}~Peracaula et al.~(2019)~\cite{Peracaula19}, where CMB temperature and lensing measurements from Planck 2015 combined with low-redshift cosmological datasets were used to constrain $-6.0\times10^{-3} < \omega_{\rm BD}^{-1} < 3.5\times10^{-4}$ ($95\%$~CL). Hence, recent constraints on the JBD coupling constant fluctuate not only due to the specific datasets used but also given the specific configuration of the JBD model considered.

Beyond the CMB and post-recombination epochs considered above, the gravitational constant can be constrained during Big Bang nucleosynthesis (BBN). The production of light elements is highly sensitive to the expansion of the universe when its temperature is around an MeV; the JBD scalar field will affect the expansion rate ($H \propto \sqrt{G/\phi}$) and thus the theory can be constrained through measurements of light element abundances in distant quasars (e.g.~\cite{Casas:1990fz,Copi:2003xd,pettorino05,clifton05,Iocco:2008va,alvey20}; also see the review in \cite{uzan11}). In the GR limit, the most recent constraint is $G_{\rm BBN}/G = 0.98^{+0.06}_{-0.06}$ ($95\%$~CL), which by assuming a linear time-dependence can be translated into a constraint on the time variation of the gravitational constant, ${\dot G}/G = 1.4^{+4.4}_{-4.7} \times 10^{-12} \, {\rm yr}^{-1}$ ($95\%$~CL)~\cite{alvey20}. Naturally, these constraints might degrade in the context of JBD theory (as the scalar field will also source the background dynamics, thereby affecting the expansion rate, as well as being responsible for the time variation of $G$). Earlier self-consistent analyses find $\omega_{\rm BD} \gtrsim 300$~($95\%$~CL) from BBN alone \cite{Casas:1990fz,clifton05}.

While astrophysical constraints on the JBD coupling constant are more powerful than the constraints from existing cosmological datasets, one expects nonlinear corrections arising in {\it generalized} JBD gravity (e.g.~Horndeski gravity) to come into play that might weaken these constraints. Indeed, if on cosmological scales, JBD gravity is merely the long wavelength limit of Horndeski gravity~\cite{Avilez:2013dxa}, then on smaller scales, screening mechanisms may completely shield astrophysical systems from fifth forces arising from the presence of scalar fields (for example, through the Vainshtein~\cite{vainshtein72, Babichev:2013usa}, Chameleon~\cite{kw04a, kw04b} or Symmetron~\cite{hk10} mechanisms). This provides additional motivation to constrain JBD gravity (and other gravitational theories) on cosmic scales across the history of the Universe.

As the amount and quality of data increases with the next generation of cosmological surveys (e.g.~\cite{euclid11,lsst09,wfirst15,desi16,simons18}), we expect to significantly improve constraints on fundamental parameters, in particular as related to the gravitational Universe (e.g.~\cite{amendola12,Alonso:2016suf,ballardini19,heinrich20}). To achieve this, it will be important to understand and control a range of observational and theoretical systematic uncertainties. Concretely, it will be important to accurately account for observational systematics such as baryonic feedback, intrinsic galaxy alignments (IA), photometric redshift uncertainties, shear calibration uncertainties, galaxy bias, and pairwise velocity dispersion (see for instance Joudaki et al.~2018~\cite{Joudaki:2017zdt} and references therein); to accurately account for theoretical systematics arising from the modeling of new physics such as neutrino mass, dark matter, and dark energy/modified gravity; and to understand the role of degeneracies between the different parameters (cosmological, astrophysical, gravitational, and instrumental) affecting the observables. We will account for these systematic uncertainties and parameter degeneracies, and will consider the JBD model as a case study for constraining extensions to GR with current and future cosmological data. 

In constraining the JBD model, we primarily consider the Planck CMB~\cite{planck2018} and the combined data vector of cosmic shear, galaxy-galaxy lensing, and overlapping redshift-space galaxy clustering from KiDS$\times$\{2dFLenS+BOSS\}~\cite{Joudaki:2017zdt}. In order to improve the parameter constraints, we further include complementary information from measurements of BAO distances, the Alcock-Paczynski effect, and growth rate (final consensus BOSS DR12~\cite{alam17}), distances to type IA supernovae (SNe;~Pantheon compilation~\cite{pantheon18}), and the small-scale CMB (ACT~\cite{aiola20}). We include the key systematic uncertainties that affect these measurements, and pay particular attention to the interplay between modified gravity, neutrino mass, and baryonic feedback. We also explore the ability of the extended model to improve the concordance between cosmological datasets, and assess the extent to which it might be favored in a model selection sense relative to $\Lambda$CDM.

We structure the paper as follows. In Section~\ref{theorysec}, we describe the background expansion and linear perturbations in JBD gravity, highlighting its impact on probes of the expansion history, the large-scale structure, and the CMB. In Section~\ref{nonlinearsec}, we capture the impact of JBD gravity on the nonlinear corrections to the matter power spectrum by performing a hybrid suite of $N$-body simulations (using modified versions of COLA~\cite{Tassev:2013pn,Winther:2017jof} and \texttt{RAMSES}~\cite{Teyssier:2001cp}) and subsequently modifying the prescription for the \hmcode~\cite{Mead15,Mead16} fitting function. In Section~\ref{analysissec}, we outline the analysis techniques, cosmological datasets, and treatment of systematic uncertainties. In Sections~\ref{cmbresultssec},~\ref{wlgcsec},~\ref{distsec}, and~\ref{fullsec}, we provide the cosmological constraints on JBD gravity, and discuss their dependence on the datasets considered, the complexity of the cosmological model, and the analysis choices. We highlight the parameter degeneracies (in particular with massive neutrinos and baryonic feedback), model selection preferences (mainly JBD gravity against GR), and changes in the concordance between datasets (between Planck and KiDS, and between Planck and Riess et al.~2019~\cite{riess2019}). In Section~\ref{discussionsec}, we conclude with a summary of the findings.

\section{Theory: background cosmology and the linear regime}
\label{theorysec}

\subsection{Background equations}
\label{backeqsec}

The line element in the Newtonian gauge for small scalar perturbations, as captured by the scalar potentials $\Psi$ and $\Phi$, is given by
\begin{equation}{\label{pertmetric}}
ds^{2} = -(1 + 2\Psi)dt^2 + a^{2}(t)(1 - 2\Phi)\delta_{ij}dx^i dx^j,
\end{equation}
where $\delta_{ij}$ is the Kronecker delta function, $x$ is the comoving position coordinate, $t$ refers to physical time, $a$ is the scale factor (equal to unity at the present time), and we have implicitly assumed that the speed of light in vacuum $c = 1$. We perturb the stress-energy tensor to linear order so that, for instance, the matter density $\rho_{\rm m}({\vec x},t) = {\bar \rho}_{\rm m}(t)\left(1+\delta_{\rm m}({\vec x},t)\right)$, where the overbar denotes the mean of the matter density, and $\delta_{\rm m}$ encodes the perturbations about the mean. In the case of JBD theory, the perturbations in the scalar field are given by $\delta\phi$, such that $\phi({\vec x},t)={\bar \phi}(t)+\delta\phi({\vec x},t)$.

Following the standard approach, the JBD equations of motion are obtained by varying the action (Eq.~\ref{JBD2}) with respect to the metric and scalar field (see e.g.~\cite{Brans:1961sx, Clifton:2011jh, Lima:2015xia}). The former gives the Einstein equations,
\begin{eqnarray}{\label{einstein}}
 G_{\mu \nu} = \frac{1}{M^2_{\rm Pl}\phi} T_{\mu \nu} &+& \frac{\omega_{\rm{BD}}}{\phi^2}\left[\nabla_\mu\phi\nabla_\nu\phi - \frac{1}{2}g_{\mu\nu}\nabla_\alpha\phi\nabla^\alpha\phi\right] \nonumber \\
 &+& \frac{1}{\phi}\left[\nabla_\mu\nabla_\nu\phi-g_{\mu \nu} \left(\Box \phi+V\right)\right], \label{einsteineq}
\end{eqnarray}
\noindent where $T_{\mu \nu}$ is the total matter stress-energy tensor and $\Box$ denotes the d'Alembertian. The latter gives the scalar field's equation of motion,
\begin{equation}{}
 \Box \phi = \frac{1}{M^2_{\rm Pl}}\left({\frac{T}{3 + 2 \omega_{\rm{BD}}}}\right) - \frac{4V-2\phi V_{\phi}}{3+2\omega_{\rm{BD}}}, \label{sfeq}
\end{equation}
where $T$ is the trace of the stress-energy tensor and $V_{\phi}=dV/d\phi$ (which vanishes in the case of the constant potential that we consider).

We begin by considering the contribution from the homogeneous background (i.e.~no perturbations). Eq.~(\ref{einsteineq}) gives the two modified Friedmann equations,
\begin{eqnarray}
\label{friedmann}
3 H^{2} &=& \frac{\rho}{M^2_{\rm Pl} \phi } - 3H \frac{\dot{\phi}}{ \phi } + \frac{\omega_{\rm{BD}}}{2}\frac{\dot{\phi}^{2}}{\phi^2} +\frac{V}{\phi} \\
2\dot{H} + 3H^{2} &=& -\frac{P}{M^2_{\rm Pl}\phi} - \frac{\omega_{\rm{BD}}}{2}\frac{\dot{\phi}^{2}}{\phi^2} - 2H \frac{\dot{\phi}}{\phi} - \frac{\ddot{\phi}}{\phi}+\frac{V}{\phi}, \nonumber
\end{eqnarray}
where $\rho$ and $P$ are the total energy density and pressure of all components except the scalar field, respectively, the Hubble parameter $H \equiv \dot a / a$, and the $N$ raised dots represent $N$th-order time derivatives. We will consider a cosmology that incorporates the usual components of the stress-energy of the Universe (photons, baryons, neutrinos, dark matter) along with the scalar field (which includes the constant potential $V = \Lambda$). In this more general case, the density parameter of each component ($\mathcal{X}$) includes $\phi$ through 
\begin{equation}
\Omega^{*}_{\mathcal{X}}=\frac{\Omega_{\mathcal{X}}}{\phi} = \frac{\rho_{\mathcal{X}}}{3H^2M^2_{\rm Pl}\phi} , 
\label{densityparam}
\end{equation}
such that $\sum \Omega^{*}_{\mathcal{X}} = 1$ in a flat Universe (in other words, when defining the density parameter of each component using the critical density in GR, they do not add to unity in a flat Universe -- we will further discuss the implication of the choice between $\Omega^{*}_{\mathcal{X}}$ and $\Omega_{\mathcal{X}}$ on the concordance between datasets in Sec.~\ref{sumsec}). Moreover, in a flat Universe, $\Lambda = 3H_0^2 (1-\Omega^{*}_{\rm {m,0}})$ where the ``0'' subscripts refer to the present time.
Reading off the first line in Eq.~(\ref{friedmann}), the energy density of the scalar field is
\begin{equation}
\rho_{\phi} = M_{\rm Pl}^2\left(\frac{\omega_{\rm BD}}{2}\frac{{\dot \phi}^2}{\phi} - 3H{\dot \phi} + V\right).
\label{rhophi}
\end{equation}
The pressure of the scalar field is similarly read off the second line in Eq.~(\ref{friedmann}), such that the effective equation of state of the scalar field is given by
\begin{equation}
w_{\phi} \equiv P_{\phi}/\rho_{\phi} = \frac{\dot{\phi}^2\omega_{\rm BD} + 4H\dot{\phi}\phi + 2\ddot{\phi}\phi - 2V\phi}{\dot{\phi}^2\omega_{\rm BD} - 6H\dot{\phi}\phi + 2V\phi}.
\label{wphi}
\end{equation}
Hence, the scalar field (including a constant potential) is responsible for the accelerated expansion of the Universe, and $w_{\phi} \rightarrow -1$ when $\phi \rightarrow {\rm constant}$. The evolution of the scalar field is given by Eq.~(\ref{sfeq}), which can be expressed as
\begin{equation}
\ddot{\phi} + 3 H \dot{\phi} = \frac{1}{M^2_{\rm Pl}} \left({\frac{\rho-3P}{3+2 \omega_{\rm{BD}}}}\right) + \frac{4V-2\phi V_{\phi}}{3+2\omega_{\rm{BD}}}.
\label{fieldfrw}
\end{equation}
Here, the left-hand side of the equation can further be expressed as $a^{-3}\frac{{\rm d}}{{\rm d}t}(\dot{\phi}a^3)$. We immediately see that there are two effects at play: the scalar field will affect the way that the energy-momentum tensor of the rest of the Universe drives the expansion rate by modifying the effective gravitational constant (i.e.~$M_{\rm Pl}^2 \rightarrow M_{\rm Pl}^2\phi$), and it will also itself be a source of energy and pressure. As we shall further see below, GR is recovered in the limit $\omega_{\rm BD}\rightarrow \infty$; the corresponding density parameter satisfies $\Omega_{\phi} = \Omega_{\Lambda} + \mathcal{O}(\omega_{\rm BD}^{-1})$ and the effective equation of state satisfies $w_{\phi} = -1 + \mathcal{O}(\omega_{\rm BD}^{-1})$ which for large $\omega_{\rm BD}$ reduces to that of a cosmological constant in GR.

\begin{figure}
\includegraphics[width=1.01\columnwidth]{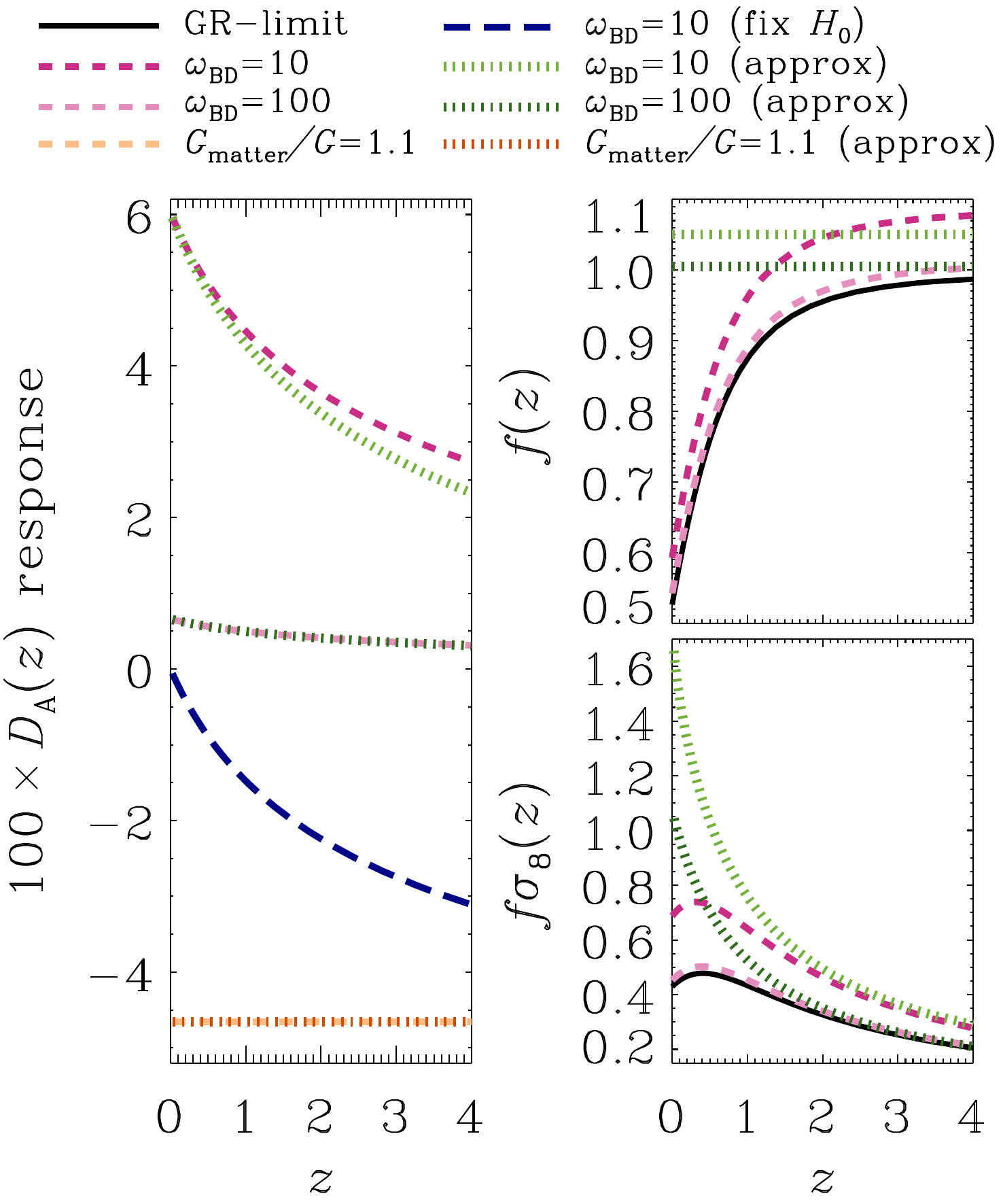}
\vspace{-1.9em}
\caption{The angular diameter distance response (left), quantified as $(D_{\rm A}^{\rm JBD} - D_{\rm A}^{\rm GR})/D_{\rm A}^{\rm GR}$, the growth rate $f(z)$ (upper right), and the product of the growth rate and the root-mean-square of the linear matter overdensity field on $8~h^{-1}~{\rm Mpc}$ scales, $f\sigma_8(z)$ (bottom right). We consider the GR-limit (solid black, corresponding to $\omega_{\rm BD} \rightarrow \infty$) and cosmologies with $\omega_{\rm BD} = \{10, 100\}$ from \eftcamb (dashed pink and light dashed pink, respectively), to be compared with the corresponding analytical matter-domination approximations for $\Delta D_{\rm A}(z)$, $f(z)$, and $f\sigma_8(z)$ in Eqs.~(\ref{deltadH0}, \ref{growthapprox}, \ref{fsig8approx}) (light dotted green and dotted green for $\omega_{\rm BD} = \{10, 100\}$, respectively). For the angular diameter distance, to assess the behavior of Eq.~(\ref{deltadH0}), we assume matter-domination for both the \eftcamb and approximate analytical solutions, and we consider a setup where the Hubble constant is kept fixed between JBD and GR (long dashed blue;~instead of fixing the matter density $\Omega_{\rm m} h^2$). We also allow the present effective gravitational constant to deviate from the GR expectation, with $G_{\rm matter}/G = 1.1$ (dashed orange and dotted red for the \eftcamb and approximate analytical solutions, respectively). For visual clarity, we only show the impact of a deviation in $G_{\rm matter}/G$ for $D_{\rm A}(z)$ as its impact on $f(z)$ and $f\sigma_8(z)$ is negligible. We note that our use of ``$G_{\rm matter}/G$'' here is shorthand for $(G_{\rm matter}/G)|_{a=1}$ (as defined in Eq.~\ref{jbdpsi}).
}
\label{figth}
\end{figure}

In Eq.~(\ref{fieldfrw}), we find that the scalar field begins to evolve after the end of the radiation-dominated era (i.e.~$\phi$ is constant and $a \propto \sqrt{t/\phi}$ during radiation domination). During the matter-dominated regime, where $\rho\propto a^{-3}$, the scalar field evolves as a power law of the scale factor \cite{nariai68,gurevich73,Clifton:2011jh},
\begin{equation}
\phi=\phi_0 a^{\frac{1}{1+\omega_{\rm{BD}}}}. 
\label{phi}
\end{equation}
Here, the subscript ``$0$'' refers to the present time, such that $\phi \leq \phi_0 \equiv \phi(a=1)$ as $a \leq 1$ (with non-negative $\omega_{\rm BD}$), and the scale factor is given by
\begin{equation}
a(t)=\left(\frac{t}{t_0}\right)^\frac{2+2\omega_{\rm BD}}{4+3\omega_{\rm BD}}.
\label{scalejbd}
\end{equation}
In the limit $\omega_{\rm BD}\rightarrow \infty$, one recovers $\phi\rightarrow \phi_0$ (a constant) and $a\propto t^{2/3}$, i.e.~the standard GR result in the matter-dominated era. The effect of the JBD coupling constant itself is to slow down the expansion rate, i.e.~the exponent in Eq.~(\ref{scalejbd}) is bounded from above by the GR value ($2/3$). Moreover, $t_0$ is related to the Hubble constant through $t_0H_0=(2+2\omega_{\rm BD})/(4+3\omega_{\rm BD})$. 

In a ``restricted'' JBD cosmology, we fix $\phi_0$ to be given by $\phi(a=1)|_{\rm restricted} = \frac{4+2\omega_{\rm BD}}{3+2\omega_{\rm BD}}$ by requiring that the effective gravitational constant is the same on local and cosmological scales at present (e.g.~\cite{Avilez:2013dxa}). We also consider an ``unrestricted'' JBD cosmology, where we allow the data to determine $\phi_0$ independently. Further in this section, we will show that this corresponds to allowing for the effective gravitational constant at present, $(G_{\rm matter}/G)|_{a=1} \propto \phi_0^{-1}$, to vary freely. At late times, the constant potential in the JBD action gives rise to the cosmic accelerating epoch, such that the effective equation of state of the scalar field at present is given by $w_{\phi}(a=1) \simeq -1$ (to increasing precision as $\omega_{\rm BD}$ increases;~evolving towards more negative values with decreasing scale factor~\cite{Lima:2015xia}). In this epoch, $\phi \propto a^{\frac{4}{1+2\omega_{\rm BD}}}$~\cite{mj84,ls89}, such that the scalar field increases marginally more rapidly with time than in the earlier matter dominated regime (for instance, by $\lesssim0.3\%$ at present for $\omega_{\rm BD} = 100$~\cite{Lima:2015xia}). 

\subsection{Toy model: modifications to distances due to JBD gravity}

In a matter-dominated Universe with zero curvature, given the impact of JBD gravity on its expansion (Eq.~\ref{scalejbd}), the angular diameter distance is given by 
\begin{eqnarray}
\label{daeqn}
D_{\rm A}(z)&=&\frac{2+2\omega_{\rm BD}}{2+\omega_{\rm BD}}\frac{cH_0^{-1}}{1+z}\left[1-\left(1+z\right)^{-\frac{2+\omega_{\rm BD}}{2+2\omega_{\rm BD}}}\right],
\end{eqnarray}
where $z = a^{-1} - 1$ is the redshift in terms of the scale factor. For an improved qualitative understanding, for small $\omega_{\rm BD}^{-1}$, this differs from the GR angular diameter distance by 
\begin{eqnarray}
\label{deltadH0}
\Delta D_{\rm A}(z) &\simeq& -\frac{2c}{\omega_{\rm BD}H_0^{\rm JBD}(1+z)}\left[\left(1-\frac{1}{\sqrt{1+z}}\right)-\frac{\ln(1+z)}{2\sqrt{1+z}}\right] \nonumber \\
&+& \frac{2c}{1+z}\left(1-\frac{1}{\sqrt{1+z}}\right)\left(\frac{1}{H_0^{\rm JBD}}-\frac{1}{H_0^{\rm GR}}\right).
\end{eqnarray}
We note that aside from the explicit $\omega_{\rm BD}$ dependence in the angular diameter distance, there is also an implicit dependence on both $\omega_{\rm BD}$ and $(G_{\rm matter}/G)|_{a=1}$ in the Hubble constant (when it is taken to be a derived parameter;~seen in Eq.~\ref{friedmann}). In Fig.~\ref{figth}, keeping the densities $\Omega_\mathcal{X}h^2$ fixed (for $\mathcal{X}$ given by baryons, cold dark matter, scalar field, and massive neutrinos), we find that the angular diameter distance is enhanced as we consider JBD instead of GR, to a decreasing extent towards higher redshifts.

This overall enhancement is driven by the second term in Eq.~(\ref{deltadH0}), as the Hubble constant in a JBD cosmology is suppressed relative to that in GR (as also seen in Fig.~\ref{figcls}), while the first term in Eq.~(\ref{deltadH0}) drives the decrease in the enhancement towards higher redshifts. Indeed, for fixed $H_0$ across cosmologies, the second term in Eq.~(\ref{deltadH0}) vanishes, such that at the same redshift, objects instead appear closer than in the corresponding GR Universe. Instead of a decrease in the difference between the distances in JBD and GR cosmologies with redshift, here the difference between the distances increases with redshift. Hence, as is well known, in comparing observables between cosmologies, it is imperative to have clarity in the specific parameters kept fixed.

As expected, given the assumption of small $\omega_{\rm BD}^{-1}$ in deriving Eq.~(\ref{deltadH0}), the analytical approximate solution for $\Delta D_{\rm A}(z)$ improves as $\omega_{\rm BD}$ increases, and the difference between the angular diameter distances in JBD and GR cosmologies decreases as $\omega_{\rm BD}$ increases. Indeed, we recover $\Delta D_{\rm A} \rightarrow 0$ as $\omega_{\rm BD} \rightarrow \infty$ and $(G_{\rm matter}/G)|_{a=1} \rightarrow 1$. Moreover, as expected from Eqs.~(\ref{friedmann}) and (\ref{jbdpsi}) below, and shown in Fig.~\ref{figth}, the angular diameter distance scales inversely with the square root of the present effective gravitational constant (i.e. $D_{\rm A} \propto 1/\sqrt{(G_{\rm matter}/G)|_{a=1}}$).

\subsection{Linear perturbations}

We now turn to linear perturbations in JBD theory. While $\delta\phi$ can be thought of as the potential for a putative fifth force (see Sec.~\ref{nonlinearsec}), 
it is often useful to consider the Einstein field equations on sub-horizon scales, i.e.~$k^2\gg(aH)^2$ where $k$ is the wavenumber. In this ``quasistatic'' regime (which is moreover characterized by the condition that time derivatives of the metric and scalar-field perturbations are negligible relative to their respective spatial derivatives), the field equations reduce to a modified Poisson equation\footnote{Here, we have defined our effective gravitational constant to be dimensionful. However, we note that it is also common to define the effective gravitational constant as dimensionless by normalizing with the bare gravitational constant, i.e.~$G_{\rm matter}/G \rightarrow G_{\rm matter}$.} (e.g.~\cite{Alonso:2016suf,DeFelice:2011hq}),
\begin{equation}
\frac{k^2}{a^2} \Psi \simeq -4 \pi G_{\rm{matter}} \rho_{\rm{m}} \delta_{\rm{m}},
\label{jbdpoisson}
\end{equation}
and a gravitational slip equation that depends only on the JBD coupling constant, 
\begin{equation}{\label{etabd}}
\gamma \equiv \frac{\Psi}{\Phi} \simeq \frac{2 + \omega_{\rm{BD}} }{1 + \omega_{\rm{BD}} }, 
\end{equation}
where the time-varying gravitational constant is given by
\begin{eqnarray}\label{geqn}
\frac{G_{\rm{matter}}}{G} \simeq \frac{1}{\bar{\phi}}\frac{4+2\omega_{\rm BD}}{3+2\omega_{\rm BD}}.
\end{eqnarray}
The motion of non-relativistic particles is thereby dictated by the modified potential
\begin{equation}
\Psi = \frac{G_{\rm matter}}{G} \Psi_{\rm GR} \simeq \left[{\frac{\bar{\phi}(a=1)}{\bar{\phi}}\left(\frac{G_{\rm{matter}}}{G}\right)\bigg\rvert_{a=1}}\right] \Psi_{\rm GR}. 
\label{jbdpsi}
\end{equation}
Hence, the effective gravitational constant evolves with time even when it is set to unity at present in the restricted JBD model. Given $\bar{\phi}(a)$ increases with $a$ this implies $G_{\rm matter}/G$ is larger in the past. A crucial point here is that we can in principle allow {\it both} $\omega_{\rm BD}$ and the value of $G_{\rm matter}/G$ at the present time to be independent parameters of the theory (corresponding to an unrestricted JBD model as discussed earlier). We will consider such a setup in this paper to obtain general constraints on JBD gravity.\footnote{{We emphasize that the value of $G_{\rm matter}/G$ at the present time is determined by the choice of initial condition when solving Eq.~(\ref{sfeq}) for $\bar{\phi}$. Allowing this to be a free parameter is what Ref.~\cite{Avilez:2013dxa} calls the {\it unrestricted} JBD model. This is to be contrasted to the {\it restricted} JBD model which corresponds to setting the initial conditions such that $\bar{\phi}(a=1) = (4+2\omega_{\rm BD})/(3+2\omega_{\rm BD})$ implying $G_{\rm{matter}}/G = 1$ at the present time.}} In addition to the Poisson equation, $G_{\rm matter}/G$ enters the Friedmann equations via ${\bar \phi}$. As a result, the $(G_{\rm matter}/G)|_{a=1}$ parameter that we vary in our analysis affects the expansion rate and any other physical process where the gravitational constant appears relative to other physical constants. 

On sub-horizon scales, pressureless matter at late times obeys an evolution equation of the form (e.g.~\cite{DeFelice:2011hq})
\begin{eqnarray}{\label{mattpert}}
\delta_{\rm{m}}^{\prime \prime} + \left[1 + \frac{{\cal H}^{\prime}}{\cal H}\right]\delta_{\rm{m}}^{\prime}- \frac{3}{2}\left( \frac{4+2\omega_{\rm BD}}{3+2\omega_{\rm BD}}\right)\Omega^{*}_{\rm{m}}(a)\delta_{\rm{m}} \simeq 0, \\ \nonumber
\end{eqnarray}
where ${\cal H}=aH$ and primes are derivatives with $\ln a$ (note that $\Omega^{*}_{\rm m}(a)$ here includes a $1/\phi$ term given its definition in Eq.~\ref{densityparam}). This can be expressed in terms of the growth rate, $f\equiv\frac{d\ln \delta_{\rm m}}{d\ln a}$, such that (e.g.~\cite{Baker:2013hia})
\begin{eqnarray}
f'+\left[1 + \frac{{\cal H}^{\prime}}{\cal H}\right]f+f^2=\frac{3}{2}\left( \frac{4+2\omega_{\rm BD}}{3+2\omega_{\rm BD}}\right)\Omega^{*}_{\rm m}(a).
\label{feqn}
\end{eqnarray}
There are two distinct effects on the growth of structure: the expansion rate (which is lower compared to GR while the first derivative is more negative) enters the friction term and enhances growth, while the source term that sets the strength of the gravitational response also boosts growth. Unlike many proposals for scalar tensor theories, where deviations from GR are synchronized with the onset of $\Lambda$-domination (and thus only kick in at late times), in JBD theory the scalar field has an effect throughout the full evolution of the Universe. 

\subsection{Toy model: modified growth rate in JBD gravity}

Again, let us focus on a pure matter-dominated cosmology, where we know that $\delta_{\rm m}\sim a$ and $f=1$ in GR. If we consider the effects of the scalar field to be small, 
an analytic approximation to the growth rate is given by 
\begin{eqnarray}
\label{growthapprox}
f\simeq 1+\frac{1}{2\omega_{\rm BD}}.
\end{eqnarray}

The impact of $\omega_{\rm BD}$ is enhanced when considering the density weighted growth rate, $f\sigma_8\sim d\delta_{\rm m}/d\ln a$, where $\sigma_8$ is the root-mean-square of the linear matter overdensity field on $8~h^{-1}~{\rm Mpc}$ scales, given its integrated effect over a long time scale (essentially since the beginning of matter domination). This can be seen via the approximate solution for $f$ in the matter era, which when integrated gives
\begin{eqnarray}
\label{fsig8approx}
f\sigma_8\simeq f\sigma_8|_{\rm GR}\left[1+\frac{1}{2\omega_{\rm BD}}\left(1+\ln \frac{z_{{\rm eq}}}{z}\right)\right].
\end{eqnarray}
In Fig.~\ref{figth}, we illustrate this enhancement in $f(z)$ and $f\sigma_8(z)$, along with the agreement between the \eftcamb and approximate analytical solutions in the matter-dominated regime. The insensitivity of the growth rate to the present effective gravitational constant also agrees with that obtained from \eftcamb.

\begin{figure*}
\includegraphics[width=\hsize]{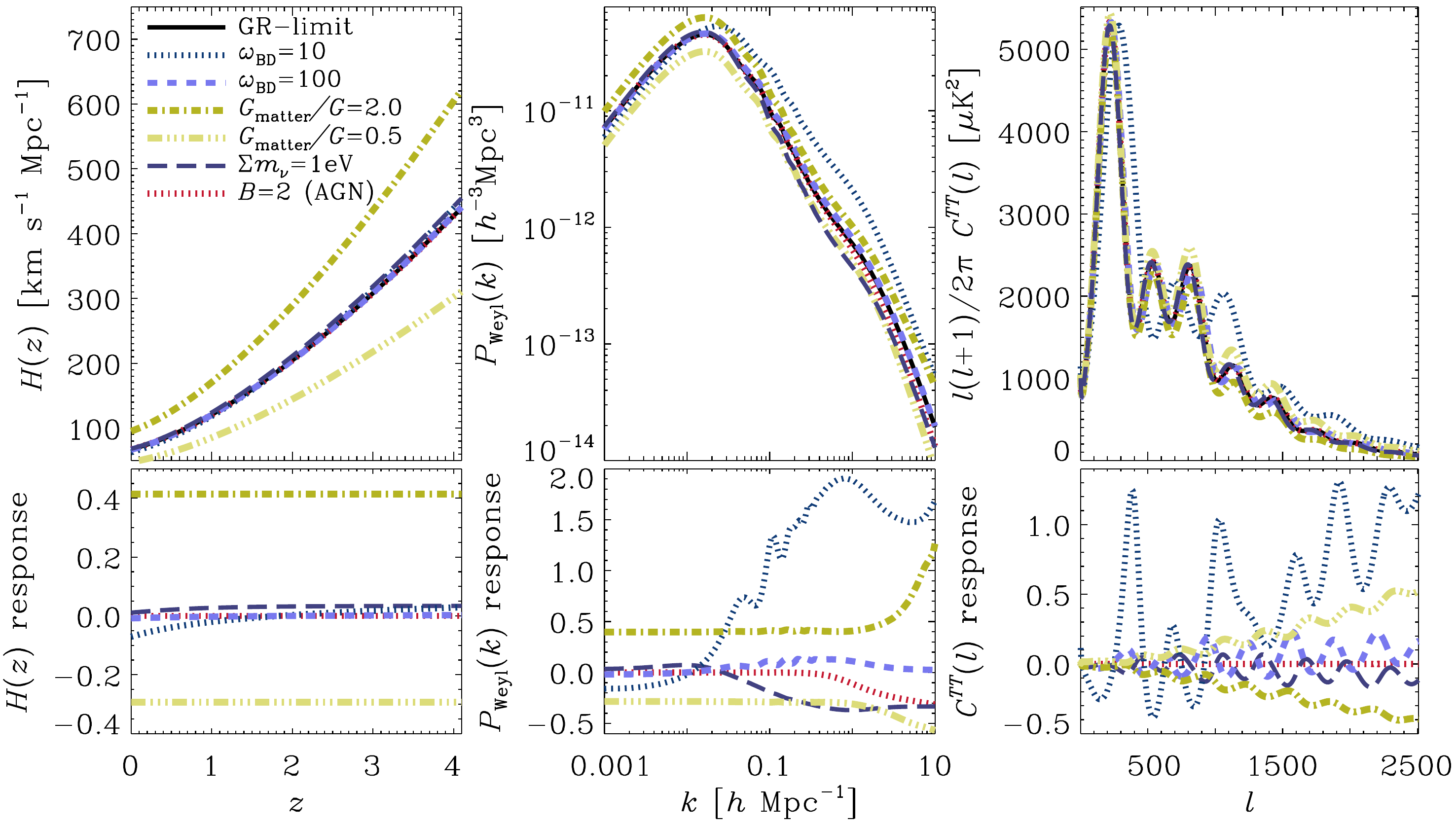}
\vspace{-1.8em}
\caption{The expansion history, $H(z)$, Weyl power spectrum, $P_{\rm Weyl} \equiv P_{\frac{\Psi+\Phi}{2}}(k)$, and CMB temperature power spectrum, $C^{TT}(\ell)$, along with their respective responses, defined as $(A^{\rm JBD} - A^{\rm GR})/A^{\rm GR}$, where $A \in \{H(z), P(k), C(\ell)\}$. Here, the Weyl power spectrum is considered at $z = 0$. For our GR limit, we have effectively imposed $\omega_{\rm BD} \rightarrow \infty$ and $G_{\rm matter}/G = 1$. For the JBD model, we show the four cases $\omega_{\rm BD} = 10$, $\omega_{\rm BD} = 100$, $G_{\rm matter}/G = 0.5$, and $G_{\rm matter}/G = 2.0$ (such that $\omega_{\rm BD} \rightarrow \infty$ when $G_{\rm matter}/G \neq 1$, and $G_{\rm matter}/G = 1$ when $\omega_{\rm BD} \neq \infty$). We emphasize that our use of ``$G_{\rm matter}/G$'' here is shorthand for $(G_{\rm matter}/G)|_{a=1}$ (as defined in Eq.~\ref{jbdpsi}), and that we have kept the density parameters $\Omega_{\mathcal{X}} h^2$ fixed rather than $\Omega_{\mathcal{X}}^{*} h^2$, where ``$\mathcal{X}$'' denotes the matter components and the scalar field (see distinction in Eq.~\ref{densityparam}). For comparison, we show the impact of the sum of neutrino masses, fixed to $\sum m_{\nu} = 1~{\rm eV}$, along with baryonic feedback corresponding to the ``AGN'' case of the OWL simulations (given by $B = 2$ in \hmcode). We moreover show the impact of JBD gravity on the polarization power spectra in Appendix~\ref{cmbpolapp} (Fig.~\ref{figclspol}).
}
\label{figcls}
\end{figure*}

\subsection{Impact of JBD gravity on the propagation of light}

Lastly, we consider the impact of JBD gravity on the propagation of light. The geodesic equation for relativistic particles (e.g.~photons) is sensitive to the sum of the metric potentials $(\Psi+\Phi)$. Given the quasistatic expressions (Eqs.~\ref{jbdpoisson}--\ref{jbdpsi}), 
\begin{eqnarray}
\label{sumpoteqn}
\Psi+\Phi&=&\left(1+\gamma^{-1}\right)\Psi
=-\left(1+\gamma^{-1}\right)G_{\rm matter}\frac{4\pi a^2\rho_{\rm m} \delta_{\rm m}}{k^2} \nonumber \\ &=& -\frac{2}{\bar{\phi}}\frac{4\pi G a^2\rho_{\rm m} \delta_{\rm m}}{k^2}=\frac{2}{\bar{\phi}}\Psi_{\rm GR},
\end{eqnarray}
which differs from the GR expectation by a factor of $1/{\bar \phi}$ (i.e.~$G_{\rm light} = \left({(1+\gamma)/(2\gamma)}\right) \, G_{\rm matter} = G/{\bar \phi}$, where the first relation is general and the second relation is specific to JBD theory). We moreover consider the ratio of the metric potential $\Psi$ (probed by e.g.~redshift-space distortions) and the sum of the potentials $(\Psi+\Phi)/2$ (probed by e.g.~weak lensing). This ratio of the potentials corresponds to the ratio of $G_{\rm matter}$ and $G_{\rm light}$, and is targeted by measurements of the ``$E_{\rm G}$ parameter'' (e.g.~\cite{zhang07,reyes10,blake20}). The ratio is only sensitive to the gravitational slip and thereby only to $\omega_{\rm BD}$ in JBD theory (which holds in both the restricted and unrestricted JBD scenarios in the quasistatic regime):
\begin{equation}
\label{potrateqn}
2\frac{\Psi}{\Psi+\Phi} = \frac{2\gamma}{1+\gamma} \simeq \frac{4+2\omega_{\rm BD}}{3+2\omega_{\rm BD}}.
\end{equation}
For a given $\omega_{\rm BD}$, this ratio does not evolve with time (such that any time variation observed would rule out JBD gravity in addition to GR and a range of other models entirely). As expected, the ratio approaches $1$ in the GR limit ($\omega_{\rm BD} \rightarrow \infty$), and the largest deviation is given by $4/3$ as $\omega_{\rm BD} \rightarrow 0$. Since $\gamma$ deviates from the GR expectation by less than a percent already for $\omega_{\rm BD} \gtrsim 100$ (where such a small $\omega_{\rm BD}$ is disfavored by current cosmological measurements, as discussed in Sec.~\ref{introsec}, but also see Sec.~\ref{modelingsec}), we do not expect even future measurements of the gravitational slip alone (where $\sigma(\gamma) \sim 0.05$ at best for Stage-IV surveys~\cite{scott09,amendola14,jpas20}) to powerfully constrain the space of viable JBD models (noting that a similar argument holds for the $E_{\rm G}$ parameter). Instead, we need to measure the expansion history and both of the potentials distinctly in order to place the strongest constraints on the underlying cosmology, which is the approach taken here.

\subsection{Connecting JBD theory to effective field theory}
\label{eftsec}

As a side note, we highlight that JBD theory can be connected to effective field theory (EFT) via the $\alpha_i$ parameters of Bellini \& Sawicki \cite{Horndeski:1974wa,Bellini:2014fua} 
(which are all zero in GR), where the ``Planck-mass run rate'' $\alpha_{\rm M} = d \ln \phi / d \ln a$, the ``braiding'' $\alpha_{\rm B} = -\alpha_{\rm M}$, the ``kineticity'' $\alpha_{\rm K} = \omega_{\rm BD} \alpha_{\rm M}^2$, and the ``tensor speed excess'' $\alpha_{\rm T} = 0$~\cite{Bellini:2017avd}. Here, $\alpha_T$ encapsulates the zero deviations to the speed of gravitational waves relative to the speed of light in JBD theory, thereby satisfying the LIGO-Virgo bound~\cite{ligovirgo}, The other $\alpha_i$ parameters are described in terms of only the coupling constant and the time-variation of the scalar field, which is itself uniquely determined by the coupling constant in standard JBD theory. Hence, the theory can be reduced to a single independent $\alpha_i$ parameter (along with the expansion rate which can be expressed in terms of the same $\alpha_i$ and the present effective gravitational constant in the unrestricted scenario). 

Concretely, as $\alpha_{\rm M}$ is only sensitive to the logarithmic derivative of $\phi$ (and thereby the logarithmic derivative of $G_{\rm matter}/G$), it has no sensitivity to the overall normalization, given by $\phi_0 \equiv \phi(a=1)$. As a result, $\alpha_{\rm M}$ has no sensitivity to deviations in the present effective gravitational constant, $(G_{\rm matter}/G)|_{a=1}$. Instead, considering the redshift dependence of $\phi$ (Sec.~\ref{backeqsec}), $\alpha_{\rm M} = 0$ during radiation domination, $\alpha_{\rm M} = (1+\omega_{\rm BD})^{-1}$ during matter domination, and $\alpha_{\rm M} = 4(1+2\omega_{\rm BD})^{-1}$ during the epoch of cosmic acceleration. This directly determines the braiding and kineticity parameters given the relations with $\alpha_{\rm M}$ above. 
For completeness, $\alpha_{\rm B} = \{0, -(1+\omega_{\rm BD})^{-1}, -4(1+2\omega_{\rm BD})^{-1}\}$ and $\alpha_{\rm K} = \{0, \omega_{\rm BD} (1+\omega_{\rm BD})^{-2}, 16\omega_{\rm BD}(1+2\omega_{\rm BD})^{-2}\}$ during the radiation, matter, and cosmic accelerating epochs, respectively.

\begin{figure*}
\includegraphics[width=1.0\hsize]{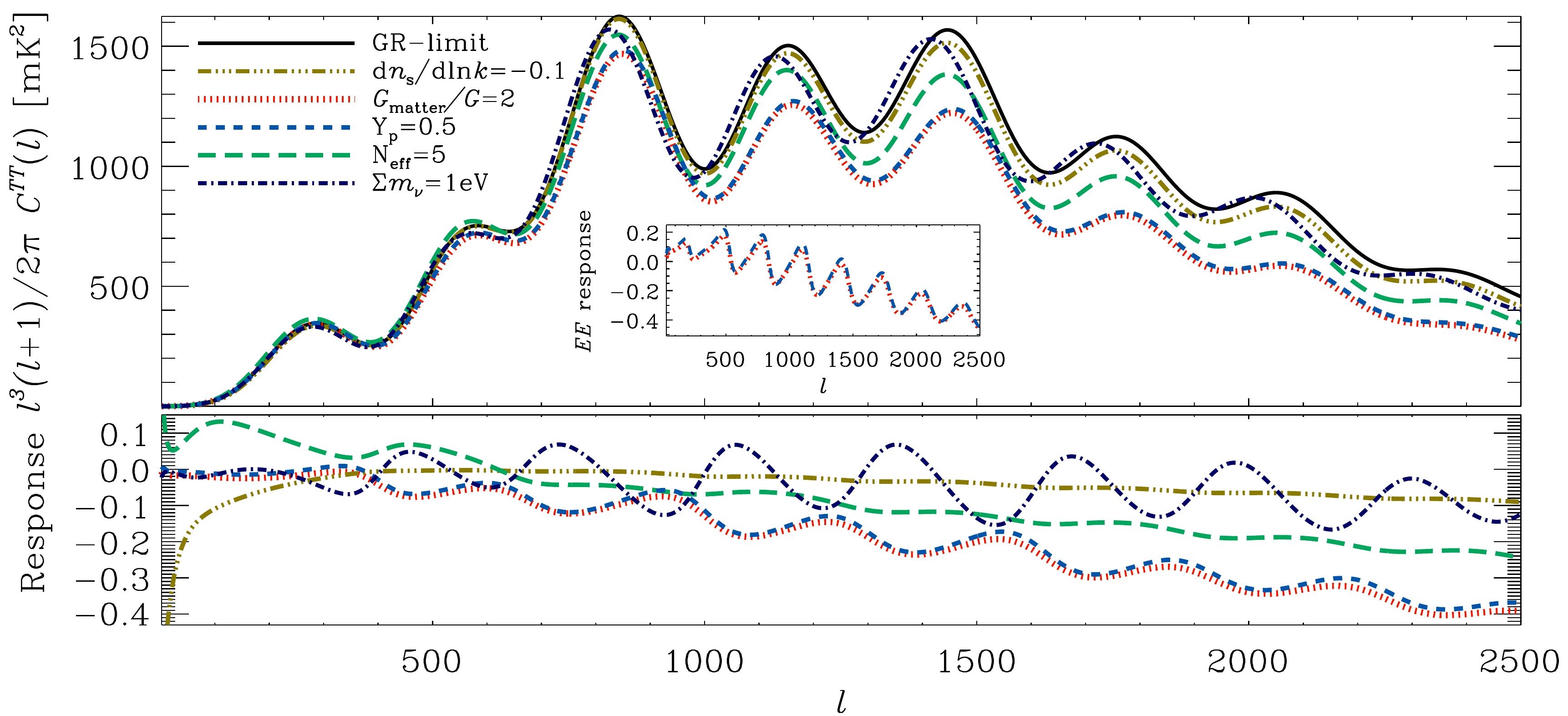}
\vspace{-1.8em}
\caption{CMB temperature power spectra in extended cosmological parameter spaces along with their respective responses, defined as $C^{\rm extended}(\ell)/C^{\rm {\Lambda}CDM}(\ell) - 1$. We consider deviations in the running of the spectral index, ${\rm d} n_{\rm s} / {\rm d} \ln k$, the effective number of neutrinos, $N_{\rm eff}$, the sum of neutrino masses, $\sum m_{\nu}$, the primordial helium abundance, $Y_{\rm P}$, and the present effective gravitational constant, $G_{\rm matter}/G$. We also highlight the $E$-mode polarization power spectrum response for $Y_{\rm P}$ and $G_{\rm matter}/G$ in the inlet (for further polarization details, see Appendix~\ref{cmbpolapp}). 
}
\label{figclsdamping}
\end{figure*}

\subsection{Correlations between modified gravity, massive neutrinos, and baryonic feedback in cosmological observables}
\label{corrsec}

We now have the required ingredients for a qualitative understanding of the impact of JBD gravity on cosmological observables. The scalar field has an effect throughout the history of the Universe, decreasing the expansion rate (hence increasing BAO and SN distances) and largely increasing the Weyl power spectrum $\smash{\left({P_{\rm Weyl} \equiv P_{\frac{\Psi+\Phi}{2}}(k)}\right)}$, to which lensing is sensitive, as $\omega_{\rm BD} \rightarrow 0$ (i.e.~{\it away} from the GR limit). This is illustrated in Fig.~\ref{figcls}, where the suppression in the expansion rate diminishes towards higher redshift (asymptoting to the GR rate), and where the Weyl power spectrum has a turnover from suppression to enhancement at $k \sim 10^{-2}\,h\,{\rm Mpc}^{-1}$ (to within $10\%$ for $\omega_{\rm BD} \in [10, 100]$) with an amplitude that increases with the strength of JBD gravity. As a result, as $\omega_{\rm BD}$ decreases, it will counter any degrees of freedom that might suppress the Weyl power spectrum on scales below this turnover (such as baryonic feedback\footnote{Baryonic feedback is incorporated in \hmcode \cite{Mead15} through calibration to the OverWhelmingly Large (OWL) hydrodynamical simulations~\cite{Schaye10,Daalen11,semboloni11}, as further discussed in Sec.~\ref{nonlinearsec} and Sec.~\ref{systsec}. In Fig.~\ref{figcls}, we consider a feedback amplitude that corresponds to the ``AGN'' case of these simulations.} and massive neutrinos\footnote{We note that the Weyl power spectrum response for the sum of neutrino masses (Fig.~\ref{figcls}) is slightly positive for $k \sim 10^{-2}~h~{\rm Mpc}^{-1}$ because the fiducial GR cosmology here has the sum of neutrino masses fixed to $\sum m_{\nu} = 0.06$~eV, which implies that the suppression in the Weyl power spectrum begins on larger scales, even though the strength of the suppression is smaller on increasingly nonlinear scales, relative to a cosmology with $\sum m_{\nu} = 1$~eV. In other words, the free streaming length of each neutrino species is inversely proportional to its mass \cite{bes80,lp12} (which can in principle be used as a distinct feature with which to probe the neutrino mass hierarchy~\cite{jimenez10}).}). 

Meanwhile, as $\omega_{\rm BD} \rightarrow 0$, the response of the CMB temperature power spectrum, $C^{TT}(\ell)$, is oscillatory along the multipoles $\ell$ (due to shifts in the locations of the peaks), and is both positive and negative below $\ell \sim 10^3$, above which it gradually increases relative to GR (further see Fig.~\ref{figclspol} for the CMB polarization power spectrum and polarization-temperature cross-spectrum). Here, we note the largely opposing effects of $\omega_{\rm BD}$ and the sum of neutrino masses, $\sum m_{\nu}$, on the CMB power spectra.

We further show the impact of changes in $G_{\rm matter}/G$ (note that as a primary parameter we always implicitly refer to its value at present), where a ratio below unity decreases the expansion rate, albeit at a constant level with redshift (given Eq.~\ref{friedmann} where $H^2_{\rm JBD}/H^2_{\rm GR} \simeq 1/\phi$), such that there is an overall renormalization of the expansion history when $G_{\rm matter}/G \neq 1$. Similarly, for the Weyl power spectrum, $G_{\rm matter}/G < 1$ provides a constant suppression on linear and mildly nonlinear scales (down to $k\sim1\,h\,{\rm Mpc}^{-1}$), where it is of the same magnitude as for the expansion rate (given Eq.~\ref{sumpoteqn} where $G_{\rm light}/G = 1/\phi$ in the quasistatic regime). This suppression is enhanced on highly nonlinear scales, where we have modified \hmcode to match the numerical simulations in Sec.~\ref{nonlinearsec}. However, in contrast to $H(z)$ and $P_{\rm Weyl}(k)$, the CMB temperature power spectrum is enhanced as $G_{\rm matter}/G < 1$, such that the response increases with $\ell$, making it a particularly suitable target for probes of the CMB damping tail (and correlated with other physics such as the running of the spectral index, neutrino mass, primordial helium abundance, and the effective number of neutrinos that affect the small-scale CMB).

As a result, there is a particular correlation between the effects of modified gravity, the sum of neutrino masses, and baryonic feedback on the Weyl power spectrum, along with distinct correlations between the effects of modified gravity and the sum of neutrino masses on the expansion rate and CMB power spectrum. A notable difference between the three is that the effective gravitational constant can in principle take on values on both ends of the fiducial expectation, where $G_{\rm matter}/G < 1$ provides an enhancement of power (rather than suppression) and thereby allows for even greater neutrino masses and baryonic feedback. The fact that the effective gravitational constant allows for both suppression and enhancement of the cosmological quantities (depending on whether $G_{\rm matter}/G$ is greater or smaller than unity) gives it greater flexibility than $\omega_{\rm BD}$ which only allows for ``one-sided'' modifications (i.e.~either suppression or enhancement). This implies that $G_{\rm matter}/G$ is better suited to alleviating possible discordances between datasets, but also to be correlated with the other aforementioned physics.

We note that the responses to $\omega_{\rm BD}$ are smaller for the expansion rate as compared to the Weyl and CMB power spectra. While distinct physics might be correlated or even degenerate for a single physical observable, they often have different signatures for distinct observables, as seen in their impact on the responses for $\{H(z), P_{\rm Weyl}(k), C^{TT}(\ell)\}$. For instance, the baryonic feedback suppresses the Weyl power spectrum but has no impact on the expansion history and CMB power spectra, while the sum of neutrino masses affects both the Weyl power spectrum and CMB power spectra but only negligibly the expansion history, and by contrast JBD gravity has a non-negligible impact on all three of these cosmological quantities. Hence, we expect that a combination of multiple complementary probes is required to robustly constrain the underlying cosmology of the Universe.

\subsection{Degeneracies with the effective gravitational constant in the CMB damping tail}
\label{dampsec}

In Fig.~\ref{figclsdamping}, we continue to highlight the impact of the effective gravitational constant in the CMB damping tail, and its possible degeneracies with other physics, such as the primordial helium abundance, $Y_{\rm P}$ (i.e.~the mass fraction of baryons in $^4$He), the running of the spectral index, ${\rm d} n_{\rm s} / {\rm d} \ln k$, and the effective number of neutrinos, $N_{\rm eff}$ (along with the sum of neutrino masses again for comparison). As discussed in Sec.~\ref{corrsec}, we obtain a suppression in the damping tail for positive perturbations in $G_{\rm matter}/G$, which is correlated with the expected suppression due to \{positive, negative, positive\} perturbations in $\{Y_{\rm P}, {\rm d} n_{\rm s} / {\rm d} \ln k, N_{\rm eff}\}$, respectively. In Ref.~\cite{hou13}, the ratio of the angular scales of the diffusion length and sound horizon, $\theta_{\rm d} / \theta_{\rm s}$, is shown to be the primary quantity that governs modifications to the damping tail and is responsible for the correlations between $Y_{\rm P}$ and $N_{\rm eff}$. 

In detail, $\theta_{\rm d} = r_{\rm d}/D_{\rm A}$ and $\theta_{\rm s} = r_{\rm s}/D_{\rm A}$, where $r_{\rm d}$ is the comoving diffusion length at recombination, $r_{\rm s}$ is the comoving size of the sound horizon at recombination, and $D_{\rm A}$ is the angular diameter distance to recombination, such that the $D_{\rm A}$ terms cancel in the ratio $\theta_{\rm d} / \theta_{\rm s} = r_{\rm d} / r_{\rm s}$. We define
\begin{equation}
\label{rseq}
r_{\rm s}(a_*) = \int_{0}^{a_*} \frac{c_{\rm s}(a) da}{a^2 H(a)},
\end{equation}
where $c_{\rm s}(a) = c/\sqrt{3\left({1+R(a)}\right)}$, $R(a) = 3\rho_{\rm b}(a)/\left({4\rho_{\gamma}(a)}\right)$, $\rho_{\rm b}$ and $\rho_{\gamma}$ are the energy densities of baryons and photons, respectively, and $a_*$ is the recombination scale factor for which the optical depth equals unity~\cite{hou13,husugiyama95}. The diffusion length is moreover given by
\begin{equation}
\label{rdeq}
r_{\rm d}(a_*) = {\pi \over 6} \left(\int_0^{a_*}{g(a) da \over {a^3 \sigma_{\rm T}n_{\rm e}H(a)}}\right)^{1/2},
\end{equation}
where $g(a) = c\left[{R^2(a) + {16\over15}\left({1+R(a)}\right)}\right] / \left({1+R^2(a)}\right)$, $\sigma_{\rm T}$ is the Thomson cross-section, and $n_{\rm e}$ is the number density of free electrons~\cite{hou13,husugiyama95}. As a result, the ratio $\theta_{\rm d} / \theta_{\rm s} \propto \sqrt{H(a)/n_{\rm e}}$, where $H(a)$ refers to the pre-recombination expansion rate. This can in turn be shown to be proportional to $(1+\mathcal{C} N_{\rm eff})^{1/4} / \sqrt{1-Y_{\rm P}}$~\cite{hou13}. The first term in the proportionality follows for fixed matter-radiation equality redshift as $H^2 \propto \rho_{\rm r} \propto 1 + \rho_{\nu} / \rho_{\gamma}$, where $\rho_{\rm r}$ is the energy density of radiation which includes a contribution from the neutrino energy density $\rho_{\nu} = \mathcal{C} N_{\rm eff} \rho_{\gamma}$, given the constant $\mathcal{C} = (7/8) \left({4/11}\right)^{4/3}$~\cite{mangano02}. The second term follows from $n_{\rm e} = (1-Y_{\rm P})\rho_{\rm b}/m_{\rm p}$, where $m_{\rm p}$ is the proton mass, and reflects the fact that helium recombines earlier than hydrogen which changes the free electron density at last scattering~\cite{hu95}.

Turning to the effective gravitational constant, we realize that its impact through the expansion rate is $H^2 \propto G_{\rm matter}/G$, such that $\theta_{\rm d} / \theta_{\rm s} \propto  (G_{\rm matter}/G)^{1/4} (1+\mathcal{C} N_{\rm eff})^{1/4} / \sqrt{1-Y_{\rm P}}$. As pointed out in Ref.~\cite{hou13}, increasing the expansion rate decreases $a_*$ and increases $n_{\rm e}(a)$~\cite{zz03}, which together slightly modify the power of $1/4$ in $(1+\mathcal{C} N_{\rm eff})^{1/4}$, and the same applies in the case of $G_{\rm matter}/G$. We note that the correlation between the helium abundance and baryon density is broken through the measurement of the latter on larger scales in the CMB (specifically by the first-to-second peak ratio relative to the first-to-third peak ratio in the temperature power spectrum)~\cite{komatsu11}. Meanwhile, the degeneracy between $Y_{\rm P}$ and $N_{\rm eff}$ (and between $G_{\rm matter}/G$ and $N_{\rm eff}$) is partly broken by the early integrated Sachs-Wolfe (ISW) effect, the potential high baryon fraction as $N_{\rm eff}$ increases, and the  phase shift in the acoustic oscillations due to neutrino perturbations~\cite{hou13} (for the neutrino phase shift, also see Ref.~\cite{bs04}). However, these and other physical effects do not help to break the degeneracy between $Y_{\rm P}$ and $G_{\rm matter}/G$, which persists as shown in Fig.~\ref{figclsdamping}.

We note that for $\omega_{\rm BD} \rightarrow \infty$, which is the ``no-slip gravity'' limit~(e.g.~\cite{linder18}) considered as $G_{\rm matter}/G$ is varied in Fig.~\ref{figclsdamping}, the effective gravitational constant does not evolve with time (hence, does not directly contribute to the ISW effect). Even as we consider an unrestricted JBD model where $\omega_{\rm BD}$ and $G_{\rm matter}/G$ are simultaneously constrained, in forthcoming sections we find that our constraints are sufficiently strong that the evolution is $\lesssim1\%$ (from the present to the BBN epoch), in agreement with the BBN constraint in Ref.~\cite{alvey20} (see Sec.~\ref{bbnta}). Focusing on the degeneracy between $G_{\rm matter}/G$ and $Y_{\rm P}$, we have also explicitly checked that the CMB temperature and polarization power spectra remain invariant to sub-percent level as we modify these two parameters (here, up to a factor of two) but keep $\theta_{\rm d} / \theta_{\rm s}$ fixed according to the relation we provide above.

We further note that $G_{\rm matter}/G$ will in principle also modify $Y_{\rm P}$ itself, as $G_{\rm matter}/G > 1$ enhances the expansion rate, which leads to an earlier freeze out of the weak and nuclear interactions in the early Universe, and thereby an over-production of $^4$He~\cite{clifton05,bambi05}. In Ref.~\cite{fields20}, this dependence of the helium abundance on the effective gravitational constant is shown to take on the form $Y_{\rm P} \propto (G_{\rm matter}/G)^{0.36}$ (also see Ref.~\cite{scherrer04}). We do not account for this effect in Fig.~\ref{figclsdamping}, but note that this would further enhance the impact of $G_{\rm matter}/G$ on the CMB ($\sim1\%$ change in $Y_{\rm P}$ for the strongest constraint, approximately $0.03$, that we obtain on $G_{\rm matter}/G$ in forthcoming sections).\footnote{As similar-factor changes in $G_{\rm matter}/G$ and $Y_{\rm P}$ have comparable effects on the CMB, this implies that the impact of $G_{\rm matter}/G$ on the CMB would be enhanced by $\sim30\%$. Put differently, our constraints on $G_{\rm matter}/G$ in forthcoming sections can either be viewed as somewhat conservative (i.e. $\lesssim30\%$ weaker), or alternatively in the context of a Universe where GR is enforced during BBN.}

While the CMB polarization is useful in constraining $G_{\rm matter}/G$ (given the qualitatively different signature of the effective gravitational constant on the polarization power spectrum, as pointed out in Ref.~\cite{zz03} and explicitly shown in Sec.~\ref{cmbresultssec} and Appendix~\ref{cmbpolapp}), it exhibits a similar degeneracy between $Y_{\rm P}$ and $G_{\rm matter}/G$ (shown in Fig.~\ref{figclsdamping}). This level of degeneracy also applies to the temperature-polarization cross-spectrum as shown in Appendix~\ref{cmbpolapp}. We therefore note that the uncertainty in the underlying gravitational theory (or the expansion rate more generally) has the potential to complicate inferences of small-scale physics targeted by CMB surveys such as AdvACT~\cite{advact}, SPT-3G~\cite{spt3g}, and the Simons Observatory~\cite{simonsobs}. While the direct measurement of $Y_{\rm P}$ from observations of low-metallicity extragalactic H{\sc ii} regions~\cite{aver15,peimbert16} is able to break its degeneracies with other parameters, the correlation of $G_{\rm matter}/G$ with parameters such as $N_{\rm eff}$ and ${\rm d} n_{\rm s} / {\rm d} \ln k$ would still remain to be disentangled (and for the CMB would be similar in nature to the correlations of a freely-varying $Y_{\rm P}$ with $N_{\rm eff}$ and ${\rm d} n_{\rm s} / {\rm d} \ln k$ in GR). 

In summary, the JBD scalar field will have an impact on a multitude of cosmological observables that we will consider in our analysis, such as the cosmic microwave background temperature and polarization, along with lower-redshift probes of the expansion history and the growth of structure, such as supernova distances, the weak lensing of galaxies, the weak lensing of the CMB, and the clustering of galaxies in redshift space. In testing JBD gravity with the latest cosmological observations, we have implemented this theory in the Einstein-Boltzmann solver \eftcamb~\cite{LCL,hu14eft}, and have performed an extensive comparison with four distinct codes~\cite{Bellini:2017avd}. The level of agreement between the codes is found to be at the sub-percent level for both the matter power spectrum and CMB temperature, polarization, and lensing power spectra, well within the precision required for current observations.

\section{Theory: Nonlinear Regime and $N$-Body Implementation}
\label{nonlinearsec}

\subsection{Background: numerical simulations with JBD gravity}

In order to more fully utilize current cosmological data, we proceed to model the density perturbations in the nonlinear regime. We revisit the equations of motion, and now consider the effect of the scalar field as that of a fifth force. Given $\phi={\bar \phi}+\delta\phi$ in the quasistatic regime (such that $\dot{\delta\phi} / \nabla \delta\phi \ll 1$ and $k^2/(aH)^2 \gg 1$), and considering a constant potential, the scalar field equation of motion (Eq.~\ref{sfeq}) is well approximated by 
\begin{eqnarray}
\frac{1}{a^2}\nabla^2\delta\phi \simeq -\frac{1}{M^{2}_{\rm Pl}}\left({\frac{\delta\rho_{\rm m}}{3+2\omega_{\rm BD}}}\right).
\label{nreqn}
\end{eqnarray}
As $\delta\rho_{\rm m}={\bar \rho}_{\rm m}\delta_{\rm m}$, this implies $\delta\phi/\bar{\phi} = \Psi/(2+\omega_{\rm BD})$. We note that $\Psi \lesssim 10^{-4}$ in a cosmological simulation and given that we are interested in the $\omega_{\rm BD}\gg 1$ regime, we can neglect terms of order $(\nabla \phi)^2/\bar{\phi}^2$ in the Einstein equations. In other words, the standard contribution of the energy density of the scalar field is insignificant as compared to the clustering component of the overall energy density.

As a result, in the {\it N}-body simulations we evolve the non-relativistic geodesic equation (e.g.~\cite{bouchet96}),
\begin{eqnarray}
{\ddot {\bf x}}+2H{\dot {\bf x}}&=& -\frac{1}{a^2}{\bf \nabla} \Psi ,  
\end{eqnarray}
where ${\bf x}$ is the position of each particle and the raised dots are, as before, derivatives with physical time. The geodesic equation is evolved along with the modified Poisson equation, re-expressed here in the form 
\begin{eqnarray} 
\nabla^2\Psi&=&\frac{3}{2}\Omega_{\rm m,0}H^2_0a^{-1}\frac{G_{\rm matter}}{G}\delta_{\rm m} ,
\end{eqnarray}
where we emphasize that the effective gravitational constant, $G_{\rm matter}$, is time-dependent (Eq.~\ref{geqn}). The initial conditions for the particles were generated with the MG-PICOLA code \cite{Winther:2017jof,howlett15} using second-order Lagrangian perturbation theory (2LPT) given a power-spectrum $P(k,z=0)$ from EFTCAMB \cite{hu14eft}. The first and second order growth factors of the density contrast in 2LPT, denoted $D_1$ and $D_2$, are determined by the equations~\cite{Winther:2017jof} 
\begin{eqnarray}
\label{growtheq}
{\ddot D}_1+2H{\dot D}_1&=&\frac{3}{2}\frac{\Omega_{\rm m,0}}{a^3}\frac{G_{\rm matter}}{G} H^2_0D_1, \nonumber\\
{\ddot D}_2+2H{\dot D}_2&=&\frac{3}{2}\frac{\Omega_{\rm m,0}}{a^3}\frac{G_{\rm matter}}{G} H^2_0\left({D_2 - D_1^2}\right),
\end{eqnarray}
which are of the same form as in $\Lambda$CDM \cite{bouchet95} aside from the $G_{\rm matter}/G$ factor.

\subsection{Hybrid suite of $N$-body simulations: COLA and \texttt{RAMSES}}

We have modified two {\it N}-body codes to obtain an accurate measurement of $P(k)$ beyond the linear regime.\footnote{A patch with the modifications to \texttt{RAMSES} can be found in \url{https://github.com/HAWinther/RamsesPatchApproxMGSolver} and the COLA code used can be found in \url{https://github.com/HAWinther/MG-PICOLA-PUBLIC}.} For $k<0.5~h~{\rm Mpc}^{-1}$, we use a modified version of the COmoving Lagrangian Acceleration (COLA) code~\cite{Tassev:2013pn,Winther:2017jof}, which solves for perturbations around paths predicted from 2LPT, and has been shown to be accurate and fast on large scales. This enables the generation of a large enough ensemble of realizations to substantially reduce sample variance on large scales: we generate $50$ realizations with $N=1024^3$ particles in a box of size $L = 1000~h^{-1}~{\rm Mpc}$ (to cover large scales) and $100$ realizations with $N=512^3$ particles in a box of size $L = 250~h^{-1}~{\rm Mpc}$ (to cover small scales). We also use a large number of steps to increase the accuracy on smaller scales ($\sim100$ steps, an order of magnitude more than typical COLA simulations). On very small scales, to probe wavenumbers out to $k=10~h~{\rm Mpc}^{-1}$, we use the \texttt{RAMSES} grid-based hydrodynamical solver with adaptive mesh refinement~\cite{Teyssier:2001cp}, modified to include JBD gravity. For each $\omega_{\rm BD}$, we have generated a higher resolution \texttt{RAMSES} simulation with $N=512^3$ particles in a box of size $L = 250~h~^{-1}~{\rm Mpc}$. 

\begin{figure}
\vspace{-0.05cm}
\includegraphics[width=1.03\columnwidth]{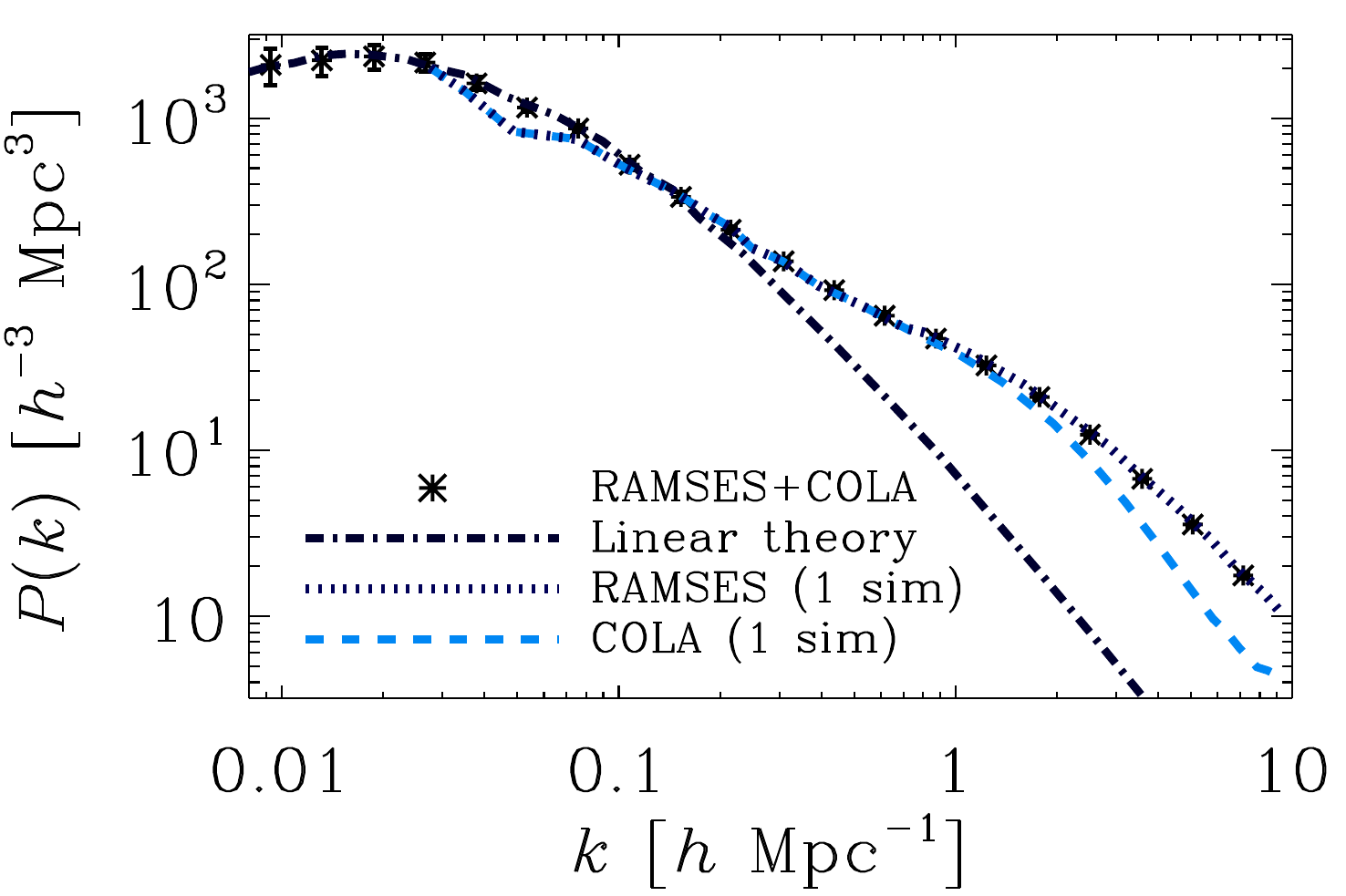}
\vspace{-0.6cm}
\caption{Matter power spectrum $P(k, z=0)$ for $\omega_{\rm BD} = 100$. Here, ``1 sim'' refers to a single realization of the initial conditions for which we run a high-resolution \texttt{RAMSES} simulation (dotted line) in addition to the COLA simulations (dashed line). 
As a result, the dip at $k \sim 0.05~h~{\rm Mpc}^{-1}$ is due to cosmic variance (for both \texttt{RAMSES} and COLA given the same initial conditions). Here, \texttt{RAMSES}+COLA incorporates all of the simulations (some with a larger box size and thereby a smaller minimum $k$) and the error bars denote the 68\% confidence level. For comparison, we also show the linear theory prediction (dot-dashed line) which expectedly agrees with the simulations on large scales but visibly deviates for $k \gtrsim 10^{-1}~h~{\rm Mpc}^{-1}$.
}
\label{fig:pofk}
\end{figure}

The \texttt{RAMSES} simulation is run with the same seed as one of the COLA simulations, chosen by the requirement that it has a $P(k,z)$ as close as possible to the mean of the ensemble of COLA simulations, which ensures that it is not an outlier realization. The COLA simulations are found to agree to 1\% with the \texttt{RAMSES} simulation for $k < 0.5~h~{\rm Mpc}^{-1}$ at $z=0$, and with an improved accuracy towards higher redshifts. For the largest wavenumbers considered here $(k_{\rm max} = 10~h~{\rm Mpc}^{-1})$, with our simulation setup, the \texttt{RAMSES} simulation is accurate to $\sim 5$--$10\%$.\footnote{For a study of the accuracy of \texttt{RAMSES} compared to other {\it N}-body codes, see e.g.~Schneider et al.~(2016)~\cite{schneider16}.} The ratio of the \texttt{RAMSES} and the COLA $P(k,z)$ for the same seed are then used to correct the COLA simulations $P(k,z)$ out to its maximum wavenumber. These simulations are carried out for $\omega_{\rm BD} = \{50,100,500,1000\}$, and we use outputs at $z=\{0,0.5,1.0\}$ as the basis for producing our modifications to the nonlinear matter power spectrum (see Fig.~\ref{fig:pofk}).

\begin{figure*}
\includegraphics[width=1.01\hsize]{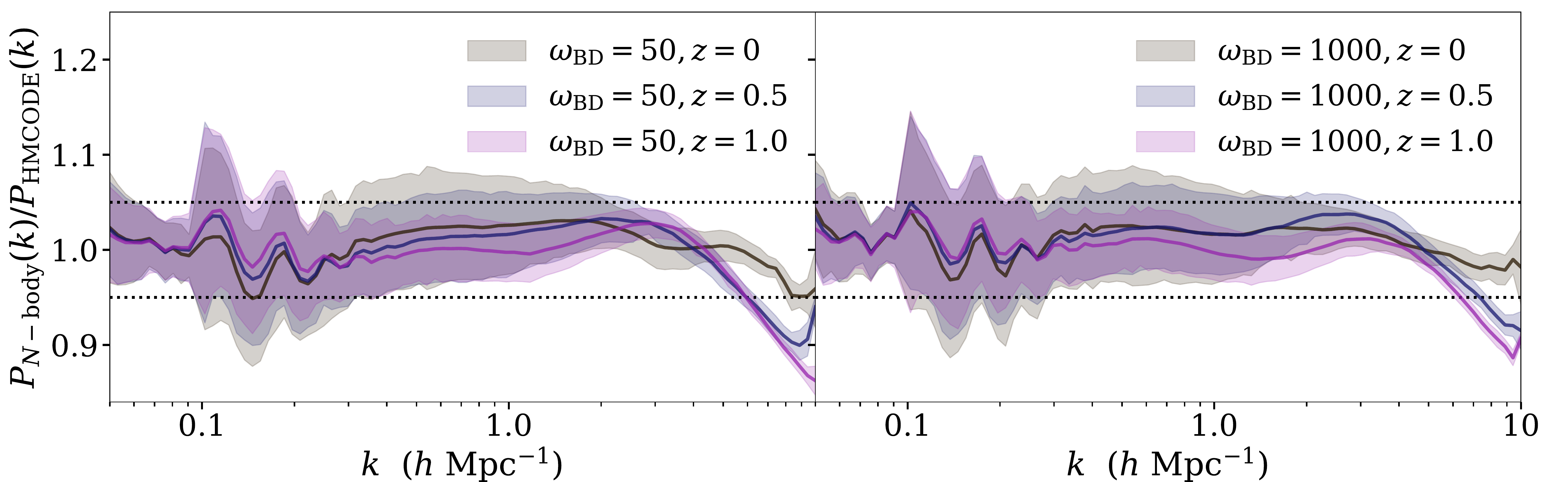}
\vspace{-0.5cm}
\caption{The ratio of matter power spectra between the $N$-body simulations and \hmcode, $P_{N-{\rm body}}(k)/P_{\hmcode}(k)$, for $\omega_{\rm BD} = 50$ (left) and $\omega_{\rm BD} = 1000$ (right) illustrating the accuracy of our modifications to \hmcode. The shaded bands encapsulate the 68\% confidence intervals given the variations within the simulations (for each of the considered redshifts, $z=0$ in grey, $z=0.5$ in light blue, and $z=1.0$ in light violet), the thin colored lines inside the bands correspond to the ratio with \hmcode given the mean of the simulations, and the dotted (black) horizontal lines mark $\pm 5\%$ deviations from unity. The agreement between \hmcode and the simulations improves towards lower redshift by our construction.
}
\label{fig:hmcode}
\end{figure*}

\subsection{JBD gravity modifications to \hmcode}

We include these modifications to the matter power spectrum in \hmcode \cite{Mead15,Mead16} (also see Refs.~\cite{hmcode20,Mead20hmx}), which is a fitting function for the nonlinear matter power spectrum based on the halo model (reviewed in e.g.~\cite{Cooray:2002dia}). For the dark matter power spectrum, \hmcode has been calibrated to the Coyote $N$-body simulations \cite{Coyote4} and is accurate at the level of $5$--$10\%$ for $z \leq 2$ and $k \leq 10~h~{\rm Mpc}^{-1}$ (improving towards lower neutrino mass, where massive neutrinos suppress the clustering of matter below the neutrino free-streaming scale). A benefit of \hmcode over other fitting functions such as \halofit~\cite{Smith03,Takahashi12} is that it has also been calibrated to the OverWhelmingly Large (OWL) hydrodynamical simulations~\cite{Schaye10,Daalen11,semboloni11}. As a result, it is able to capture the impact of baryonic feedback in the nonlinear $P(k,z)$ (at $\simeq5\%$ level precision for the same redshifts and scales) and to marginalize over the uncertainty in the modeling. \hmcode moreover improves its modeling of the impact of massive neutrinos in the nonlinear $P(k,z)$ by calibrating to the massive neutrino simulations of Massara et al.~(2014)~\cite{massara14} (agreement achieved at the few percent level for $m_{\nu} \leq 0.6~{\rm eV}$, $z \leq 1$, $k \leq 10~h~{\rm Mpc}^{-1}$~\cite{Mead16}).

To account for JBD gravity within the \hmcode framework, in addition to modifying the expansion rate, density parameters, and growth function, we follow the approach in Mead et al.~(2016)~\cite{Mead16} and adjust the virialized halo overdensity, $\Delta_{\rm V}$, defined through
\begin{eqnarray}
r_{\rm V} = \left(\frac{3M}{4\pi\Delta_{\rm V}{{\bar \rho}_{\rm m}}}\right)^{\frac{1}{3}} ,
\end{eqnarray}
where $r_{\rm V}$ is the corresponding virial radius and $M$ is the halo mass. We find that a good fit that accounts for JBD gravity is given by 
\begin{eqnarray}
\Delta_{\rm V} &=& \Omega_{\rm m}(z)^{-0.352}\times\\
&&\left(d_0 + (418.0 - d_0) \arctan\left((0.001|\omega_{\rm BD}-50.0|)^{0.2} \right)\frac{2}{\pi}\right),\nonumber
\end{eqnarray}
where $d_0 = 320.0 + 40.0z^{0.26}$. Our fitting function has the desired property that as $\omega_{\rm BD}\to\infty$ we recover the usual GR prescription in \hmcode.\footnote{We also considered $\omega_{\rm BD}$-dependent modifications to the linear collapse threshold, $\delta_{\rm c}$, and the linear spectrum damping factor, $f_{\rm damp}$, and found that they only marginally improve the \hmcode fit to the JBD simulations.} We note that baryonic feedback does not modify the fitting function for the virialized halo overdensity in \hmcode; instead it modifies the halo concentration-mass relation and the amount of halo ``bloating''~\cite{Mead15}. However, to account for the impact of massive neutrinos, the fitting function is further multiplied by a factor of $1+0.916f_{\nu}$~\cite{Mead16}, where $f_{\nu} = \Omega_{\nu}/\Omega_{\rm m}$.\footnote{In accounting for the impact of massive neutrinos on the nonlinear matter power spectrum, we note that \hmcode further modifies the linear collapse threshold, $\delta_{\rm c}$, by a factor of $1+0.262f_{\nu}$~\cite{Mead16}.} 

\subsection{JBD matter power spectrum}

In Fig.~\ref{fig:hmcode}, we show the ratio of $P(k)$ measured from our simulations to the modified \hmcode for $\omega_{\rm BD}$ of $50$ and $1000$ considering $z\in\{0, 0.5, 1.0\}$ and $k\leq10~h~{\rm Mpc}^{-1}$. We find that the agreement is within $\sim10\%$ for the full range of nonlinear scales and redshifts, and across the wide range of coupling strengths, which is sufficient for our purposes given the precision of current data. The agreement is achieved through the \hmcode modifications, and would substantially degrade in their absence (for instance, by up to a factor of three for $\omega_{\rm BD} = 100$ over the range of scales and redshifts considered).

As $P_{\omega_{\rm BD} = 1000}(k,z)/P_{\rm GR}(k,z) \approx 1$ (at the sub-percent level across scales and redshifts, as also indicated in Fig.~\ref{figcls}), together the $\omega_{\rm BD} = 50$ and $\omega_{\rm BD} = 1000$ cases effectively show that our \hmcode implementation has the same level of accuracy in JBD gravity as in GR (which is the optimal outcome). We also employed simulations with $(G_{\rm matter}/G)|_{a=1} \in [0.5, 1.5]$; as \hmcode includes the effects of this degree of freedom via the linear matter power spectrum (through an overall rescaling of the amplitude, as seen in Fig.~\ref{figcls}), we found that no further \hmcode modifications were needed to fit the simulations on nonlinear scales (with agreement due to changing $(G_{\rm matter}/G)|_{a=1}$ to within $5\%$ for $k\leq10~h~{\rm Mpc}^{-1}$).

\subsection{Alternative approaches to capturing nonlinear corrections}

For the precision needs of future surveys, this approach to capturing the nonlinear corrections can be improved through both a greater number of simulations and higher-resolution simulations taking into account the full cosmology dependence (instead of fitting for a fixed cosmology, and in the \hmcode context modifying parameters in addition to $\Delta_{\rm V}$ entering the halo model). We note that an alternative approach consists of creating a full simulation-based emulator as has been achieved in $\Lambda$CDM (e.g \cite{lawrence10,knabenhans19,angulo20}). Given the simplicity of the JBD model (where a JBD simulation has the same computational cost as a standard GR simulation), an emulator accounting for JBD gravity is feasible but computationally expensive. 

In this regard, the ``reaction approach'' of Cataneo et al.~(2019) \cite{cataneo19} (also see Bose et al.~(2020) \cite{bose20}) is promising, as it only requires a computation of halo model and 1-loop perturbation theory power spectra in the modified gravity cosmology together with an emulator for $\Lambda$CDM, and has already demonstrated percent-level accuracy on highly nonlinear scales for more complicated modified gravity models. Ultimately, an emulator needs to be able to {\it simultaneously} account for a wide range of physics, such as cold dark matter, massive neutrinos, modified gravity, and baryonic feedback, which is increasingly within reach.\footnote{We note that a separate emulator for baryonic feedback based on the `baryonic correction model'' \cite{schneider19} has been considered in Refs.~\cite{schneider20a, schneider20b}.}

\subsection{Including physical effects: independently versus combined}

While we have considered changes to \hmcode by calibrating to simulations that only include a single extension to $\Lambda$CDM (i.e.~simulations that either account for baryonic feedback, neutrino mass, or modified gravity, but not simultaneously), we note that their effects are propagated in a coherent way in \hmcode through changes to the expansion rate, density parameters, growth function, virialized halo overdensity, the linear collapse threshold, concentration-mass relation, and halo bloating. In a sense, therefore, \hmcode is simultaneously accounting for the impact of baryonic feedback, neutrino mass, and modified gravity (also see the discussion in Ref.~\cite{Mead16}). 

We note that the differences are small ($\lesssim5\%$ even for $k=10~h~{\rm Mpc}^{-1}$) compared to the case where the effects of baryonic feedback, neutrino mass, and modified gravity are separately propagated in the nonlinear matter power spectrum~\cite{Mead16}. This is in agreement with a range of simulations where baryons and modified gravity~\cite{puchwein13,al19,ha20}, baryons and massive neutrinos~\cite{mummery17,mccarthy18}, and modified gravity and massive neutrinos~\cite{baldi14,giocoli18,wkwz19} have been simultaneously considered. Lastly, we emphasize that the question of separability of the physical effects is distinct from the question of degeneracies between the effects; as shown in Fig.~\ref{figcls} it might indeed be a challenge to distinguish the physical imprints of baryonic feedback, massive neutrinos, and modified gravity from one another, which we further explore in this analysis.

\section{Analysis: Model requirements, datasets, systematic uncertainties, priors, \\data concordance}
\label{analysissec}

In our cosmological analysis, we perform Markov Chain Monte Carlo (MCMC) calculations with a modified version of \cosmolss~\cite{Joudaki:2016mvz,Joudaki:2017zdt}, which is based on \camb~\cite{LCL,hu14eft} and \cosmomc~\cite{Lewis:2002ah}. We use a convergence criterion that obeys $R-1 < 2 \times 10^{-2}$, where the Gelman-Rubin $R$-statistic~\cite{Gelman92} is the variance of chain means divided by the mean of chain variances. Here, we describe the cosmological datasets used, systematic uncertainties included, parameter priors enforced, approaches for performing model selection and assessing dataset concordances, and our requirements on extended models.

\subsection{Requirements on extended models: what must be satisfied to replace the standard model?}
\label{requirementssec}

As we explore the viability of JBD gravity with current cosmological datasets, we highlight three requirements that need to be {\it simultaneously} satisfied for an extended cosmological model to be considered as a genuine alternative to the standard $\Lambda$CDM model. In devising these requirements, we take the approach that the ``burden of proof'' is on the extended model. We list these requirements below:

\begin{enumerate}

\item{{\it The extended model needs to be strongly favored in a model selection sense.} This can be assessed by considering statistical measures such as the Bayesian evidence, goodness of fit, and deviance information criterion (see Sec.~\ref{modelselec}). The criterion for what constitutes ``strongly favored'' can be determined through for instance Jeffreys' scale (Sec.~\ref{modelselec}).}

\item{{\it The extended model needs to exhibit a $5\sigma$ or greater deviation in the additional parameters that are introduced} (with respect to their values in the standard model limit). As an example, this imposes the requirement that the effective gravitational constant $G_{\rm matter}/G$ rules out the GR value of unity at this statistical significance. The choice of $5\sigma$ is motivated by the particle physics gold standard.}

\item{{\it The extended model needs to stay robust when considering additional data.} In other words, it is not sufficient for the extended model to satisfy the two requirements above for only a subset of established cosmological data (such as the CMB alone), but rather it needs to satisfy the requirements as additional established datasets are simultaneously considered in the analysis (such as the CMB together with BAO and supernova distances, and in other regimes probed by for example black holes and gravitational waves). 
As part of this condition we include the ideal scenario in which the extended model makes a successful observational prediction that differs from the standard model expectation at $5\sigma$ or greater statistical significance (i.e.~going beyond the second requirement above by predicting the change for a new probe).}

\end{enumerate}

Further, there is an optional but strongly desired feature that we seek in an extended model:

\begin{enumerate}\setcounter{enumi}{3}

\item{{\it The underlying cosmology favored by different probes are in agreement within the extended model.} This is a particularly desired feature in the event of discordances between cosmological probes in the standard model. The ability of the extended model to bring about concordance among probes ideally, but not necessarily, holds for all probes (in other words, the ability to solve both the $H_0$ and $S_8 \equiv \sigma_8 \sqrt{\Omega_{\rm m}/0.3}$ tensions rather than only one of the two). We consider this to be an optional feature of the extended model to allow for the possibility that remaining discordances between probes are caused by unaccounted systematic uncertainties in the data (which admittedly requires careful consideration as unaccounted systematic uncertainties would also have an effect on model selection and extended model parameter preferences). Conversely, if the extended model worsens tensions among probes, this would raise strong concerns about its viability (especially if the first three conditions are not already satisfied).
}

\end{enumerate}

As an example of an extended model that illustrates these conditions, we refer to evolving dark energy as expressed by the equation of state $w(a) = w_0 + (1-a)w_a$, where $w_0$ and $w_a$ are the additional free parameters of the model. In considering cosmic shear data from the Kilo Degree Survey (KiDS) and CMB temperature and polarization data from the Planck satellite, in Ref.~\cite{joudaki17} this model was found to completely resolve the tension between these datasets (as further manifested through the $S_8$ parameter). This allowed for the combined analysis of KiDS and Planck which favored a deviation in the extended parameters at $3\sigma$ statistical significance. The combined data moreover found the extended model to be moderately favored relative to $\Lambda$CDM from a model selection standpoint. 

However, in Ref.~\cite{Joudaki:2017zdt}, this promising model was revisited in the context of additional galaxy-galaxy lensing and redshift-space galaxy clustering data from KiDS overlapping with the 2-degree Field Lensing Survey and the Baryon Oscillation Spectroscopic Survey. Here, the additional probes reintroduced the $S_8$ tension and thereby prevented the combined analysis of these datasets with the Planck CMB. This in turn lowered the statistical significance of the deviation of the extended ($w_0, w_a$) parameters from their standard model expectation and removed the model selection preference relative to $\Lambda$CDM. Hence, the extended model visibly failed the third requirement above and was no longer considered to be a promising candidate. 

Our past experiences are reflected in the conditions described here, which illustrate that it is not straightforward to replace the standard model (by construction as the burden of proof is on the extended model and the burden itself is high). We use these conditions to assess the viability of JBD gravity in forthcoming sections.

\subsection{Cosmological datasets}
\label{datasec}

We consider the following datasets in our cosmological analysis, either separately or in combination.

\subsubsection{Cosmic microwave background}
\label{cmbsec}

We consider the CMB temperature, polarization, and lensing reconstruction angular power spectra from the Planck satellite. We consider the 2018 dataset of Planck~\cite{planck2018}, and in some cases also contrast the differences in the parameter constraints with the 2015 dataset of Planck~\cite{planck2015}. 

We distinguish between two setups for Planck 2018, one that we denote ``Planck18'' which includes the CMB temperature and polarization data (TT,TE,EE+lowE;~where the low-multipole polarization is obtained from the High Frequency Instrument, HFI), and another that we denote ``All-Planck18'' which additionally includes the lensing reconstruction (TT,TE,EE+lowE+lensing). The corresponding two cases for Planck 2015 are ``Planck15'' which includes the CMB temperature and low-$\ell$ polarization data (TT+lowP;~where the low-multipole polarization is obtained from the Low Frequency Instrument, LFI), along with ``All-Planck15'' which additionally includes the $TE$ and $EE$ spectra at high multipoles together with lensing reconstruction (TT,TE,EE+lowP+lensing).

In addition to Planck, we consider the improvements in the parameter constraints from a combined analysis with the small-scale CMB temperature and polarization measurements from the $2008$--$2018$ observing seasons of the Atacama Cosmology Telescope (ACT;~primarily DR4~\cite{aiola20} but also contrasting against DR3~\cite{actpol17}). We note that the covariance between Planck and ACT is not included, which is estimated to be adequate at the $\lesssim5\%$ level in $\Lambda$CDM (and certain one-parameter extensions) following the scale cuts imposed on the ACT temperature power spectrum~\cite{aiola20}. As a precaution, therefore, we only consider the combined analysis of Planck and ACT in distinct cases.

For the primary CMB anisotropies, JBD gravity provides a shift in the locations of the peaks due to the change in the expansion history (as seen in Fig.~\ref{figcls} and Fig.~\ref{figclspol}; where the JBD expansion history directly modifies quantities such as the epoch of matter-radiation equality and angular size of the sound horizon at recombination~\cite{ck99,Clifton:2011jh, liddle98}). By modifying the time-variation in the gravitational potentials, ${{\rm d} \over {\rm d}t}(\Psi+\Phi)$, JBD gravity in principle also affects the lowest multipoles in the CMB temperature power spectrum (and cross-correlations) via the integrated Sachs-Wolfe (ISW) effect \cite{isw67}. The impact of JBD gravity on the CMB due to the ISW effect, however, is statistically diminished due to cosmic variance (along with the generally mild evolution of the scalar field). The CMB lensing potential power spectrum~\cite{planck2015lensing,planck2018lensing}, $C_{\ell}^{\phi\phi}(\theta)$, is affected by JBD gravity through modifications to the sum of the metric potentials $\left({\Psi+\Phi}\right)$ and the expansion history, and to lesser extent through the growth function, as described in Sec.~\ref{theorysec} (also see Ref.~\cite{planckmg15} on the impact of modified gravity on the CMB).

\subsubsection{Weak gravitational lensing tomography and overlapping redshift-space galaxy clustering}
\label{sec3x2pt}

Following the analysis of Joudaki et al.~(2018)~\cite{Joudaki:2017zdt}, we consider measurements of cosmic shear, galaxy-galaxy lensing, and redshift-space galaxy clustering from the Kilo Degree Survey (KiDS-450)~\cite{kuijken15,Hildebrandt16} overlapping with the 2-degree Field Lensing Survey~\cite{blake16} (2dFLenS) and the Baryon Oscillation Spectroscopic Survey (BOSS DR10\footnote{We note that subsequent BOSS and eBOSS data releases~\cite{ross20} do not contain additional observations in the KiDS footprint.})~\cite{dawson13}. The measurements are given by $\{\xi_+^{ij}, \xi_-^{ij}, \gamma_{\rm t}^j, P_0, P_2\}$, where $\xi_{\pm}^{ij}(\theta)$ are the tomographic two-point shear correlation functions (for bins $i$ and $j$, and angular scales $\theta$), $\gamma_{\rm t}^i(\theta)$ is the tomographic galaxy-galaxy lensing angular cross-correlation function, and $P_{0/2}(k)$ are the monopole and quadrupole power spectra \cite{Joudaki:2017zdt}. We emphasize that the galaxy-galaxy lensing and multipole power spectrum measurements are only considered in the overlapping areas with KiDS.

This data vector is constructed from four tomographic bins of source galaxies in the redshift range $0.1<z_{\rm B}<0.9$, where $\Delta z_{\rm B}$ = 0.2 and $z_{\rm B}$ is the best-fit redshift by the Bayesian photometric redshift code \bpz~\cite{benitez2000}. Moreover, there are four samples of lens galaxies: \{2dFLOZ, BOSS~LOWZ\} which cover the redshift range $0.5 < z < 0.43$ and \{2dFHIZ, BOSS~CMASS\} which cover the redshift range $0.43 < z < 0.7$. We include the full covariance between these observables using numerical simulations, as detailed in Ref.~\cite{Joudaki:2017zdt}. This covariance assumes a fixed $\Lambda$CDM cosmology, which is a sufficient approximation to current data~\cite{kodwani19}.\footnote{As shown in Ref.~\cite{kodwani19}, the bias due to the parameter dependence of the covariance decreases as the number of modes in a given survey increases, and is found to be at most $0.2\sigma$ for combined analyses of weak lensing and galaxy clustering even for surveys targeting less than 1\% of the sky or very large scales (multipoles $\ell<20$). One approach to account for the ``wrong'' cosmology of the covariance is to infer the best-fit cosmology, and then use it to obtain an updated covariance that enters a new run, iteratively until there is convergence between the two cosmologies. For the particular case of KiDS, this has been shown to impact the parameter constraints at the level of $<0.1\sigma$~\cite{hildebrandt19, asgari20b}. While the impact of the fixed covariance might increase in the case of modified gravity, we still expect it to be insignificant (additionally given the overall small deviations from GR).}

KiDS-450~\cite{Hildebrandt16} encompasses $360$~deg$^2$ on the sky, contains an effective number density $n_{\rm eff} = 8.5$ galaxies arcmin$^{-2}$, possesses a median source redshift $z_{\rm m} = 0.53$, and yields similar cosmological constraints to the subsequent analysis of the KiDS+VIKING-450 cosmic shear dataset~\cite{hildebrandt19}. The cosmological constraints are further in close agreement, both in terms of the posterior mean and uncertainty, with those obtained from the cosmic shear dataset of the Dark Energy Survey (DES-Y1~\cite{desy1shear}) when considered in a homogenized analysis setup as shown in Ref.~\cite{joudaki20}. We consider both KiDS$\times$2dFLenS and KiDS$\times$\{2dFLenS+BOSS\}~\cite{Joudaki:2017zdt} as described here, in order to explore the capabilities of a ``$3 \times 2{\rm pt}$'' dataset where the galaxy clustering is restricted to the overlapping regions with KiDS, as compared to a combined analysis of KiDS with the full spectroscopic datasets (where the impact of galaxy-galaxy lensing becomes negligible given the substantial difference in current imaging and spectroscopic observing areas). We do not consider the recent KiDS-1000 cosmic shear dataset~\cite{asgari20b} as it was not available during the course of this work, and we do not consider the combined cosmic shear dataset of KiDS+VIKING-450 and DES-Y1~\cite{joudaki20}, with similar constraining power to KiDS-1000.

The wide spectroscopic surveys of \{BOSS, 2dFLenS\} contain \{$1\times10^6$, $4\times10^4$\} galaxies over \{$1\times10^4$, $7\times10^2$\} deg$^2$ on the sky, respectively \cite{dawson13,blake16}. We restrict both of these surveys to the overlapping regions with KiDS-450, such that in these regions LOWZ contains 5044 lens galaxies over 125.0 deg$^2$, CMASS contains 20476 lens galaxies over 221.7 deg$^2$, 2dFLOZ contains 2214 lens galaxies over 122.4 deg$^2$, and 2dFHIZ contains 3676 lens galaxies over 122.4 deg$^2$ (also see below for use of the full BOSS dataset)~\cite{Joudaki:2017zdt}.

To avoid nonlinearities in the galaxy bias, we restrict the galaxy-galaxy lensing and redshift-space galaxy clustering measurements to linear scales, such that $\theta_{\rm min} = 12~{\rm arcmin}$ for galaxy-galaxy lensing and $k_{\rm max} = 0.125~h~{\rm Mpc}^{-1}$ for the multipole power spectra in accordance with the ``fiducial'' analysis in Ref.~\cite{Joudaki:2017zdt} (see Table 2 therein). The cosmic shear measurements are allowed to extend into the nonlinear regime, and correspond to the same scale cuts as in the fiducial KiDS-450 analyses~\cite{Hildebrandt16,Joudaki:2017zdt}.

Analogous to CMB lensing, the galaxy lensing $\xi^{ij}_{\pm}(\theta)$ and galaxy-galaxy lensing $\gamma^j_{\rm t}(\theta)$ measurements are mainly sensitive to JBD gravity via the modifications to the expansion history and the auto and cross-spectrum of $\Psi+\Phi$ (i.e.~we integrate over the Weyl power spectrum rather than the matter power spectrum for the weak lensing and galaxy-galaxy lensing calculations; there is also a minor contribution to the intrinsic galaxy alignments through the growth function). In turn, the multipole power spectra $P_{0/2}$ are sensitive to JBD gravity via the expansion history, matter power spectrum, and the growth rate (and thereby the potential $\Psi$), discussed in Sec.~\ref{theorysec}.

\subsubsection{Growth rate, baryon acoustic oscillations, and Alcock-Paczynski effect}
\label{bossdatasec}

In order to more fully utilize the statistical power of BOSS, we also consider the ``final BAO+FS consensus'' constraints on $\{f\sigma_8, D_{\rm V}/r_{\rm d}, F_{\rm AP}\}$ from the BOSS DR12 dataset~\cite{alam17} (i.e.~not restricted to the overlapping regions with KiDS-450).\footnote{We do not use the eBOSS DR16 dataset \cite{ross20,alam20} as the likelihood has not yet become publicly available. This dataset will allow for higher redshifts to be probed than considered here (i.e.~$z>0.7$), in particular through the higher-redshift growth rate measurements (as the impact of JBD gravity on BAO distances diminishes with redshift).} Here, we use the distance scale, $D_{\rm V}(z) = \left[D_{\rm M}^2(z) {{cz} \over {H(z)}}\right]^{1/3}$,~where $D_{\rm M}(z)$ is the comoving angular diameter distance, along with the comoving size of the sound horizon, $r_{\rm d} = \int_{z_{\rm d}}^{\infty} c_s(z)/H(z) dz$, at the end of the baryon-drag epoch, $z_{\rm d}$, where $c_s(z)$ is the sound speed, and the Alcock-Paczynski parameter $F_{\rm AP}(z) = D_{\rm M}(z) H(z)/c$.

In combining these measurements with the cosmic shear, galaxy-galaxy lensing, and overlapping redshift-space galaxy clustering measurements of Sec.~\ref{sec3x2pt}, we avoid double-counting BOSS by only including the KiDS$\times$2dFLenS measurements (i.e.~excluding the KiDS$\times$BOSS galaxy-galaxy lensing and multipole power spectrum measurements). Given the significantly larger amount of BOSS data outside of the overlapping regions with KiDS (in our case by more than a factor of $50$), we expect this trade-off to increase the cosmological constraining power. We do not include a covariance between the BOSS $\{f\sigma_8, D_{\rm V}/r_{\rm d}, F_{\rm AP}\}$ measurements with the galaxy-galaxy lensing and galaxy clustering measurements involving 2dFLenS as the two datasets encompass distinct areas on the sky. To distinguish these BOSS measurements from those restricted to the overlapping regions with KiDS, we will refer to these as ``All-BOSS'' (given their use of the full BOSS dataset).

As discussed in Sec~\ref{theorysec}, JBD gravity directly modifies these observables through its impact on the growth rate and expansion history (and thereby distances and sound horizon).

\subsubsection{Supernovae}
\label{snsec}

To further improve the cosmological constraining power, we consider Type Ia supernova (SN) distance measurements from the Pantheon compilation~\cite{pantheon18} in some of our calculations. This compilation contains a total of 1048 SNe between $0.01 < z < 2.3$ from Pan-STARRS1, SDSS, SNLS, various low-$z$ and HST samples, and constitutes the largest combined sample of SN Ia. We use these SN distances to constrain JBD gravity through the impact of the expansion history (described in Sec.~\ref{theorysec}).

\subsection{Systematic uncertainties}
\label{systsec}

\subsubsection{Cosmic microwave background (temperature, polarization, lensing reconstruction)}

In obtaining unbiased cosmological results, we account for the systematic uncertainties affecting the measurements. For the Planck \{2018, 2015\} CMB temperature and polarization power spectra, this includes marginalizing over $\{21, 27\}$ astrophysical foreground and instrumental modeling parameters in the MCMC analysis, respectively ($15$ nuisance parameters for TT alone and $\{6, 12\}$ additional parameters when further including TE+EE; in particular due to galactic dust emission, the cosmic infrared background, the thermal and kinetic Sunyaev-Zel'dovich effects, radio and infrared point sources, and power spectrum calibration uncertainties)~\cite{planck2018like,planck2015like}. The difference in the number of nuisance parameters between the 2018 and 2015 datasets of Planck is that the dust amplitudes in EE are fixed from the cross-correlations with the 353-GHz maps in the 2018 analysis~\cite{planck2018like}. When including the Planck CMB lensing power spectrum, the map-based calibration parameter is varied in the analysis, which already belongs to the above set of CMB nuisance parameters~\cite{planck2015lensing,planck2018lensing}.

Moreover, in combining Planck with ACT, one additional calibration parameter is varied in the analysis, which linearly scales the estimated ACT temperature-polarization cross-spectrum and quadratically scales the polarization auto-spectrum~\cite{actpol17,aiola20}.

\subsubsection{3$\times$2{\rm pt} (cosmic shear, galaxy-galaxy lensing, multipole power spectra)}

For the KiDS$\times$\{2dFLenS+BOSS\} cosmic shear, galaxy-galaxy lensing, and redshift-space galaxy clustering observables, we include uncertainties due to the intrinsic galaxy alignments, baryonic feedback, photometric redshift distributions, multiplicative shear calibration, galaxy bias, pairwise velocity dispersion, and non-Poissonian shot noise in our analysis in accordance with the treatment in Joudaki et al.~(2018)~\cite{Joudaki:2017zdt}. This introduces 14 additional parameters that are varied in the MCMC calculations (1 from intrinsic galaxy alignments, 1 from baryonic feedback, 4 from galaxy bias, 4 from velocity dispersion, 4 from non-Poissonian shot noise). 

The intrinsic galaxy alignments encapsulate correlations of the intrinsic ellipticities of galaxies with themselves and with the shear of background sources (i.e.~the lensing two-point functions are constructed from correlations of the sum of shear and intrinsic ellipticity). We modify the strength of the intrinsic alignment signal by allowing for the amplitude $A_{\rm IA}$ in the ``nonlinear linear alignment (NLA) model''~\cite{Hirata:2004gc,Bridle:2007ft} to vary freely (noting that the results are robust to the redshift dependence~\cite{Joudaki:2017zdt}), which affects the theoretical estimates for cosmic shear and galaxy-galaxy lensing. We note that the modeling of the intrinsic alignments is assumed to be the same as in GR (i.e.~we have not modified the NLA model), and further work is required to understand how separable the modeling of this systematic uncertainty is from physics such as modified gravity and massive neutrinos. However, as our parameter constraints tend to be consistent with GR (and zero neutrino mass; shown in Secs.~\ref{wlgcsec},~\ref{distsec},~\ref{fullsec}), any correction to the NLA model is expected to be small (as in a sense it is a ``correction to a correction'').

We propagate the uncertainties in the photometric redshift distributions of the source samples by performing each MCMC calculation over 1000 bootstrap realizations of the redshift distributions (until convergence)~\cite{Hildebrandt16}, which does not introduce additional parameters and affects only the cosmic shear and galaxy-galaxy lensing estimates (we do not allow for uncertainties in the spectroscopic redshift distributions used for the lens samples). We assume a linear galaxy bias for each lens sample (2dFLOZ, 2dFHIZ, LOWZ, CMASS), motivated by our linear scale cuts for galaxy-galaxy lensing and redshift-space galaxy clustering (described in Sec.~\ref{sec3x2pt}), which introduces 4 additional parameters that are varied. The same holds for the velocity dispersion and non-Poissonian shot noise, where a parameter is introduced for each lens sample, leading to 8 additional parameters that modify the theoretical estimates for the multipole power spectra. The uncertainties in the multiplicative shear calibration are propagated via the covariance matrix as in Hildebrandt et al.~(2017)~\cite{Hildebrandt16} and does not introduce additional degrees of freedom.

Lastly, we allow for the baryonic feedback amplitude, $B$, to vary via \hmcode (as discussed in Sec.~\ref{nonlinearsec}), which modifies the halo concentration-mass relation (and further modifies the halo bloating parameter $\eta_0$ through the relationship given in Ref.~\cite{Joudaki:2017zdt}). We place particular emphasis on baryon feedback given the degeneracy with modified gravity and neutrino mass through their impact on the nonlinear matter power spectrum (as shown in Fig.~\ref{figcls};~also see~Refs.~\cite{Mead16,chisari19}), Here, we have assumed that calibrating \hmcode to simulations that separately include baryonic physics, massive neutrinos, and modified gravity provides a close approximation to one where it is calibrated to simulations that simultaneously include these effects (see Sec.~\ref{nonlinearsec} for a discussion). As different approaches to simulating baryonic feedback lead to quantitatively different predictions for the nonlinear matter power spectrum~\cite{huang19}, we aim to capture this uncertainty by our wide prior on the feedback amplitude ($1<B<4$) in accordance with Ref.~\cite{Joudaki:2017zdt} (effectively ``washing out'' some of the information in the nonlinear regime). Here, $B = 3.13$ corresponds to ``dark matter only'' (i.e.~no feedback), while $B = 2.0$ corresponds to the AGN case of the OWL simulations~\cite{Mead15,Daalen11}.\footnote{We note that the AGN case of the OWL simulations is quoted as corresponding to $B = 2.3$ in Mead et al.~(2015)~\cite{Mead15}, but it is given by $B = 2.0$ following the updated $\eta_0$--$B$ parameterization in Ref.~\cite{Joudaki:2017zdt}.} 

\subsubsection{BAO distances, Alcock-Paczynski effect, and growth rate}

For the BOSS growth rate and BAO measurements $\left({f\sigma_8, D_{\rm V}/r_{\rm d}, F_{\rm AP}}\right)$, survey-related systematics are propagated into the galaxy weights (e.g.~due to redshift failures, fibre collisions, and dependencies between the number density of observed galaxies and stellar density and seeing) and modeling systematics are propagated into the covariance matrix (e.g.~due to differences between different pre-reconstruction and post-reconstruction measurement approaches, modeling of redshift space distortions and galaxy bias, and differences in covariance matrix approaches) \cite{anderson14,vargas18,alam17}. As a result, no additional parameters are varied in the MCMC analysis when these BOSS measurements are considered.

We emphasize that the KiDS$\times$BOSS galaxy-galaxy lensing and multipole power spectrum measurements are not used together with these BOSS growth rate and BAO measurements to avoid double-counting the BOSS data (i.e.~we restrict to combining KiDS$\times$2dFLenS with only either of the two, as described in Sec.~\ref{bossdatasec}). We also note that our Jordan-Brans-Dicke gravity analysis brings about secondary systematic uncertainties, as for instance the redshift-space distortion and galaxy bias modeling and reconstruction methods have not been adequately tested in the context of modified gravity, but we expect that these uncertainties are subdominant to the present statistical uncertainties.

\subsubsection{Supernova distances}

For the Pantheon supernova distances, uncertainties due to e.g.~calibration, distance bias corrections, coherent flow corrections, and Milky Way extinction corrections are included in the analysis~\cite{pantheon18}. These uncertainties are not propagated via additional parameters in the MCMC, but instead through a correction to the covariance matrix (i.e.~by adding a systematic covariance matrix to the statistical covariance matrix).

\begin{table}
\begin{center}
\vspace{-0.55em}
\caption{Priors on the cosmological and astrophysical parameters varied in the MCMC runs (excluding the CMB nuisance parameters). Here, ``JBD potential density'' refers to the density parameter of the scalar potential (see e.g.~Eq.~\ref{rhophi}), and the Hubble constant is a derived parameter. While varied independently, we impose the same prior ranges on the galaxy bias, pairwise velocity dispersion, and shot noise for all four lens samples (2dFLOZ, 2dFHIZ, LOWZ, CMASS) in the overlapping regions with KiDS. Here, our $\Lambda$CDM limit corresponds to $\ln \omega_{\rm BD}^{-1} = -17$ and $G_{\rm matter}/G = 1.0$ (we have confirmed the agreement in our results between this choice and pure $\Lambda$CDM). We note that the effective gravitational constant parameter is defined at present and ``$G_{\rm matter}/G$'' is here shorthand for $(G_{\rm matter}/G)|_{a=1}$ (see~Eq.~\ref{jbdpsi}).
}
\begin{tabular}{lll}
\toprule
Parameter & Symbol & Prior\\
\midrule
Cold dark matter density & $\Omega_{\mathrm c}h^2$ & $[0.001, 0.99]$\\
Baryon density & $\Omega_{\mathrm b}h^2$ & $[0.013, 0.033]$\\
JBD potential density & $\Omega_{V}h^2$ & $[0.01, 0.99]$\\
Amplitude of scalar spectrum & $\ln{(10^{10} A_{\mathrm s})}$ & $[1.7, 5.0]$\\
Scalar spectral index & $n_{\mathrm s}$ & $[0.7, 1.3]$\\
Optical depth & $\tau$ & $[0.01, 0.8]$ \\
Dimensionless Hubble constant & $h$ & $[0.4, 1.0]$ \\
Pivot scale $[{\rm{Mpc}}^{-1}]$ & $k_{\rm pivot}$ & 0.05 \\
\midrule
Intrinsic alignment amplitude & $A_{\rm IA}$ & $[-6, 6]$\\
Baryonic feedback amplitude & $B$ & $[1, 4]$\\
Linear galaxy bias & $b_x$ & $[0, 4]$\\
Velocity dispersion $[h^{-1}{\rm Mpc}]$ & $\sigma_{{\rm v},x}$ & $[0, 10]$\\
Non-Poissonian shot noise $[h^{-1}{\rm Mpc}]^3$ & $N_{{\rm shot},x}$ & $[0, 2300]$\\
\midrule
Sum of neutrino masses [eV] & $\sum m_\nu$ & $[0.06, 10]$\\
JBD coupling constant & $\ln \omega_{\rm BD}^{-1}$ & $[-17, -2.3]$\\
Effective gravitational constant & $G_{\rm matter}/G$ & $[0.5, 2.0]$\\
\bottomrule
\end{tabular}
\label{tabpriors}
\end{center}
\end{table}

\subsection{Parameter priors}
\label{priorsec}

We consider uniform priors on the standard cosmological parameters, with ranges given in Table~\ref{tabpriors} (similar to Ref.~\cite{Joudaki:2017zdt}). Concretely, this model includes the 6 standard, or ``vanilla'', cosmological parameters: the present cold dark matter density parameter, $\Omega_{\rm c} h^2 \in [0.001,0.99]$, the present baryon density parameter, $\Omega_{\rm b} h^2 \in [0.013,0.033]$, the present density parameter of the flat JBD potential which constitutes a component of the full JBD density in Eq.~(\ref{rhophi}), $\Omega_{V}h^2 \in [0.01, 0.99]$ (varied in lieu of the approximation to the angular size of the sound horizon, $\theta_{\rm MC}$), the amplitude of the scalar spectrum, $\ln(10^{10}A_{\rm s}) \in [1.7,5.0]$, the scalar spectral index, $n_{\rm s} \in [0.7,1.3]$, and the optical depth to reionization, $\tau \in [0.01,0.8]$. Here, the informative uniform prior on the baryon density is motivated by measurements of the primordial deuterium abundance~\cite{cyburt16,pdg19}, where our prior range encapsulates the $10\sigma$ uncertainties.

In our fiducial model, we fix the neutrinos to be massless. We note that this massless neutrino setup yields similar results to one where the sum of neutrino masses is fixed to $\sum m_{\nu} = 0.06$~eV (for existing datasets) and is chosen for increased computational speed. However, we also consider a setup where the sum of neutrino masses is varied, where we impose a uniform prior on $\sum m_{\nu} \in [0,2.0]~\rm eV$. We use the pivot scale $k_{\rm pivot}=0.05~{\rm Mpc}^{-1}$, and note that the Hubble constant $H_0 = 100 \, h \, {\rm km} \, {\rm s}^{-1} \, {\rm Mpc}^{-1}$ is a derived parameter, for which we impose the implicit bound $h \in [0.4,1.0]$. We will also consider constraints on the derived parameter $S_8 = \sigma_8 \sqrt{\Omega_{\rm m}/0.3}$, where $\sigma_8$ is the present root-mean-square of the linear matter density contrast on $8~h^{-1}~{\rm Mpc}$ scales and $\Omega_{\rm m}$ is the present matter density parameter.

For the KiDS$\times$\{2dFLenS+BOSS\} nuisance parameters, we impose the same fiducial priors as given in Ref.~\cite{Joudaki:2017zdt}, and for the CMB nuisance parameters, we impose the same priors as used by Planck~\cite{planck2018}. Notably, as discussed in Sec.~\ref{systsec}, we allow for a wide uniform prior on the baryonic feedback amplitude $B \in [1,4]$. We moreover impose a uniform prior on $\ln \omega^{-1}_{\rm BD}$ in agreement with the approach in Avilez \& Skordis (2014)~\cite{Avilez:2013dxa}. As the GR-limit is obtained for $\omega_{\rm BD} \rightarrow \infty$, we have taken $\ln \omega^{-1}_{\rm BD} \in [-17,-2.3]$ such that we obtain the upper bound $\omega_{\rm BD} \lesssim 2.4 \times 10^7$ (noting that this limit yields close to identical MCMC results to a pure $\Lambda$CDM run) while the lower bound is given by $\omega_{\rm BD} \geq 10$. When allowing for variations in the present effective gravitational constant, $G_{\rm matter}/G$ $($written as shorthand for $(G_{\rm matter}/G)|_{a=1})$, we impose a uniform prior in the range $[0.5,2.0]$, 
designed to be non-informative (i.e.~constraints restricted to a narrower region within the range) and motivated by our numerical simulations (Sec.~\ref{nonlinearsec}).

\subsection{Model selection and data concordance assessment: connection between different estimators}
\label{modelselec}

There has been a substantial amount of recent activity in devising the optimal approach for performing model selection and assessing dataset concordances (e.g.~\cite{spiegelhalter02,mrs06,ktp06,liddle07, trotta08, spiegelhalter14, Joudaki:2016mvz, raveri15, grandis15, seehars15, lin17,raverihu19, handley19,lemos19,feeney19,kohlinger19,adhikari19,nicola19}). Here, we consider the Deviance Information Criterion (DIC) \cite{spiegelhalter02} to assess the relative preference of two distinct cosmological models. The ${\rm{DIC}}$ is given by 
\begin{equation}
\label{dicigen}
{\rm DIC} = \chi^2_{\rm eff}(\hat{\theta}) + 2p_D = 2\overline{\chi^2_{\rm eff}(\theta)} - \chi^2_{\rm eff}(\hat{\theta}),
\end{equation} 
where the first term is the best-fit effective $\chi^2$ (here defined as the effective $\chi^2$ at the posterior maximum point, $\hat{\theta}$, equivalent to the maximum likelihood point for uniform priors) and the second term is the Bayesian complexity $p_D$ (given by the difference of the mean of the $\chi^2$ over the posterior, captured by the overbar, and the best-fit effective $\chi^2$). The Bayesian complexity encapsulates the effective number of parameters of a model, such that more complex models are penalized. In this definition of the DIC, we have neglected a factor of $-2\ln \mathcal{Z}$, where $\mathcal{Z} \equiv P(D | \mathcal{M})$ is the probability ($P$) of the data~($D$) given the model ($\mathcal{M}$), known as the Bayesian evidence (e.g.~Ref.~\cite{trotta08}). For reference, we take a difference of $\Delta {\rm DIC} = {\rm DIC}_{\rm extended} - {\rm DIC}_{\Lambda{\rm CDM}} = \{5, 10\}$ to correspond to \{moderate, strong\} preference in favor of the standard cosmological model (given odds of $1/12$ and $1/148$, respectively).

In assessing the concordance between distinct cosmological datasets, we consider the $\log \mathcal{I}$ statistic~\cite{Joudaki:2016mvz}, where ``log'' denotes the common logarithm. This concordance statistic is obtained from the DIC estimates, such that 
\begin{equation}
\label{logistat}
\log {\mathcal{I}}(D_1, D_2) = \mathcal{A} \, [{{{\rm{DIC}}(D_1 \cup D_2)} - {{\rm{DIC}}(D_1) - {{\rm{DIC}}(D_2)}}}],
\end{equation} 
where $\mathcal{A} = -[2 \ln(10)]^{-1}$, while $D_1$ and $D_2$ denote two distinct datasets and ${{\rm{DIC}}(D_1 \cup D_2)}$ is the joint DIC of the two datasets. This allows us to estimate the probability of the combined data with respect to the individual data probabilities (in a similar form to the Bayes factor). The DIC has the advantages that it is symmetric (neither dataset is considered to be more fundamental than the other), it is simple to compute (in particular compared to the full evidence), and it can be used to assess possible discordances between constraints from the same dataset (allowing for improvements in the treatment of systematic uncertainties). Turning to the drawbacks, among others, its use of a point estimate can be stochastically affected by the data, and beyond brute force, no accurate and efficient method exists for computing the uncertainties in the DIC (for further details, see e.g.~Ref.~\cite{Joudaki:2016mvz}).

We will further show that our $\log {\mathcal{I}}$ statistic can be related to a combination of the Bayesian evidence, the Kullback-Leibler (KL) divergence, and the goodness of fit. To this end, we can express the deviance information criterion as
\begin{align}
\label{dicigen}
{\rm DIC} &= \chi^2_{\rm eff}(\hat{\theta}) + 2p_D = 2\overline{\chi^2_{\rm eff}(\theta)} - \chi^2_{\rm eff}(\hat{\theta}) \nonumber\\
&= -4 \int \mathcal{P}\ln \mathcal{L}d\theta - \chi^2_{\rm eff}(\hat{\theta}) \nonumber\\
&= -4 \int \mathcal{P}\ln \mathcal{L \over Z}d\theta - 4\ln\mathcal{Z} - \chi^2_{\rm eff}(\hat{\theta}) \nonumber\\
&= -4 \int \mathcal{P}\ln {\mathcal{P} \over \Pi}d\theta - 4\ln\mathcal{Z} - \chi^2_{\rm eff}(\hat{\theta}) \nonumber\\
&= -4 \left[\mathcal{D}_{\rm KL} + \ln\mathcal{Z} + \frac{1}{4} \chi^2_{\rm eff}(\hat{\theta})\right],
\end{align} 
where we have used Bayes' theorem, $\mathcal{P} = \mathcal{L} \Pi / \mathcal{Z}$, to re-express the equation in terms of the Kullback-Leibler divergence, $\mathcal{D}_{\rm KL}(\mathcal{P},\Pi) = \int \mathcal{P}\ln {\mathcal{P} \over \Pi}d\theta$, which measures the ``relative entropy'' or information gain from the prior to the posterior. Here, $\mathcal{P} \equiv P(\theta|D,\mathcal{M})$ is the posterior, $\mathcal{L} \equiv P(D|\theta,\mathcal{M})$ is the likelihood, $\Pi \equiv P(\theta|\mathcal{M})$ is prior, and we note that $\int \mathcal{P} \ln\mathcal{Z}d\theta = \ln\mathcal{Z} \int \mathcal{P} d\theta = \ln\mathcal{Z}$ because $\int \mathcal{P} d\theta = 1$ by definition. 
As a result, the DIC can be expressed as the sum of the Kullback-Leibler divergence, the Bayesian evidence, and the goodness of fit.
We can therefore express our tension statistic as
\begin{align}
\label{logiigen}
&\log {\mathcal{I}}(D_1, D_2) = \nonumber\\
&-4\mathcal{A} \, \left[\mathcal{D}_{\rm KL}(D_1 \cup D_2) - \mathcal{D}_{\rm KL}(D_1) - \mathcal{D}_{\rm KL}(D_2)\right] \nonumber\\
&-4\mathcal{A} \, \left[\ln\mathcal{Z}(D_1 \cup D_2) - \ln\mathcal{Z}(D_1) - \ln\mathcal{Z}(D_2)\right] \nonumber\\
&-\mathcal{A} \, \left[\chi^2_{\rm eff}(\hat{\theta},~D_1 \cup D_2) - \chi^2_{\rm eff}(\hat{\theta},~D_1) -\chi^2_{\rm eff}(\hat{\theta},~D_2)\right],
\end{align} 
which can then be defined in terms of the differences (see Refs.~\cite{raverihu19,handley19,lemos19})
\begin{align}
\label{logiigen}
\ln I_{\rm KL} &=  \mathcal{D}_{\rm KL}(D_1) + \mathcal{D}_{\rm KL}(D_2) - \mathcal{D}_{\rm KL}(D_1 \cup D_2), \nonumber\\
\ln R &= \ln\frac{\mathcal{Z}(D_1 \cup D_2)}{\mathcal{Z}(D_1)\mathcal{Z}(D_2)}, \nonumber\\
Q_{\rm DMAP} &= \chi^2_{\rm eff}(\hat{\theta},~D_1 \cup D_2) - \chi^2_{\rm eff}(\hat{\theta},~D_1) -\chi^2_{\rm eff}(\hat{\theta},~D_2),
\end{align} 
where $R$ is known as the Bayes ratio (and $I_{\rm KL}$ is commonly expressed as $I$, see e.g.~Refs.~\cite{handley19,lemos19}), such that
\begin{align}
\label{logiigenshort}
\log {\mathcal{I}}(D_1, D_2) = -4 \mathcal{A} \, \left[ \ln R - \ln I_{\rm KL} + \frac{1}{4} Q_{\rm DMAP} \right].
\end{align} 
Hence, we find that the $\log {\mathcal{I}}$ measure of the concordance between datasets can be expressed in terms of corresponding measures involving the Bayesian evidence, the Kullback-Leibler divergence, and the goodness of fit. In its purest form, this reduces down to a measure of the mean and best-fit effective $\chi^2$, which eliminates the prior volume dependence in the $R$ and $I_{\rm KL}$ terms (see also e.g.~Refs.~\cite{handley19,heymans20}).

\begin{figure}
\vspace{-0.3em}
\includegraphics[width=1.03\hsize]{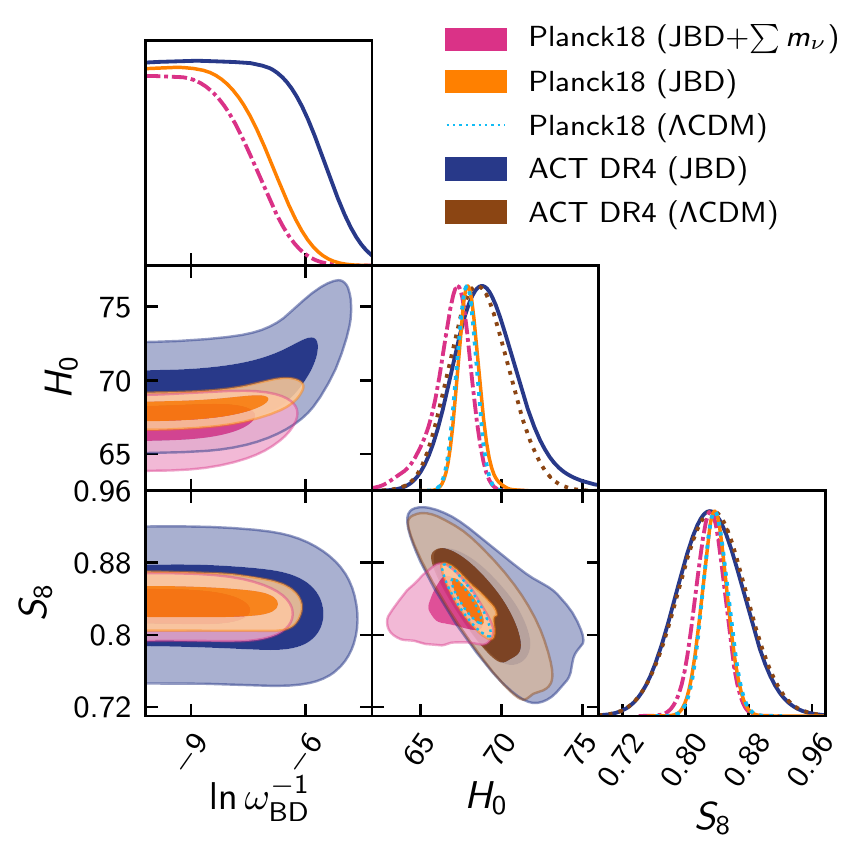}
\vspace{-2.8em}
\caption{\label{figplanck}
Marginalized posterior distributions (inner $68\%$~CL, outer $95\%$~CL) of the JBD parameter $\ln \omega_{\rm BD}^{-1}$, the Hubble constant $H_0$ (in units of ${\rm km} \, {\rm s}^{-1} {\rm Mpc}^{-1}$), and $S_8 = \sigma_8 \sqrt{\Omega_{\mathrm m}/0.3}$ from the CMB temperature and polarization measurements of Planck 2018 and ACT DR4. All other standard cosmological parameters are simultaneously varied. For visual clarity, we have zoomed in on the $\ln \omega_{\rm BD}^{-1}$ axis where the contours begin to flatten in the plane with $H_0$, as they stay flat and unbounded to the negative end of our prior (at $\ln \omega_{\rm BD}^{-1}  = -17$). 
}
\end{figure}

\begin{figure}
\vspace{-0.3em}
\includegraphics[width=1.03\hsize]{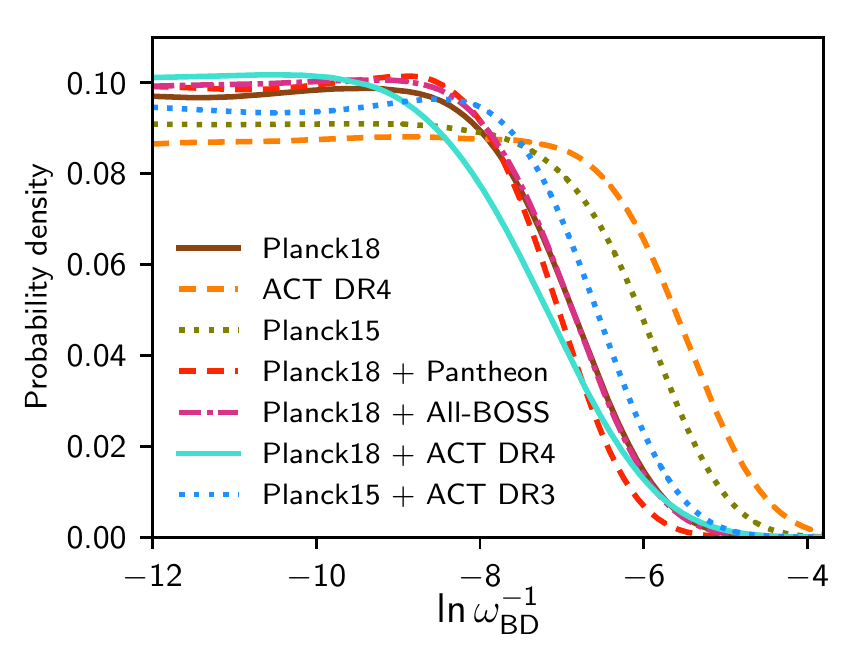}
\vspace{-3em}
\caption{\label{figwbd} 
Marginalized posterior distributions for the JBD parameter $\ln \omega_{\rm BD}^{-1}$. We simultaneously vary all standard cosmological parameters (but keep $\sum m_{\nu}$ and $G_{\rm matter}/G$ fixed). For visual clarity, we have zoomed in on the $\smash{\ln \omega_{\rm BD}^{-1}}$ axis, as the distributions continue to stay flat towards the GR limit at $\smash{\ln \omega_{\rm BD}^{-1} = -\infty}$ (in practice to the negative end of our prior range at $\ln \omega_{\rm BD}^{-1}  = -17$).
}
\end{figure}

For reference, we follow Jeffreys' scale \cite{jeffreys,kr95} and take $|\log \mathcal{I}| \gtrsim \{1/2,1,2\}$ to reflect \{substantial, strong, decisive\} concordance between the datasets (given probability ratios in excess of 3.2, 10, and 100, respectively). Similar negative values correspond to discordance, or ``tension'', between datasets. We also consider the tension between datasets for distinct cosmological parameters, such as $H_0$ and $S_8$, by evaluating 
\begin{equation}
\label{tstat}
T(p) = \left|\overline{p^{D_1}} - \overline{p^{D_2}}\right| / \sqrt{\sigma^2\left(p^{D_1}\right) + \sigma^2\left(p^{D_2}\right)} 
\end{equation}
for a given parameter $p$, where overbar denotes the posterior mean, the vertical bars denote absolute value, and $\sigma$ refers to the symmetric $68\%$ confidence interval about the posterior mean (we note that the posterior maximum is another popular choice, with similar results for current data;~e.g.~\cite{joudaki20,heymans20}).

Given the computational simplicity, we restrict the model selection and dataset concordance assessments to the DIC and $\log \mathcal{I}$ statistic, respectively. However, we note that similar results are expected for analogous computations with the Bayesian evidence and Bayes factor~\cite{trotta08}, respectively (see for example the agreement between the approaches in~Ref.~\cite{Joudaki:2016mvz}). Further alternative approaches for data concordance assessments include, among others, the ``surprise''~\cite{grandis15,seehars15}, the ``index of inconsistency''~\cite{lin17}, the ``debiased evidence ratio''~\cite{raverihu19}, the $Q_{\rm DMAP}$ statistic~\cite{raverihu19}, the $Q_{\rm UDM}$ statistic~\cite{raverihu19}, the ``suspiciousness''~\cite{handley19,lemos19}, and posterior predictive distributions (e.g.~\cite{feeney19,kohlinger19}). As noted in Refs.~\cite{raverihu19,heymans20}, and shown in this section, these concordance statistics can often be directly related; of particular relevance here 
\begin{equation}
\label{logirelate}
\log \mathcal{I} = [2 \ln(10)]^{-1} [Q_{\rm DMAP} + 4 \ln S], 
\end{equation}
where $S \equiv R/I_{\rm KL}$ is the suspiciousness.

\begin{table*}
\vspace{-1.1em}
\caption{\label{subtab1} Marginalized posterior means and $68\%$ confidence intervals for a subset of the cosmological parameters. For the JBD parameter $\omega_{\rm BD}$ and the sum of neutrino masses $\sum m_{\nu}$, we quote the $95\%$ confidence lower and upper bounds, respectively. The Hubble constant $H_0$ is in units of km\,s$^{-1}$\,Mpc$^{-1}$, $\sum m_{\nu}$ is in units of eV, and $S_8=\sigma_8\sqrt{\Omega_{\rm m}/0.3}$. A table element with ``$\cdots$'' implies that the parameter is not varied in the analysis. Here, ``$\Lambda$CDM'' includes massless neutrinos and is taken as the limiting case of the JBD model, in principle as $\omega_{\rm BD} \rightarrow \infty$, and in practice as $\ln \omega_{\rm BD}^{-1} = -17$ (along with $G_{\rm matter}/G = 1$). We note that minor deviations in the $\Lambda$CDM results from those reported by Planck~\cite{planck2015,planck2018} and ACT~\cite{actpol17,aiola20} are due to this limit and the specific priors used (e.g.~we do not add an external Gaussian prior on $\tau$ as in ACT, and all three neutrinos are massless in our fiducial model). The tensions $T(H_0)$ and $T(S_8)$ are against Riess et al.~2019~\cite{riess2019} and KiDS$\times$\{2dFLenS+BOSS\}, respectively (in the latter case only against KiDS$\times$2dFLenS when Planck is combined with BOSS), and are evaluated in a consistent manner where the datasets are considered in the same cosmology relative to one another (in practice only applies for $S_8$).
}
\renewcommand{\footnoterule}{} 
\begin{tabularx}{\textwidth}{l@{\extracolsep{\fill}}ccccc|cc}
\hline
Probe setup & $\omega_{\rm BD}$ & $G_{\rm matter}/G$ & $\sum m_{\nu}$ & $H_0$ & $S_8$ & $T(H_0)$ & $T(S_8)$\\
\hline
${\rm Planck18} \, \left({\Lambda {\rm CDM}}\right)$ & $\cdots$ & $\cdots$ & $\cdots$ & $67.84^{+0.60}_{-0.61}$  & $0.838^{+0.016}_{-0.016}$  & $4.0$ & $2.4$\\
${\rm Planck18} \, \left({\Lambda {\rm CDM}+\sum m_{\nu}}\right)$ & $\cdots$ & $\cdots$ & $0.38$  & $66.73^{+1.48}_{-0.69} $ & $0.830^{+0.018}_{-0.018} $  & $4.1$ & $2.5$\\
${\rm Planck18} \, \left({{\rm JBD}}\right)$ & $1150$ & $\cdots$ & $\cdots$ & $68.00^{+0.60}_{-0.71}$  & $0.837^{+0.016}_{-0.016}$  & $3.9$ & $2.4$\\
${\rm Planck18} \, \left({{\rm JBD}+\sum m_{\nu}}\right)$ & $1710$ & $\cdots$ & $0.37$  & $ 66.79^{+1.44}_{-0.76} $  & $ 0.831^{+0.018}_{-0.018} $  & $4.0$ & $2.4$\\
${\rm Planck18} \, \left({{\rm JBD}+G_{\rm matter}+\sum m_{\nu}}\right)$ & $1120$ & $0.993^{+0.026}_{-0.038}$ & $0.43$  & $ 66.40^{+1.99}_{-1.97}$ & $0.835^{+0.023}_{-0.023}$ & $3.1$ & $1.5$\\
${\rm Planck18}+{\rm ACT~DR4} \, \left({\Lambda {\rm CDM}}\right)$ & $\cdots$ & $\cdots$ & $\cdots$ & $68.00^{+0.55}_{-0.56}$  & $0.841^{+0.015}_{-0.015}$  & $4.0$ & $2.5$\\
${\rm Planck18}+{\rm ACT~DR4} \, \left({{\rm JBD}}\right)$ & $1380$ & $\cdots$ & $\cdots$ & $67.88^{+0.55}_{-0.65}$  & $0.840^{+0.015}_{-0.015}$  & $3.9$ & $2.5$\\
${\rm ACT~DR4} \, \left({\Lambda {\rm CDM}}\right)$ & $\cdots$ & $\cdots$ & $\cdots$ & $68.68^{+1.69}_{-1.92}$  & $0.833^{+0.042}_{-0.042}$ & $2.3$ & $1.6$\\
${\rm ACT~DR4} \, \left({{\rm JBD}}\right)$ & $330$ & $\cdots$ & $\cdots$ & $69.25^{+2.35}_{-1.67}$  & $0.831^{+0.042}_{-0.042}$  & $1.9$ & $1.6$\\
${\rm All}$-${\rm Planck18} \, \left({\Lambda{\rm CDM}}\right)$ & $\cdots$ & $\cdots$ & $\cdots$ & $67.97^{+0.55}_{-0.54}$ & $0.834^{+0.012}_{-0.013}$ & $4.0$ & $2.4$ \\
${\rm All}$-${\rm Planck18} \, \left({{\rm JBD}}\right)$ & $810$ & $\cdots$ & $\cdots$ & $68.15^{+0.50}_{-0.75}$ & $0.833^{+0.013}_{-0.013}$ & $3.8$ & $2.3$ \\
${\rm Planck18}+{\rm Pantheon} \, \left({{\rm JBD}}\right)$ & $1440$ & $\cdots$ & $\cdots$ & $68.05^{+0.57}_{-0.64} $  & $0.835^{+0.015}_{-0.015} $  & $3.9$ & $2.3$\\
${\rm Planck18}+{\rm All}$-${\rm BOSS} \, \left({{\rm JBD}}\right)$ & $1170$ & $\cdots$ & $\cdots$ & $68.18^{+0.45}_{-0.54}$  & $0.832^{+0.013}_{-0.013} $  & $3.9$ & $2.3$\\
${\rm Planck15} \, \left({\Lambda {\rm CDM}}\right)$ & $\cdots$ & $\cdots$ & $\cdots$  & $67.94^{+1.00}_{-0.98} $ & $0.853^{+0.025}_{-0.025} $  & $3.5$ & $2.5$\\
${\rm Planck15} \, \left({{\rm JBD}}\right)$ & $530$ & $\cdots$ & $\cdots$ & $68.27^{+0.93}_{-1.29} $  & $ 0.850^{+0.025}_{-0.025} $  & $3.2$ & $2.4$\\
${\rm Planck15} \, \left({{\rm JBD}+G_{\rm matter}}\right)$ & $850$ & $1.024^{+0.046}_{-0.053}$ & $\cdots$ & $69.86^{+2.57}_{-2.88} $  & $ 0.842^{+0.034}_{-0.034} $  & $1.6$ & $1.6$\\
${\rm Planck15}+{\rm ACT~DR3} \, \left({{\rm JBD}}\right)$ & $900$ & $\cdots$ & $\cdots$ & $67.97^{+0.92}_{-1.10} $  & $0.851^{+0.023}_{-0.024} $  & $3.5$ & $2.5$\\
\hline
\end{tabularx}
\end{table*}

\section{Results: cosmic microwave background}
\label{cmbresultssec}

We now methodically present the cosmological results of the JBD model, with a particular focus on the impact of different data combinations, modeling choices, and possible degeneracies between modified gravity, massive neutrinos, and astrophysics (especially baryonic feedback). Henceforth, we interchangeably refer to ``$G_{\rm matter}$'' and ``$G_{\rm matter}/G$'' when considering the normalized effective gravitational constant at present, $(G_{\rm matter}/G)|_{a=1}$, as a free parameter in our analysis. In the assessment of the concordance between datasets, this is self-consistently performed under the same cosmological model.

We consider the following six distinct cosmologies: $a)$ ``$\Lambda$CDM'' which refers to our $\Lambda$CDM limit of the JBD model (see Table~\ref{tabpriors}), $b)$ ``$\Lambda$CDM + $\sum m_{\nu}$'' where we allow the sum of neutrino masses to be further varied, $c)$ ``${\rm JBD}$'' where the modified gravity parameter $\ln \omega_{\rm BD}^{-1}$ is varied in addition to the $\Lambda$CDM parameters (referred to as the standard, or restricted, JBD model), $d)$ ``${\rm JBD}+\sum m_{\nu}$'' where the sum of neutrino masses is varied in the restricted JBD model, $e)$ ``${\rm JBD}+G_{\rm matter}$'' where the present effective gravitational constant, $(G_{\rm matter}/G)|_{a=1}$, is varied together with $\ln \omega_{\rm BD}^{-1}$ in addition to the $\Lambda$CDM parameters (referred to as the unrestricted JBD model), and $f)$ ``${\rm JBD}+G_{\rm matter}+\sum m_{\nu}$'' where the sum of neutrino masses is varied together with the two modified gravity parameters in the unrestricted JBD model.

\subsection{Planck and ACT}
\label{planckactsec}

It is well established that the CMB currently constrains the standard cosmological model more powerfully than any other cosmological probe (e.g.~\cite{planck2018}). Here, we find the strongest independent constraints on the JBD model from the Planck CMB, in particular through the impact of JBD gravity on the location of the CMB peaks and the damping tail (as discussed in Secs.~\ref{theorysec} and~\ref{cmbsec}). 

\subsubsection{Constraining the JBD coupling constant}

As shown in Table~\ref{subtab1} and Figs.~\ref{figplanck} and \ref{figwbd}, our constraints are sensitive to the specific details of the CMB data and cosmological model, but are generally at the $\omega_{\rm BD} \gtrsim 10^{3}$ level (at $95\%$~CL, where the inequality symbol is here taken to denote ``greater than approximately''). Concretely, $\omega_{\rm BD} > \{1150, 1710, 1120\}$ at $95\%$~CL from the Planck 2018 CMB temperature and polarization measurements when considering the restricted JBD model, the restricted JBD model where the sum of neutrino masses is allowed to vary, and the unrestricted JBD model where the sum of neutrino masses is allowed to vary, respectively. 
For comparison, by further including the Planck CMB lensing reconstruction in our ``All-Planck18'' setup, the lower bound in the restricted JBD model with fixed neutrino masses weakens to $\omega_{\rm BD} > 810$ ($95\%$~CL).

We emphasize that larger values of $\omega_{\rm BD}$ imply greater consistency with GR. Moreover, while the uncertainty on $\omega_{\rm BD}$ improves as we add more data, the one-sided bound will either weaken or strengthen depending on how much JBD gravity is favored by the additional data (i.e.~it reflects a shift in the posterior rather than a narrowing of the posterior, similar to the one-sided bound for the sum of neutrino masses). Our results can be contrasted with the Planck constraints on $\omega_{\rm BD}$ in Refs.~\cite{Avilez:2013dxa, lwc13,Ballardini:2016cvy,Ooba:2017gyn,Peracaula19,ballardini20}, where the upper bound fluctuates between approximately $10^2$ to $10^3$ at $95\%$~CL depending on the specific datasets and details of the JBD modeling. We will return to this comparison in Secs.~\ref{sumsec} and \ref{modelingsec}.

We moreover consider the high-multipole CMB temperature and polarization measurements from ACT DR4, which on its own constrains $\omega_{\rm BD} > 330$~($95\%$~CL) in the restricted JBD model with fixed neutrino masses. By combining Planck 2018 (temperature and polarization) and ACT DR4 in this model, we find a strengthening in the lower bound on $\omega_{\rm BD}$ from $1150$ to $1380$ (see Table~\ref{subtab1}). We again note that this does not imply a correspondingly significant improvement in the uncertainty on $\omega_{\rm BD}$, but rather a shift in the amount of modified gravity favored by the data (given the one-sided bound). In Fig.~\ref{figwbd}, we show the marginalized posteriors for the coupling constant for Planck and ACT, along with other datasets such as BOSS and Pantheon, where the lower bound on $\omega_{\rm BD}$ is largely unchanged for Planck combined with BOSS and it strengthens to $1440$ for Planck combined with Pantheon.

\subsubsection{Impact of restricted JBD gravity on $H_0$ and $S_8$}

In Fig.~\ref{figplanck}, we show the Planck and ACT constraints for the restricted JBD model in the subspace $\{\ln \omega_{\rm BD}^{-1}, H_0, S_8\}$, where the Hubble constant increases as $\omega_{\rm BD}$ decreases (i.e.~as the strength of JBD gravity increases). The upturn in $H_0$ is stronger for ACT because its constraint on the coupling constant is weaker, and it is less pronounced for Planck when the sum of neutrino masses is varied (as compared to fixed neutrino masses;~given the stronger constraint on the coupling constant). As expected, the contours in the plane of $\ln \omega_{\rm BD}^{-1}$ and $H_0$ flatten as we approach the GR limit, here in practice as $\ln \omega_{\rm BD}^{-1} \lesssim -9$. Meanwhile, the contour in the plane of $S_8$ and $\ln \omega_{\rm BD}^{-1}$ has a negligile downturn in $S_8$ as JBD gravity increases, and appears largely flat.

The Hubble constant is $H_0 = 68.00^{+0.60}_{-0.71}~{\rm km} \, {\rm s}^{-1} {\rm Mpc}^{-1}$ and $H_0 = 69.25^{+2.35}_{-1.67}~{\rm km} \, {\rm s}^{-1} {\rm Mpc}^{-1}$ in the restricted JBD model for Planck and ACT, respectively. These constraints are both approximately $10\%$ weaker and positively shifted by $0.3\sigma$ relative to the corresponding constraints in $\Lambda$CDM. As compared to the direct measurement of the Hubble constant by Riess et al.~2019~\cite{riess2019}, where $H_0 = 74.03 \pm 1.42 ~ {\rm km} \, {\rm s}^{-1} {\rm Mpc}^{-1}$, this implies $\{0.1\sigma, 0.4\sigma\}$ decreases in the tension for \{Planck, ACT\}, respectively. The combined Planck+ACT constraint on the Hubble constant is $H_0 = 67.88^{+0.55}_{-0.65}~{\rm km} \, {\rm s}^{-1} {\rm Mpc}^{-1}$, which corresponds to a marginal increase in the uncertainty and decrease of the tension (by $8\%$ and $0.1\sigma$, respectively) compared to $\Lambda$CDM. Turning to the $S_8$ parameter, the marginalized constraints are effectively unchanged relative to $\Lambda$CDM for these datasets.

\begin{figure}
\vspace{-0.34em}
\includegraphics[width=1.02\hsize]{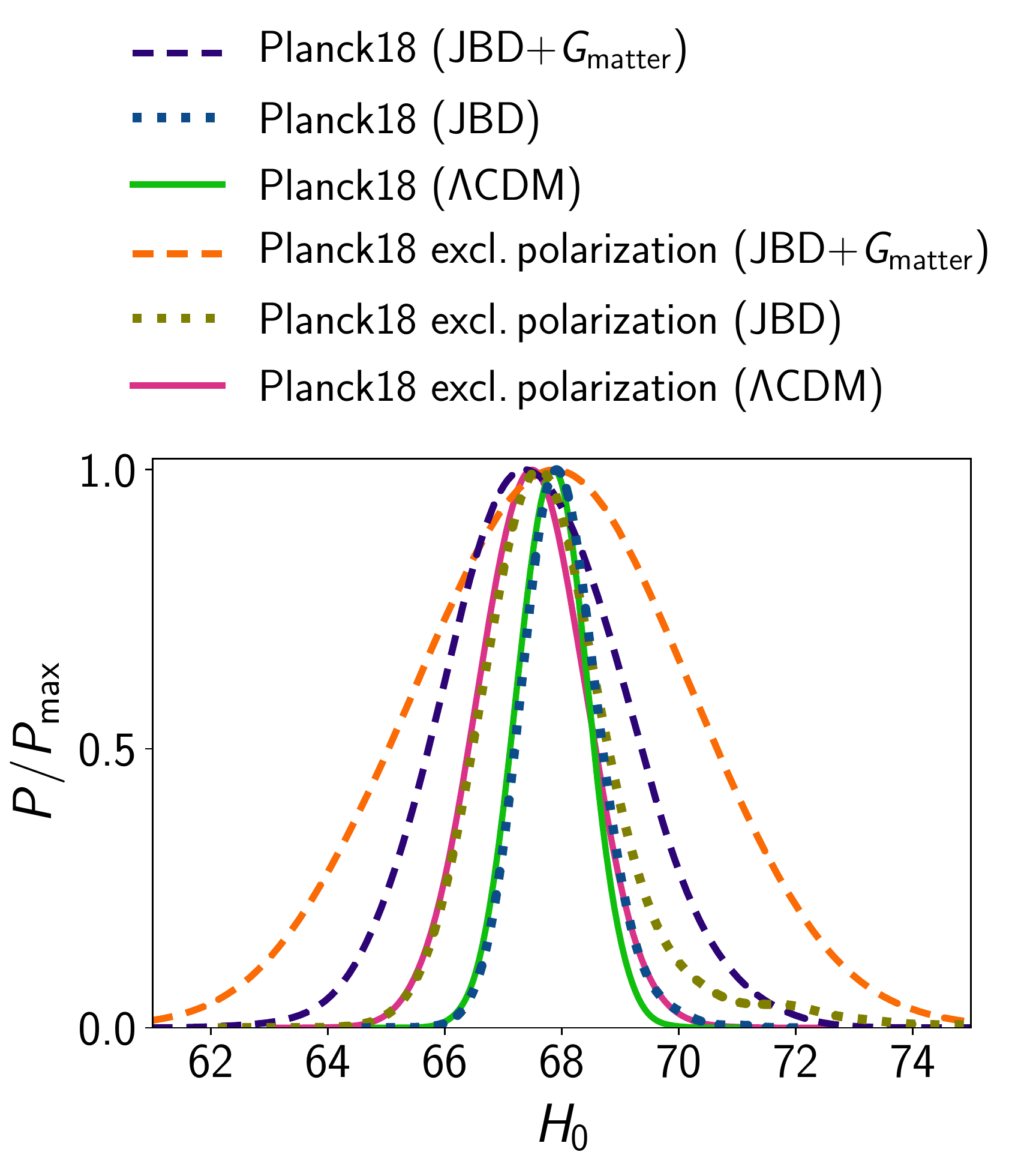}
\vspace{-2.4em}
\caption{\label{figpol} 
Marginalized posterior distributions for the Hubble constant, $H_0$ (in units of $\smash{{\rm km} \, {\rm s}^{-1} {\rm Mpc}^{-1}}$), from the Planck 2018 dataset. The different distributions either include or exclude the small-scale polarization data (i.e.~``TT,TE,EE+lowE'' as compared to ``TT+lowE'') and consider either the unrestricted JBD model, the restricted JBD model, or $\Lambda$CDM (in all cases with $\sum m_{\nu}$ fixed).
}
\end{figure}

\subsubsection{The effective gravitational constant and its impact on tensions}

Focusing on Planck 2018 alone, in addition to the coupling constant, in the unrestricted JBD model with massive neutrinos, we constrain the present effective gravitational constant at few-percent level precision, such that $G_{\rm matter}/G = 0.993^{+0.026}_{-0.038}$, in full agreement with GR. As shown in Table~\ref{subtab1} (and illustrated in the forthcoming sections), when the present effective gravitational constant is varied, the uncertainties in the Hubble constant and $S_8$ increase substantially. Concretely, $H_0 = 66.40^{+1.99}_{-1.97} ~ {\rm km} \, {\rm s}^{-1} {\rm Mpc}^{-1}$ and $S_8 = 0.835^{+0.023}_{-0.023}$. As compared to the restricted JBD model with massive neutrinos (i.e.~where $G_{\rm matter}/G = 1$), the posterior means are marginally shifted (by $0.2$--$0.3\sigma$), while the uncertainties on $H_0$ and $S_8$ increase by $80\%$ and $25\%$, respectively. As a result, the discordance with the Riess et al.~(2019) measurement of $H_0$ alleviates to $3.1\sigma$ in this model, while the difference in $S_8$ with low-redshift probes such as  ${\rm KiDS}\times\{{\rm 2dFLenS}+{\rm BOSS}\}$ (discussed further in Sec.~\ref{wlgcsec}) narrows down to $1.5\sigma$. As compared to the $\Lambda$CDM model with massive neutrinos, this implies a $1.0\sigma$ decrease in both tensions.

We can understand these results in the context of Fig.~\ref{figcls}, where $G_{\rm matter}/G$ has a greater flexibility in its modification of the CMB power spectra (i.e.~both suppression and enhancement of the responses depending on whether $G_{\rm matter}/G < 1$ or  $G_{\rm matter}/G > 1$, while $\omega_{\rm BD} \rightarrow 0$ largely enhances the fluctuations). Indeed, larger $H_0$ is favored by $G_{\rm matter}/G > 1$. For a given response in the CMB temperature power spectrum by either $\omega_{\rm BD}$ or $G_{\rm matter}/G$, the corresponding response in the expansion rate is moreover substantially larger for the effective gravitational constant. We note that this is qualitatively similar in the case of $S_8$. In forthcoming sections, we will revisit the unrestricted JBD model as additional datasets are considered and will explicitly show the positive correlation between $G_{\rm matter}/G$ and $H_0$, and the negative correlation between $G_{\rm matter}/G$ and $S_8$ (and, correspondingly, the negative correlation between $H_0$ and $S_8$). 

\subsubsection{The sum of neutrino masses}

The constraint on the sum of neutrino masses is robust, changing from $0.38$~eV in $\Lambda$CDM to $0.37$~eV in the restricted JBD model and to $0.43$~eV in the unrestricted JBD model. However, we note that the marginalized constraints (on the sum of neutrino masses, along with other parameters such as the present effective gravitational constant) possess some dependence on the choice of parameterization of the coupling constant, which we explore further for the full combination of datasets in Sec.~\ref{modelingsec}.

\subsubsection{The significance of the small-scale CMB polarization for the parameter constraints}
\label{cmbpolsec}

In Appendix~\ref{cmbpolapp}, we discuss how the different signatures of JBD gravity on the CMB polarization power spectrum (and temperature-polarization cross spectrum) can be used to improve the constraints on this theory and the physics that correlates with it.
In assessing the impact of the CMB polarization on the parameter constraints, we thereby consider a case where it is removed in the analysis of the Planck 2018 dataset on small scales (i.e.~we consider ``TT+lowE'' instead of the fiducial case of ``TT+TE+EE+lowE''). For concreteness, we fix the sum of neutrino masses in carrying this out.

In the restricted JBD model, the $95\%$ lower bound on the coupling constant degrades from $1150$ to $430$ as the small-scale CMB polarization is excluded. 
As a result, the exclusion of the polarization allows for higher values of $H_0$ and lower values of $S_8$, with a more pronounced ``hook-shaped'' turn of the respective contours in the plane with $\omega_{\rm BD}$ (as compared to that seen in Fig.~\ref{figplanck}). This in turn is manifested in wider tails in the high-$H_0$ and low-$S_8$ regime of the respective one-dimensional marginalized posteriors. In Fig.~\ref{figpol}, we provide an illustration of this for the $H_0$ posterior, which in addition to the restricted JBD model includes the corresponding cases (i.e.~Planck 2018 including or excluding small-scale polarization) for the unrestricted JBD model and $\Lambda$CDM.

For all three models, the constraints on $H_0$ and $S_8$ substantially improve as the small-scale CMB polarization is included in the cosmological analysis. In the restricted JBD model, the constraint on $H_0 = 67.89^{+0.83}_{-1.35} ~ {\rm km} \, {\rm s}^{-1} {\rm Mpc}^{-1}$ improves by $40\%$ to $H_0 = 68.00^{+0.60}_{-0.71} ~ {\rm km} \, {\rm s}^{-1} {\rm Mpc}^{-1}$, while the constraint on $S_8 = 0.840^{+0.025}_{-0.025}$ improves by $35\%$ to $S_8 = 0.837^{+0.016}_{-0.016}$. Hence, the small-scale CMB polarization improves the $H_0$ and $S_8$ constraints to agree more closely with the corresponding results in $\Lambda$CDM. This allows for a strengthening of the $H_0$ tension with Riess et al.~(2019)~\cite{riess2019} from $3.4\sigma$ to $3.9\sigma$ and a marginal strengthening of the $S_8$ tension with KiDS$\times$\{2dFLenS$+$BOSS\} from $2.2\sigma$ to $2.3\sigma$. The discordances are not substantially modified as the CMB temperature uncertainties on $H_0$ and $S_8$ are already subdominant to those from Riess et al.~(2019)~\cite{riess2019} and KiDS$\times$\{2dFLenS$+$BOSS\}, respectively.

In the unrestricted JBD model, the small-scale CMB polarization improves the constraint on $G_{\rm matter}/G$ by $35\%$. As expected by the degeneracy with the Hubble constant, the constraint on $H_0 = 67.85^{+2.36}_{-2.36} ~ {\rm km} \, {\rm s}^{-1} {\rm Mpc}^{-1}$ in turn improves by the same percentage to $H_0 = 67.58^{+1.47}_{-1.58} ~ {\rm km} \, {\rm s}^{-1} {\rm Mpc}^{-1}$. This corresponds to a strengthening of the tension with Riess et al.~(2019)~\cite{riess2019} from $2.2\sigma$ to $3.1\sigma$. In Fig.~\ref{figpol}, we find that the small-scale polarization narrows the $H_0$ posterior on both ends, while primarily the low-$H_0$ end of the posterior is narrowed in $\Lambda$CDM. Similarly, the small-scale CMB polarization improves the constraint on $S_8 = 0.840^{+0.033}_{-0.036}$ by 40\% to $S_8 = 0.841^{+0.021}_{-0.022}$, which marginally strengthens the tension with KiDS$\times$\{2dFLenS$+$BOSS\} from $1.6\sigma$ to $1.7\sigma$. Here, the improvement in the $S_8$ posterior is approximately symmetric, as compared an asymmetric improvement in the high end of the posterior in $\Lambda$CDM. Hence, while the $S_8$ tension is largely unaffected (as the CMB temperature uncertainty on $S_8$ is already subdominant), the small-scale CMB polarization is a driving force behind the persistence of the $H_0$ tension and complicates the ability of an extension such as JBD gravity to resolve it. 

\begin{figure*}
\vspace{-0.7em}
\includegraphics[width=0.89\hsize]{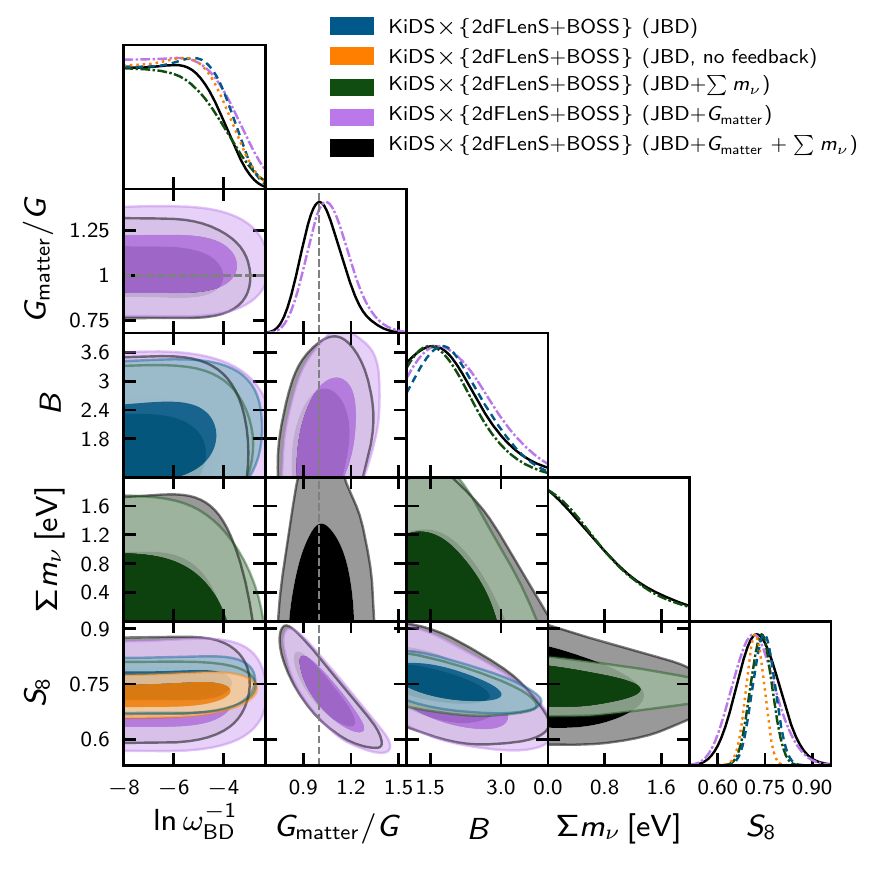}
\vspace{-2em}
\caption{\label{figkids}
Marginalized posterior distributions (inner $68\%$~CL, outer $95\%$~CL) of the JBD parameter $\ln \omega_{\rm BD}^{-1}$, the present effective gravitational constant, $\smash{G_{\rm matter}/G}$, the baryonic feedback amplitude, $B$, the sum of neutrino masses, $\smash{\sum m_{\nu}}$, and $S_8 = \sigma_8 \sqrt{\Omega_{\mathrm m}/0.3}$ from KiDS$\times$\{2dFLenS+BOSS\}. All other standard cosmological and systematics parameters are simultaneously varied. For visual clarity, we have zoomed in on the $\smash{\ln \omega_{\rm BD}^{-1}}$ axis where the distributions begin to flatten, as they stay flat and unbounded in the direction of the GR limit (i.e.~towards the negative end of our prior range at $\smash{\ln \omega_{\rm BD}^{-1}  = -17}$). We do not include the $\Lambda$CDM case for visual clarity, given the similar constraints on \{$\smash{B, S_8}$\} to the JBD case.
}
\end{figure*}

\begin{table*}
\vspace{-1.1em}
\caption{\label{subtab2} Marginalized posterior means and $68\%$ confidence intervals for a subset of the cosmological parameters. For the JBD parameter, $\omega_{\rm BD}$, and the sum of neutrino masses, $\sum m_{\nu}$ (in units of eV), we quote the $95\%$ confidence lower and upper bounds, respectively. We also quote the $95\%$ confidence upper bound for the feedback amplitude, $B$. The Hubble constant, $H_0$, is in units of km\,s$^{-1}$\,Mpc$^{-1}$ and we define $S_8=\sigma_8\sqrt{\Omega_{\rm m}/0.3}$. The tensions $T(H_0)$ and $T(S_8)$ are against Riess et al.~(2019)~\cite{riess2019} and Planck18, respectively. The symbol ``$\diamond$'' implies that the parameter is effectively unconstrained by the data, and the symbol ``$\circ$'' implies that the tension $T$ is not meaningful to quote (i.e.~$T \sim 0$). See Table~\ref{subtab1} for further~details.
}
\renewcommand{\footnoterule}{} 
\begin{tabularx}{\textwidth}{l@{\extracolsep{\fill}}ccccccc|cc}
\hline
Probe setup & $\omega_{\rm BD}$ & $G_{\rm matter}/G$ & $B$ & $A_{\rm IA}$ & $\sum m_{\nu}$ & $H_0$ & $S_8$ & $T(H_0)$ & $T(S_8)$\\
\hline
${\rm KiDS}\times{\rm 2dFLenS} \left({\Lambda{\rm CDM}}\right)$ & $\cdots$ & $\cdots$ & $3.5$ & $1.23^{+0.69}_{-0.55}$ & $\cdots$ & $\diamond$ & $0.736^{+0.039}_{-0.038}$ & $\circ$ & $2.4$ \\
${\rm KiDS}\times\{{\rm 2dFLenS}+{\rm BOSS}\} \left({\Lambda{\rm CDM}}\right)$ & $\cdots$ & $\cdots$ & $3.2$ & $1.71^{+0.48}_{-0.47}$ & $\cdots$ & $\diamond$  & $ 0.746^{+0.034}_{-0.035} $  & $\circ$ & $2.4$ \\
${\rm KiDS}\times{\rm 2dFLenS}+{\rm All}$-${\rm BOSS} \left({\Lambda {\rm CDM}}\right)$ & $\cdots$ & $\cdots$ & $3.3$ & $1.12^{+0.52}_{-0.46}$ & $\cdots$ & $70.6^{+5.8}_{-4.4}$ & $0.745^{+0.029}_{-0.028}$ & $0.65$ & $2.8$ \\
${\rm KiDS}\times\{{\rm 2dFLenS}+{\rm BOSS}\} \left({\Lambda{\rm CDM}+\sum m_{\nu}}\right)$ & $\cdots$ & $\cdots$ & $3.1$ & $1.65^{+0.48}_{-0.44}$ & $1.7$ & $\diamond$  & $ 0.736^{+0.033}_{-0.033} $  & $\circ$ & $2.5$\\
${\rm KiDS}\times\{{\rm 2dFLenS}+{\rm BOSS}\} \left({{\rm JBD}}\right)$ & $56$ & $\cdots$ & $3.3$ & $1.64^{+0.49}_{-0.48}$ & $\cdots$ & $\diamond$  & $ 0.747^{+0.035}_{-0.035} $  & $\circ$ & $2.4$ \\
${\rm KiDS}\times\{{\rm 2dFLenS}+{\rm BOSS}\} \left({{\rm JBD,~no~feedback}}\right)$ & $63$ & $\cdots$ & $\cdots$ & $1.78^{+0.49}_{-0.49}$ & $\cdots$ & $\diamond$  & $ 0.718^{+0.028}_{-0.028} $  & $\circ$ & $3.7$ \\
${\rm KiDS}\times\{{\rm 2dFLenS}+{\rm BOSS}\} \left({{\rm JBD}+\sum m_{\nu}}\right)$ & $80$ & $\cdots$ & $3.2$ & $1.67^{+0.46}_{-0.46}$ & $1.7$ & $\diamond$ & $ 0.739^{+0.034}_{-0.034} $  & $\circ$ & $2.4$ \\
${\rm KiDS}\times\{{\rm 2dFLenS}+{\rm BOSS}\} \left({{\rm JBD}+\sum m_{\nu},~{\rm no~feedback}}\right)$ & $71$ & $\cdots$ & $\cdots$ & $1.76^{+0.48}_{-0.48}$ & $1.6$ & $\diamond$ & $ 0.707^{+0.028}_{-0.028} $  & $\circ$ & $3.7$ \\
${\rm KiDS}\times\{{\rm 2dFLenS}+{\rm BOSS}\} \left({{\rm JBD}+G_{\rm matter}}\right)$ & $50$ & $1.07^{+0.12}_{-0.15}$ & $3.4$ & $1.54^{+0.47}_{-0.47}$ & $\cdots$ & $\diamond$ & $ 0.717^{+0.064}_{-0.073} $  & $\circ$ & $1.7$ \\
${\rm KiDS}\times\{{\rm 2dFLenS}+{\rm BOSS}\} \left({{\rm JBD}+G_{\rm matter}+\sum m_{\nu}}\right)$ & $78$ & $1.03^{+0.11}_{-0.15}$ & $3.3$ & $1.67^{+0.47}_{-0.47}$ & $1.7$ & $\diamond$ & $ 0.731^{+0.064}_{-0.070} $  & $\circ$ & $1.6$ \\
${\rm KiDS}\times{\rm 2dFLenS}+{\rm All}$-${\rm BOSS} \left({{\rm JBD}+G_{\rm matter}+\sum m_{\nu}}\right)$ & $83$ & $1.05^{+0.18}_{-0.20}$ & $3.2$ & $1.36^{+0.49}_{-0.44}$ & $1.4$ & $73.9^{+8.1}_{-8.2}$ & $0.737^{+0.064}_{-0.085}$ & $0.02$ & $1.3$ \\
\hline
\end{tabularx}
\end{table*}

\subsubsection{Planck 2018 $+$ ACT DR4 versus Planck 2015 $+$ ACT DR3}

We can further contrast the Planck 2018 constraints to those obtained with the baseline 2015 dataset (in other words, we are here comparing Planck 2018 ``TT,TE,EE+lowE'' with Planck 2015 ``TT+lowP''). As shown in Table~\ref{subtab1} and Fig.~\ref{figwbd}, in the restricted JBD model, the lower bound on $\omega_{\rm BD}$ for Planck 2018 is more than a factor of two stronger than Planck 2015, while the lower bound for Planck~2018~+~ACT~DR4 is $50\%$ stronger than the bound from Planck~2015~+~ACT~DR3. As our Planck 2018 baseline setup includes both the CMB temperature and polarization, while the Planck 2015 setup only includes the CMB temperature (and low-$\ell$ LFI polarization), this highlights the impact of including the high-multipole CMB polarization in the analysis (along with the new HFI low-multipole polarization likelihood in lieu of the LFI likelihood). 

Moreover, by allowing for the effective gravitational constant to vary, the tension in the Hubble constant is ameliorated by $2\sigma$ for Planck 2015 (instead of $1\sigma$ for Planck 2018, where high-multipole polarization is notably included). Indeed, as the Planck 2018 small-scale polarization is excluded from the analysis in Sec.~\ref{cmbpolsec}, the parameter constraints are more comparable to those from Planck 2015. In Appendix~\ref{planckapp}, we will further contrast the constraints from these two datasets when including the CMB temperature, polarization, and lensing reconstruction in both cases.

\subsubsection{Model selection}

We quantify the model selection preference for the JBD gravity model relative to the standard model by the deviance information criterion (discussed in Sec.~\ref{modelselec}), where the additional degrees of freedom impose a penalty on the extended model through the Bayesian complexity. As shown in Appendix~\ref{extraparamsdicapp} (specifically Table~\ref{subtabdic}), for all of the models considered here, $\Delta {\rm DIC} \gtrsim 0$, indicating no model selection preference for JBD gravity relative to $\Lambda$CDM.

\section{Results: Weak lensing and overlapping redshift-space galaxy clustering}
\label{wlgcsec}

We next consider the combined analysis of KiDS overlapping with 2dFLenS and BOSS, where JBD has an impact on both the expansion history and the growth of structure (see Sec.~\ref{sec3x2pt}). We do not consider KiDS cosmic shear on its own but always in conjunction with the spectroscopic surveys to leverage the ability to constrain modified gravity. 

\subsection{KiDS$\times$\{2dFLenS$+$BOSS\}}
\label{wlgcsec1}

\subsubsection{Correlations of modified gravity, neutrino mass, baryonic feedback, and intrinsic alignments}

In the analysis of KiDS$\times$\{2dFLenS$+$BOSS\}, we constrain $\omega_{\rm BD} \gtrsim 100$ (shown in Table~\ref{subtab2}) for all of the different cosmological models considered (where we include or exclude the baryonic feedback amplitude, $B$, the sum of neutrino masses, $\sum m_{\nu}$, and the present effective gravitational constant, $G_{\rm matter}/G$, as free parameters in the model). As shown in Fig.~\ref{figkids}, this consistency in the bounds is explained by the lack of correlation between $\omega_{\rm BD}$ and $\{B, \sum m_{\nu}, G_{\rm matter}/G\}$. The effective gravitational constant is further marginally correlated with the baryonic feedback amplitude and uncorrelated with the sum of neutrino masses, while the latter two are anti-correlated with one another.

The $95\%$ upper bound on the sum of neutrino masses is within $1.6$~eV to $1.7$~eV for the different cases, and the upper bound on the baryonic feedback amplitude varies between $3.1$ to $3.4$ (near the border of the no-feedback, or ``dark matter only,'' value of $B = 3.13$~\cite{Mead15}). We find, however, that the posterior for the baryonic feedback amplitude peaks at $B\lesssim2$ suggesting a preference for strong feedback (as noted in Ref.~\cite{Joudaki:2017zdt}). Moreover, we do not find a significant correlation between the intrinsic alignment amplitude and $\{B,\sum m_{\nu}, \omega_{\rm BD}, G_{\rm matter}/G\}$, such that the posterior mean for $A_{\rm IA}$ varies between $1.54$ to $1.78$ for the different cosmological setups while the uncertainty changes by less than 5\% (from $0.47$ to $0.49$). 

Beyond $\omega_{\rm BD}$, in the unrestricted JBD model, we constrain $G_{\rm matter}/G = 1.07^{+0.12}_{-0.15}$ when the neutrinos are assumed massless and $G_{\rm matter}/G = 1.03^{+0.11}_{-0.15}$ when the sum of neutrino masses is allowed to vary (the marginally tighter constraint on $G_{\rm matter}/G$ is due to the one-sided nature of the neutrino mass bounds), both in full agreement with the GR expectation and nearly a factor of four weaker than the constraint from the Planck CMB.

\subsubsection{Impact on the $S_8$ tension}

KiDS$\times$\{2dFLenS$+$BOSS\} is not sufficiently powerful to meaningfully constrain the Hubble constant. However, this dataset does constrain $S_8$ in a way that is sensitive to the cosmological modeling. In the $\Lambda$CDM limit, $S_8 = 0.746^{+0.034}_{-0.035}$, which is only marginally affected (at the sub-percent level) by a restricted coupling to the JBD scalar field (i.e.~to $\omega_{\rm BD}$;~see Table~\ref{subtab2}). We find a 20\% decrease in the uncertainty in $S_8$ and $\sim1\sigma$ shift towards lower values when neglecting baryonic feedback, in agreement with earlier KiDS analyses (e.g.~Refs.~\cite{Hildebrandt16, Joudaki:2017zdt};~also see Refs.~\cite{hildebrandt19, asgari20b}). Moreover, as found in earlier analyses (e.g.~Refs.~\cite{joudaki17,Joudaki:2017zdt}), varying the sum of neutrino masses decreases the posterior mean of $S_8$, here by approximately $\Delta S_8 \sim -0.01$ corresponding to a $0.3\sigma$ negative shift (with marginal change in the uncertainty).

As we consider an unrestricted JBD model (where $G_{\rm matter}$ is varied alongside $\omega_{\rm BD}$), the additional degree of freedom degrades the uncertainty on $S_8$ by nearly a factor of two, such that the level of concordance with Planck improves to $T(S_8) = 1.6\sigma$ (an improvement by $|\Delta T(S_8)| = 0.8\sigma$). Moreover, the anti-correlation of $S_8$ with $G_{\rm matter}$ lowers the posterior mean by $\Delta S_8 = -0.030$ when the neutrinos are assumed massless and by $\Delta S_8 = -0.008$ when the sum of neutrino masses is varied. In other words, while a variation of $\sum m_{\nu}$ lowers the posterior mean of $S_8$ by approximately $0.01$ in $\Lambda$CDM and in a restricted JBD model, the parameter correlations in the unrestricted JBD model drive the opposite behavior (i.e.~a positive shift of $\Delta S_8 = 0.014$).

\subsubsection{Model selection}

As shown in Appendix~\ref{extraparamsdicapp} (specifically Table~\ref{subtabdic}), and in agreement with earlier findings (e.g.~\cite{Hildebrandt16, Joudaki:2017zdt}), including baryonic feedback is marginally favored in a model selection sense ($\Delta {\rm DIC} \simeq -1$). For all of the cosmological extensions considered, there is no meaningful statistical preference for the extended model (considering both restricted and unrestricted JBD gravity, with and without massive neutrinos) relative to $\Lambda$CDM, with $\Delta {\rm DIC} \gtrsim 0$. 

\subsection{KiDS$\times$2dFLenS $+$ All-BOSS}
\label{wlgcsec2}

As the BOSS dataset in Sec.~$\ref{wlgcsec1}$ is restricted to the overlapping regions with KiDS, we also consider a scenario where the full BOSS dataset is included in our analysis. 
Here, we do not consider the galaxy-galaxy lensing signal between KiDS and BOSS given the marginal improvements in the cosmological constraining power that this would provide.\footnote{As discussed in Sec.~\ref{bossdatasec}, this is because the overlapping area between KiDS-450 and BOSS DR10 is more than a factor of $50$ smaller than the full BOSS area, and does not increase for later BOSS data releases. As a result, the improvement in the BOSS galaxy bias constraints from including galaxy-galaxy lensing with KiDS is modest, as shown in e.g.~Ref.~\cite{Joudaki:2017zdt}.} Instead, we consider ``$3 \times 2{\rm pt}$'' measurements for KiDS and 2dFLenS (i.e.~cosmic shear, galaxy-galaxy lensing, and overlapping redshift-space galaxy clustering) along with the BOSS DR12 growth rate, AP parameter, and BAO distance measurements (described in sections~\ref{sec3x2pt} and \ref{bossdatasec}, respectively). We denote this setup ``KiDS$\times$2dFLenS $+$ All-BOSS''.

\subsubsection{$\Lambda$CDM comparisons: smaller overlapping spectroscopic area versus larger disconnected area}

We first compare the $\Lambda$CDM results for KiDS$\times$2dFLenS $+$ All-BOSS against KiDS$\times$\{2dFLenS$+$BOSS\}, finding a constraint on $S_8 = 0.745^{+0.029}_{-0.028}$ in our new setup as compared to $S_8 = 0.746^{+0.034}_{-0.035}$ when BOSS is restricted to the overlapping area with KiDS. This corresponds to a remarkable consistency in the posterior mean (less than $0.03\sigma$ shift) and a $20\%$ improvement in the uncertainty. As a result, the $S_8$ tension with the Planck CMB is at the level of $2.8\sigma$ (an increase by $0.4\sigma$). We further note that as compared to KiDS$\times$2dFLenS alone, where $S_8 = 0.736^{+0.039}_{-0.038}$, the posterior mean increases by $0.3\sigma$ and the uncertainty decreases by $25\%$. Given the full BOSS dataset, we are moreover able to infer the Hubble constant, $H_0 = 70.6^{+5.8}_{-4.4} ~ {\rm km} \, {\rm s}^{-1} {\rm Mpc}^{-1}$, which reflects a strong (closer than $1\sigma$) agreement with both the Planck CMB and Riess et al.~(2019)~\cite{riess2019} measurements of the Hubble constant. 

We can also compare the constraints on the astrophysical systematics, focusing on the self-calibration of the baryonic feedback and intrinsic alignments. As the constraint on the baryonic feedback amplitude, $B$, is driven by the cosmic shear (given its sensitivity to nonlinear scales in the matter power spectrum along with the linear-scale restriction for the galaxy-galaxy lensing and multipole power spectra  to avoid nonlinearities in the galaxy bias in KiDS$\times$\{2dFLenS$+$BOSS\}), we find strong consistency in the upper bound on the baryonic feedback amplitude, such that $B<3.3$ for KiDS$\times$2dFLenS $+$ All-BOSS as compared to $B<3.2$ for KiDS$\times$\{2dFLenS$+$BOSS\} and $B<3.5$ for KiDS$\times$2dFLenS alone (all at $95\%$~CL).

\begin{figure*}
\vspace{-0.7em}
\includegraphics[width=0.94\hsize]{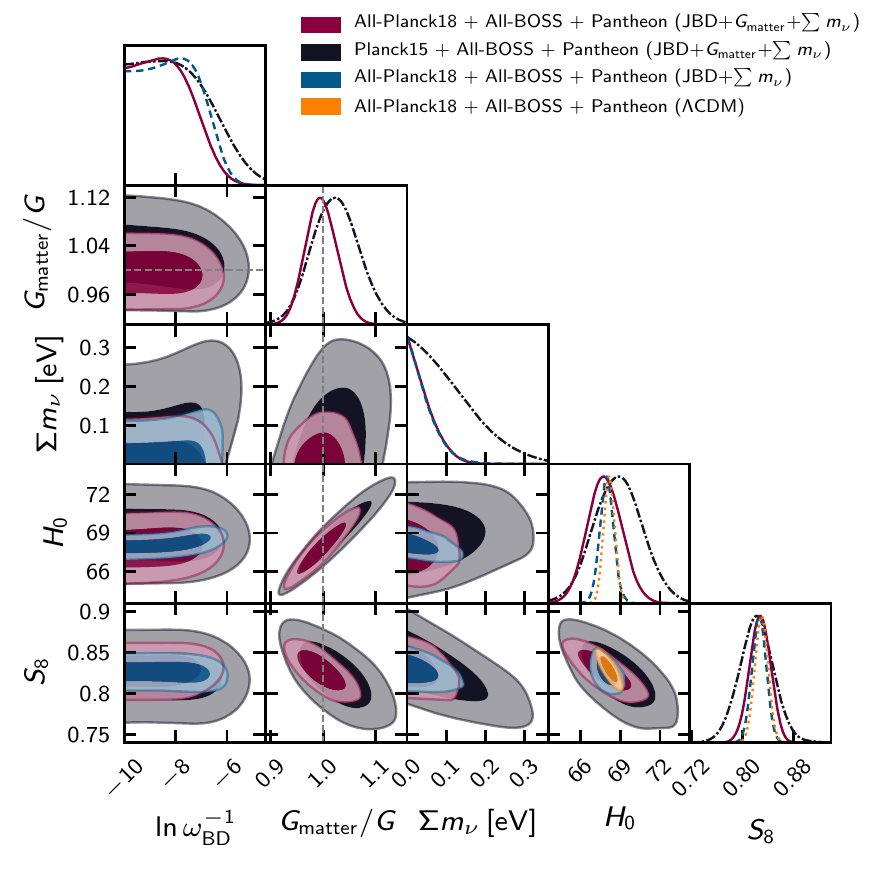}
\vspace{-2.9em}
\caption{\label{figplanckextra}
Marginalized posterior distributions (inner $68\%$~CL, outer $95\%$~CL) of the JBD parameter, $\ln \omega_{\rm BD}^{-1}$, the present effective gravitational constant, $\smash{G_{\rm matter}/G}$, the sum of neutrino masses, $\smash{\sum m_{\nu}}$, the Hubble constant, $H_0$ (in units of $\smash{{\rm km} \, {\rm s}^{-1} {\rm Mpc}^{-1}}$), and $\smash{S_8 = \sigma_8 \sqrt{\Omega_{\mathrm m}/0.3}}$ from the Planck, BOSS, and Pantheon datasets. All other standard cosmological and systematics parameters are simultaneously varied. For visual clarity, we have zoomed in on the $\smash{\ln \omega_{\rm BD}^{-1}}$ axis where the distributions 
flatten towards the GR limit at $-\infty$ (in practice towards $\smash{\ln \omega_{\rm BD}^{-1}  = -17}$).
}
\end{figure*}

\begin{table*}
\vspace{-0.7em}
\caption{\label{subtab5} Marginalized posterior means and $68\%$ confidence intervals for a subset of the cosmological parameters. For the JBD parameter, $\omega_{\rm BD}$, and the sum of neutrino masses, $\sum m_{\nu}$, we quote the $95\%$ confidence lower and upper bounds, respectively. The sum of neutrino masses is in units of eV, the Hubble constant $H_0$ is in km\,s$^{-1}$\,Mpc$^{-1}$,  $S_8=\sigma_8\sqrt{\Omega_{\rm m}/0.3}$, and ``$\cdots$'' implies that the parameter is not varied in the analysis. 
The tensions $T(H_0)$ and $T(S_8)$ are against Riess et al.~(2019)~\cite{riess2019} and KiDS$\times$2dFLenS, respectively. See Table~\ref{subtab1} for further details.
}
\renewcommand{\footnoterule}{} 
\begin{tabularx}{\textwidth}{l@{\extracolsep{\fill}}ccccc|cc}
\hline
Probe setup & $\omega_{\rm BD}$ & $G_{\rm matter}/G$ & $\sum m_{\nu}$ & $H_0$ & $S_8$ & $T(H_0)$ & $T(S_8)$\\
\hline
${\rm All}$-${\rm Planck18}+{\rm All}$-${\rm BOSS}+{\rm Pantheon} \left({\Lambda{\rm CDM}}\right)$ & $\cdots$ & $\cdots$ & $\cdots$ & $68.15^{+0.42}_{-0.41}$  & $0.830^{+0.010}_{-0.010}$ & $4.0$ & $2.3$\\
${\rm All}$-${\rm Planck18}+{\rm All}$-${\rm BOSS}+{\rm Pantheon} \left({\Lambda{\rm CDM}+\sum m_{\nu}}\right)$ & $\cdots$ & $\cdots$ & $0.11$ & $67.86^{+0.51}_{-0.44}$  & $0.826^{+0.011}_{-0.011}$ & $4.1$ & $2.7$\\
${\rm All}$-${\rm Planck18}+{\rm All}$-${\rm BOSS}+{\rm Pantheon} \left({{\rm JBD}}\right)$ & $1460$ & $\cdots$ & $\cdots$ & $68.22^{+0.42}_{-0.48}$ & $0.830^{+0.010}_{-0.010}$ & $3.9$ & $2.3$\\
${\rm All}$-${\rm Planck18}+{\rm All}$-${\rm BOSS}+{\rm Pantheon} \left({{\rm JBD}+\sum m_{\nu}}\right)$ & $970$ & $\cdots$ & $0.11$ & $67.97^{+0.51}_{-0.52}$  & $0.826^{+0.011}_{-0.011}$ & $4.0$ & $2.6$\\
${\rm All}$-${\rm Planck18}+{\rm All}$-${\rm BOSS}+{\rm Pantheon} \left({{\rm JBD}+G_{\rm matter}}\right)$ & $2040$ & $0.989^{+0.030}_{-0.030}$ & $\cdots$ & $67.80^{+1.28}_{-1.32}$ & $0.835^{+0.015}_{-0.016}$ & $3.2$ & $0.7$\\
${\rm All}$-${\rm Planck18}+{\rm All}$-${\rm BOSS}+{\rm Pantheon} \left({{\rm JBD}+G_{\rm matter}+\sum m_{\nu}}\right)$ & $1340$ & $0.997^{+0.029}_{-0.031}$ & $0.11$ & $67.82^{+1.29}_{-1.28}$  & $0.828^{+0.016}_{-0.016}$ & $3.2$ & $0.6$\\
\hline
\end{tabularx}
\end{table*}

The intrinsic alignment amplitude is constrained to $A_{\rm IA} = 1.12^{+0.52}_{-0.46}$ for KiDS$\times$2dFLenS $+$ All-BOSS, which reflects an approximately $1\sigma$ downward shift as compared to the measurement from KiDS$\times$\{2dFLenS$+$BOSS\} (where $A_{\rm IA} = 1.71^{+0.48}_{-0.47}$). The downward shift in the amplitude brings it in greater agreement with the result from KiDS alone (where $A_{\rm IA} = 1.16^{+0.77}_{-0.60}$ in Ref.~\cite{Joudaki:2017zdt}) and from KiDS$\times$2dFLenS (where $A_{\rm IA} = 1.23^{+0.69}_{-0.55}$). In spite of the overall weaker parameter constraints, we note that the constraint on $A_{\rm IA}$ is marginally stronger (by 3\%) in the case of KiDS$\times$\{2dFLenS$+$BOSS\} relative to KiDS$\times$2dFLenS $+$ All-BOSS, given the additional information from galaxy-galaxy lensing between KiDS and BOSS (this can be compared to the $30\%$ improvement in the $A_{\rm IA}$ uncertainty that galaxy-galaxy lensing provides for KiDS$\times$\{2dFLenS$+$BOSS\} as compared to KiDS alone~\cite{Joudaki:2017zdt}). 

\subsubsection{Moving beyond $\Lambda$CDM}

As we turn to an extended cosmological model with unrestricted JBD gravity and massive neutrinos, we constrain $\sum m_{\nu} < 1.4~{\rm eV}$ ($95\%$~CL), $\omega_{\rm BD} > 83$ ($95\%$~CL), and $G_{\rm matter}/G = 1.05^{+0.18}_{-0.20}$, all in agreement with the fiducial model expectation. While the $\sum m_{\nu}$ and $\omega_{\rm BD}$ bounds are marginally different to those obtained from the KiDS$\times$\{2dFLenS$+$BOSS\} dataset (where the respective bounds are $\omega_{\rm BD} > 78$ and $\sum m_{\nu} < 1.7~{\rm eV}$ at $95\%$~CL), the constraint on $G_{\rm matter}/G$ is approximately $50\%$ weaker as compared to that obtained from KiDS$\times$\{2dFLenS$+$BOSS\}. In other words, in constraining modified gravity through the unrestricted JBD model, the self-consistent $3\times2{\rm pt}$ analysis in KiDS$\times$\{2dFLenS$+$BOSS\} is here found to be more powerful than the alternative approach where growth rate and distance measurements from the full BOSS dataset are independently included in KiDS$\times$2dFLenS $+$ All-BOSS.

In the unrestricted JBD model with massive neutrinos, we continue to find a weak correlation for the baryonic feedback amplitude with the neutrino mass and modified gravity degrees of freedom (similarly for the intrinsic alignments). Concretely, $B < 3.2$ ($95\%$~CL), which agrees to within $|\Delta B| = 0.1$ with the corresponding bound in $\Lambda$CDM (and to that obtained from KiDS$\times$\{2dFLenS$+$BOSS\}). Analogous to that found for the Planck CMB, mainly as a result of the positive correlation between $G_{\rm matter}/G$ and $H_0$, the Hubble constant is shifted to larger values (by $\Delta H_0 = 3.3 ~ {\rm km} \, {\rm s}^{-1} {\rm Mpc}^{-1}$). As expected, in this extended cosmology, the uncertainty on the Hubble constant is larger than in $\Lambda$CDM (by $60\%$), and continues to be consistent with both the Planck CMB and Riess et al.~(2019)~\cite{riess2019} measurements to within $1\sigma$.

Driven by the anti-correlation with $G_{\rm matter}/G$, in the unrestricted JBD model with massive neutrinos we constrain $S_8 = 0.737^{+0.064}_{-0.085}$, which reflects a downward shift by $\Delta S_8 = -0.008$ ($0.3\sigma$ relative to the $\Lambda$CDM uncertainty) and factor of $2.6$ increase in the uncertainty relative to $\Lambda$CDM. The constraint on $S_8$ is weaker (by $10\%$) than that obtained from KiDS$\times$\{2dFLenS$+$BOSS\}, which is consistent with the correspondingly weaker constraint on $G_{\rm matter}/G$. Here, the constraint on $S_8$ agrees with the Planck CMB constraint at $1.0\sigma$. Although the concordance between probes increases for this extended cosmology, and the best-fit $\chi^2_{\rm eff}$ is marginally improved relative to $\Lambda$CDM (by $\Delta\chi^2_{\rm eff} = -1.5$), the unrestricted JBD model is not found to be favored from the standpoint of model selection (given $\Delta {\rm DIC} = 1.9$).

\section{Results: Distances and growth rates in combined analyses}
\label{distsec}

\subsection{All-Planck $+$ All-BOSS $+$ Pantheon}
\label{distsec1}

\subsubsection{Constraining JBD gravity}

We next turn to the combined analysis of the Planck 2018 CMB temperature, polarization, and lensing reconstruction (``All-Planck''), the BOSS BAO distances, AP parameters, and growth rates (``All-BOSS''), along with the Pantheon supernova distances. The combined analysis of these datasets allows for constraints on the coupling constant $\omega_{\rm BD} > 1460$~($95\%$~CL) in the restricted JBD model with massless neutrinos, and $\omega_{\rm BD} > 970$~($95\%$~CL) when including the sum of neutrino masses as an additional degree of freedom (see Table~\ref{subtab5} and Fig.~\ref{figplanckextra}). 

These constraints can be contrasted with those found for the Planck CMB temperature and polarization alone (Sec.~\ref{cmbresultssec}), where $\omega_{\rm BD} > 1150$~($95\%$~CL) in the restricted JBD model and $\omega_{\rm BD} > 1710$~($95\%$~CL) when further allowing the sum of neutrino masses to vary in the analysis. In other words, as Planck (temperature and polarization) is combined with the lower redshift probes (BOSS, Pantheon, and Planck lensing reconstruction), the lower bound on the coupling constant increases in the massless neutrino scenario and decreases as the sum of neutrino masses is allowed to vary.

The lower bound on the coupling constant further sharpens as we consider an unrestricted JBD model. Here, $\omega_{\rm BD} > 2040$~($95\%$~CL) in a model with massless neutrinos, and $\omega_{\rm BD} > 1340$~($95\%$~CL) in a model where the sum of neutrino masses is further varied (which can be compared to the bound from the Planck temperature and polarization alone, where $\omega_{\rm BD} > 1120$ at $95\%$~CL). In the unrestricted JBD model with massless neutrinos, we constrain $G_{\rm matter}/G = 0.989^{+0.030}_{-0.030}$ in concordance with the GR expectation to within $1\sigma$. This constraint is robust to the corresponding setup where the sum of neutrino masses is varied, as $\Delta G_{\rm matter}/G = +0.008$ with effectively the same uncertainty, and is driven by the Planck CMB temperature and polarization (as the uncertainty narrows by $6\%$ with the inclusion of the additional datasets).

\subsubsection{The sum of neutrino masses}

The sum of neutrino masses is constrained to $\sum m_{\nu} < 0.11$~eV ($95\%$~CL) in $\Lambda$CDM, which remains robust as we consider a cosmology with JBD gravity (i.e.~less than $0.01$~eV degradation;~both restricted and unrestricted models;~up to factor of $\sim4$ improvement compared to the Planck CMB temperature and polarization alone through the breaking of the geometric degeneracy). As seen in Fig.~\ref{figplanckextra}, there are only mild correlations between the extended cosmological parameters $\{\sum m_{\nu}, \omega_{\rm BD}, G_{\rm matter}\}$, which allows for largely decoupled constraints (with the exception of the $\omega_{\rm BD}$ bound).

On closer inspection, as the strength of JBD gravity increases (i.e.~as $\omega_{\rm BD}$ decreases), there is a rise in the sum of neutrino masses, while $G_{\rm matter}/G$ appears marginally positively correlated with $\sum m_{\nu}$ and negatively correlated with $\ln \omega_{\rm BD}^{-1}$, such that a stronger effective gravitational constant allows for larger neutrino masses and weaker coupling constant. However, as the data favors $G_{\rm matter}/G \simeq 1$ (to within $1\sigma$), these mild correlations are manifested in the posterior mean to lesser extent. 

\subsubsection{Impact on $H_0$ and $S_8$}

The two derived parameters $H_0 = 68.15^{+0.42}_{-0.41} ~ {\rm km} \, {\rm s}^{-1} {\rm Mpc}^{-1}$ and $S_8 = 0.830^{+0.010}_{-0.010}$ in $\Lambda$CDM degrade by $10\%$ and mildly shift downwards (by $0.7\sigma$ in $H_0$ and $0.4\sigma$ in $S_8$) as we allow the sum of neutrino masses to vary (such that $H_0 = 67.86^{+0.51}_{-0.44} ~ {\rm km} \, {\rm s}^{-1} {\rm Mpc}^{-1}$ and $S_8 = 0.826^{+0.011}_{-0.011}$). In the restricted JBD model, as in Sec.~\ref{planckactsec}, the Hubble constant increases as the coupling constant decreases with the characteristic ``hook-shape'' upturn of the contour in the $\ln \omega_{\rm BD}^{-1}$--$H_0$ plane (which expectedly flattens towards the GR limit). However, the marginalized constraint on $H_0$ only changes mildly relative to that in $\Lambda$CDM (positive shift in the posterior mean by $0.2\sigma$ and $\lesssim10\%$ broadening of the uncertainty;~negligible changes in the case of $S_8$), similar to that found for the Planck CMB temperature and polarization alone.

In the unrestricted JBD model, the contour in the plane of $\ln \omega_{\rm BD}^{-1}$ and $H_0$ is substantially flattened (likewise for the contour in the $\omega_{\rm BD}$--$S_8$ plane), and the correlation between these parameters becomes negligible. Instead, there is a strong correlation of $H_0$ and $S_8$ with the present effective gravitational constant 
(given its impact on the metric potentials along with the expansion and growth histories;~see Figs.~\ref{figth} and~\ref{figcls}). Here, there is a factor of $3.1$ and $1.5$ increase in the uncertainties on $H_0$ and $S_8$ relative to $\Lambda$CDM, respectively, such that $H_0 = 67.80^{+1.28}_{-1.32} ~ {\rm km} \, {\rm s}^{-1} {\rm Mpc}^{-1}$ and $S_8 = 0.835^{+0.015}_{-0.016}$. While the posterior means shift marginally relative to $\Lambda$CDM (by $\Delta H_0 = -0.35 ~ {\rm km} \, {\rm s}^{-1} {\rm Mpc}^{-1}$ and $\Delta S_8 = -0.005$), given the increase in the uncertainties, the concordances with the Riess et al.~(2019)~\cite{riess2019} and KiDS$\times$2dFLenS datasets improve to $3.2\sigma$ (from $4.0\sigma$ in $\Lambda$CDM) and $0.7\sigma$ (from $2.3\sigma$ in $\Lambda$CDM), respectively. In the unrestricted JBD model with massive neutrinos, the additional degree of freedom in $\sum m_{\nu}$ has only a marginal impact on $H_0$ and $S_8$ (the strongest impact is on $S_8$ which shifts downwards by $0.5\sigma$ compared to the unrestricted JBD model with fixed neutrino masses, such that the concordance with KiDS$\times$2dFLenS improves to $0.6\sigma$).

As compared to the corresponding constraints from the Planck CMB temperature and polarization alone (i.e.~by adding BOSS, Pantheon, and Planck lensing), the constraints on $H_0$ and $S_8$ respectively strengthen by $30\%$ and $40\%$ in both $\Lambda$CDM and the restricted JBD model, which in the case of $H_0$ increases to a $60\%$ improvement in $\Lambda$CDM and $50\%$ improvement in the restricted JBD model when additionally including massive neutrinos (the improvement in $S_8$ remains at the $40\%$ level in these cosmologies). In the unrestricted JBD model with massive neutrinos, there is a $35\%$ improvement in the uncertainty on $H_0$ and $30\%$ improvement in the uncertainty on $S_8$. Across the different cosmologies, the posterior mean of $H_0$ ``stabilizes'' with the additional datasets, covering a range of $\Delta H_0 = 0.4 ~ {\rm km} \, {\rm s}^{-1} {\rm Mpc}^{-1}$ in the ``All-Planck18 $+$ All-BOSS $+$ Pantheon'' setup, as compared to $\Delta H_0 = 1.6 ~ {\rm km} \, {\rm s}^{-1} {\rm Mpc}^{-1}$ from the Planck CMB temperature and polarization alone, which reflects an increase in the agreement with the standard model (however, noting that the spread in $\Delta S_8 \lesssim 0.01$ is approximately the same between the two data setups).

\begin{figure*}
\vspace{-0.7em}
\includegraphics[width=0.96\hsize]{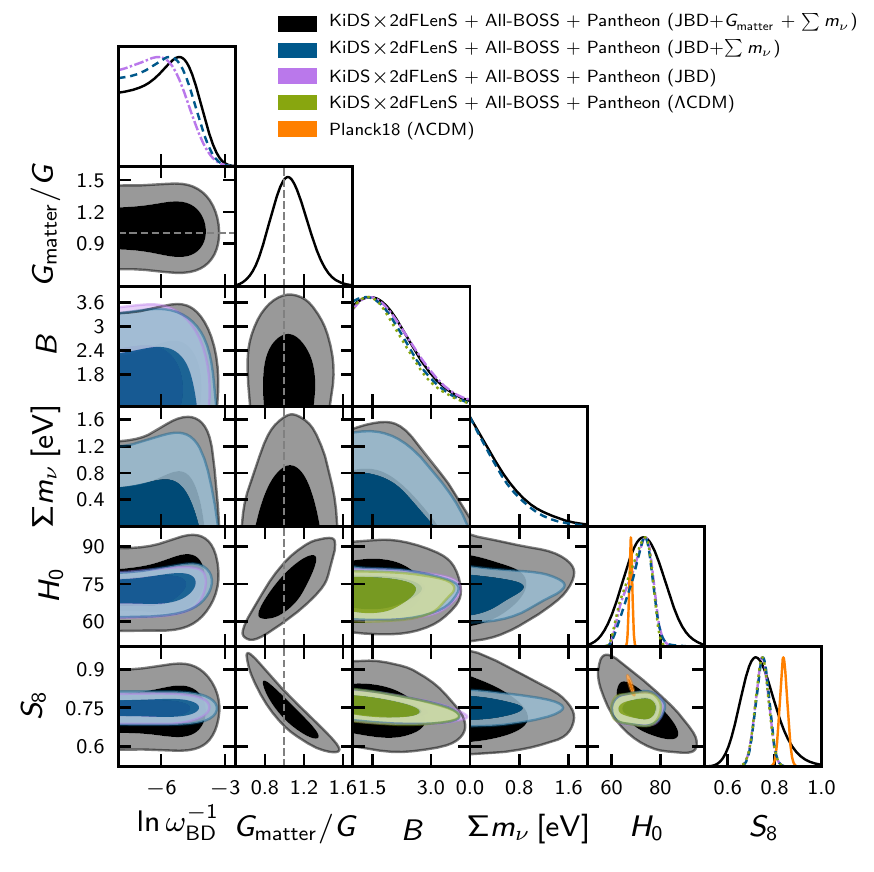}
\vspace{-2.4em}
\caption{\label{figkidsextra} 
Marginalized posterior distributions (inner $68\%$~CL, outer $95\%$~CL) of the JBD parameter, $\ln \omega_{\rm BD}^{-1}$, the present effective gravitational constant, $\smash{G_{\rm matter}/G}$, the baryonic feedback amplitude, $B$, the sum of neutrino masses, $\smash{\sum m_{\nu}}$, the Hubble constant, $H_0$ (in units of $\smash{{\rm km} \, {\rm s}^{-1} {\rm Mpc}^{-1}}$), and $S_8 = \sigma_8 \sqrt{\Omega_{\mathrm m}/0.3}$ from the KiDS, 2dFLenS, BOSS, and Pantheon datasets. We also include the Planck18 $\Lambda$CDM constraints for comparison. All other standard cosmological and systematics parameters are simultaneously varied. For visual clarity, we have zoomed in on the $\smash{\ln \omega_{\rm BD}^{-1}}$ axis where the distributions begin to flatten, as they stay flat and unbounded in the direction of the GR limit (in practice towards $\smash{\ln \omega_{\rm BD}^{-1}  = -17}$).
}
\end{figure*}

\begin{table*}
\vspace{-1.1em}
\caption{\label{subtab4} Marginalized posterior means and $68\%$ confidence intervals for a subset of the cosmological parameters. For the JBD parameter, $\omega_{\rm BD}$, and the sum of neutrino masses, $\sum m_{\nu}$ (in units of eV), we quote the $95\%$ confidence lower and upper bounds, respectively. We also quote the $95\%$ confidence upper bound for the feedback amplitude, $B$. The Hubble constant, $H_0$, is in units of km\,s$^{-1}$\,Mpc$^{-1}$ and we define $S_8=\sigma_8\sqrt{\Omega_{\rm m}/0.3}$. The tensions $T(H_0)$ and $T(S_8)$ are against Riess et al.~(2019)~\cite{riess2019} and Planck18, respectively. See Table~\ref{subtab1} for further~details.
}
\renewcommand{\footnoterule}{} 
\begin{tabularx}{\textwidth}{l@{\extracolsep{\fill}}ccccccc|cc}
\hline
Probe setup & $\omega_{\rm BD}$ & $G_{\rm matter}/G$ & $B$ & $A_{\rm IA}$ & $\sum m_{\nu}$ & $H_0$ & $S_8$ & $T(H_0)$ & $T(S_8)$\\
\hline
${\rm KiDS}\times{\rm 2dFLenS}+{\rm All}$-${\rm BOSS}+{\rm Pantheon} \left({\Lambda{\rm CDM}}\right)$ & $\cdots$ & $\cdots$ & $3.1$ & $1.13^{+0.53}_{-0.46}$ & $\cdots$ & $ 70.8^{+6.0}_{-3.7} $  & $ 0.746^{+0.028}_{-0.028} $  & $0.64$ & $2.8$ \\
${\rm KiDS}\times{\rm 2dFLenS}+{\rm All}$-${\rm BOSS}+{\rm Pantheon} \left({\Lambda{\rm CDM}+\sum m_{\nu}}\right)$ & $\cdots$ & $\cdots$ & $3.1$ & $1.34^{+0.49}_{-0.48}$ & $1.1$ & $ 71.0^{+5.7}_{-3.2} $  & $ 0.748^{+0.028}_{-0.028} $  & $0.65$ & $2.5$ \\
${\rm KiDS}\times{\rm 2dFLenS}+{\rm All}$-${\rm BOSS}+{\rm Pantheon} \left({{\rm JBD}}\right)$ & $140$ & $\cdots$ & $3.3$ & $1.12^{+0.51}_{-0.45}$ & $\cdots$ & $71.3^{+5.9}_{-4.0} $  & $ 0.746^{+0.028}_{-0.028} $  & $0.53$ & $2.8$ \\
${\rm KiDS}\times{\rm 2dFLenS}+{\rm All}$-${\rm BOSS}+{\rm Pantheon} \left({{\rm JBD}+\sum m_{\nu}}\right)$ & $100$ & $\cdots$ & $3.2$ & $1.36^{+0.51}_{-0.46}$ & $1.2$ & $71.7^{+5.4}_{-3.8}$  & $0.748^{+0.028}_{-0.028}$ & $0.48$ & $2.5$ \\
${\rm KiDS}\times{\rm 2dFLenS}+{\rm All}$-${\rm BOSS}+{\rm Pantheon} \left({{\rm JBD}+G_{\rm matter}}\right)$ & $120$ & $1.10^{+0.19}_{-0.22}$ & $3.4$ & $1.13^{+0.51}_{-0.45}$ & $\cdots$ & $74.7^{+8.5}_{-9.3}$ & $0.719^{+0.062}_{-0.085}$ & $0.08$ & $1.5$ \\
${\rm KiDS}\times{\rm 2dFLenS}+{\rm All}$-${\rm BOSS}+{\rm Pantheon} \left({{\rm JBD}+G_{\rm matter}+\sum m_{\nu}}\right)$ & $79$ & $1.05^{+0.18}_{-0.20}$ & $3.3$ & $1.36^{+0.51}_{-0.47}$ & $1.3$ & $73.4^{+8.4}_{-8.3}$ & $0.737^{+0.063}_{-0.085}$ & $0.07$ & $1.4$ \\
\hline
\end{tabularx}
\end{table*}

\subsubsection{Model selection}

As discussed in Sec.~\ref{requirementssec}, for an extended cosmological model to replace the standard model, we require that it exhibits a statistically significant deviation (by $>5\sigma$) in any extended parameters from the standard model expectation, shows a ``strong'' improvement (e.g.~given Jeffreys' scale~\cite{jeffreys}) in a model selection comparison with the standard model (employing statistical tools such as the deviance information criterion or Bayesian evidence~\cite{spiegelhalter02, trotta08}), and ideally alleviates possible discordances between datasets.

We have shown above that the last condition is partly satisfied. However, in spite of the successes of the unrestricted JBD model in improving the concordance between datasets, we note that it is not favored in a model selection comparison with $\Lambda$CDM (given $\Delta {\rm DIC} \gtrsim 0$). This also holds for all of the other extended cosmological models considered here (i.e.~models involving restricted JBD and massive neutrinos). The extended cosmological parameters $\{\omega_{\rm BD}, G_{\rm matter}, \sum m_{\nu}\}$ moreover do not exhibit a statistically significant deviation from the fiducial expectation, such that any increase in the concordance between datasets is driven by the increase in the parameter uncertainties rather than a change in the actual cosmology.

\subsubsection{Comparing against the Planck 2015 dataset}

In Fig.~\ref{figplanckextra}, we further compare the constraints from ``All-Planck18 $+$ All-BOSS $+$ Pantheon'' in the unrestricted JBD model with massive neutrinos to the corresponding constraints from ``Planck15 + All-BOSS + Pantheon'', where only the baseline Planck 2015 CMB temperature and LFI low-$\ell$ polarization is considered instead of the full Planck 2018 dataset. Here, we find substantially weaker constraints on the subspace of parameters $\{\ln \omega_{\rm BD}^{-1}, G_{\rm matter}/G, \sum m_{\nu}, H_0, S_8\}$. In particular, the tension in $H_0$ with the measurement of Riess et al.~(2019)~\cite{riess2019} decreases by $1.0\sigma$, down to $2.2\sigma$. As seen in Sec.~\ref{cmbresultssec}, this is similar to the difference in the Hubble constant tension in the unrestricted JBD model when comparing the Planck 2018 temperature and polarization (where $T(H_0) \approx 3\sigma$) with the Planck~2015 temperature and LFI low-$\ell$ polarization (where $T(H_0) \lesssim 2\sigma$). In other words, it is predominantly the Planck~2018 polarization information that is driving the $H_0$ tension higher in this model.

We have also considered constraints on the restricted JBD models where the Planck~2018 and 2015 setups both include the respective high-multipole polarization and lensing measurements (i.e.~comparing ``All-Planck18 $+$ All-BOSS $+$ Pantheon'' with ``All-Planck15 $+$ All-BOSS $+$ Pantheon''). As shown in Appendix~\ref{planckapp} (specifically Fig.~\ref{figplanck18rest}, also see Table~\ref{subtabplanck15}), we find no significant differences in the inferences in the subspace of $\{\ln \omega_{\rm BD}^{-1}, \sum m_{\nu}, \tau, H_0, S_8\}$ aside from the constraint on $\sum m_{\nu}$, driven by the greater than factor-of-two improved 2018 constraint on the optical depth to reionization, $\tau$ (as $C_{\ell} \propto A_{\rm s} e^{-2\tau}$, such that the constraint on the primordial scalar amplitude, $A_{\rm s}$ is comparably improved given the degeneracy with the optical depth~\cite{planck2018}).

Combining All-BOSS and Pantheon with All-Planck15, we obtain the lower bound on $\omega_{\rm BD} > 1050$~($95\%$~CL) in a restricted JBD model with massless neutrinos, and $\omega_{\rm BD} > 590$~($95\%$~CL) in a restricted JBD model with massive neutrinos (which in both cases corresponds to a weakening of the bound by $\Delta \omega_{\rm BD} \approx -400$ as compared to the equivalent setup with Planck 2018). The sum of neutrino masses is constrained to $\sum m_{\nu} < 0.19$~eV~($95\%$~CL), which corresponds to a $70\%$ weakening relative to the bound from the equivalent setup with Planck 2018.

Moreover, as shown in Appendix~\ref{planckapp} (see Fig.~\ref{figplanck18rest} therein), the $H_0$ and $S_8$ constraints are marginally affected by the change between the Planck 2015 and 2018 datasets, and the respective tensions in these parameters with other datasets remain at a similar level ($<0.1\sigma$ differences). For the restricted JBD model, we have also checked that we continue to find no meaningful model selection preference relative to $\Lambda$CDM for the Planck 2015 data combination (given $\Delta {\rm DIC} \gtrsim 0$ for both neutrino setups).

\begin{figure}
\vspace{-0.5em}
\includegraphics[width=1.03\hsize]{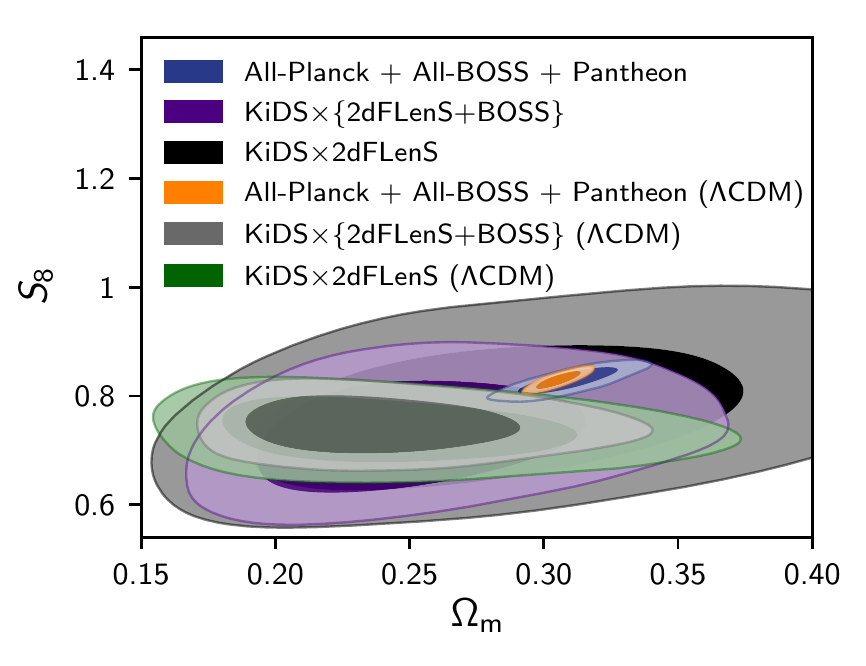}
\vspace{-3em}
\caption{\label{figs8}
Marginalized posterior distributions (inner $68\%$~CL, outer $95\%$~CL) in the $S_8$--$\Omega_{\rm m}$ plane for the dataset combinations All-Planck + All-BOSS + Pantheon, KiDS$\times$\{2dFLenS+BOSS\}, and KiDS$\times$2dFLenS in the unrestricted JBD modified gravity model with massive neutrinos (JBD+$G_{\rm matter}$+$\sum m_{\nu}$) and in $\Lambda$CDM. Here, ``All-Planck'' refers to the Planck 2018 CMB temperature, polarization, and lensing reconstruction, and we simultaneously vary all standard cosmological and systematics parameters.
}
\end{figure}

\begin{figure*}
\vspace{-0.7em}
\includegraphics[width=1.0\hsize]{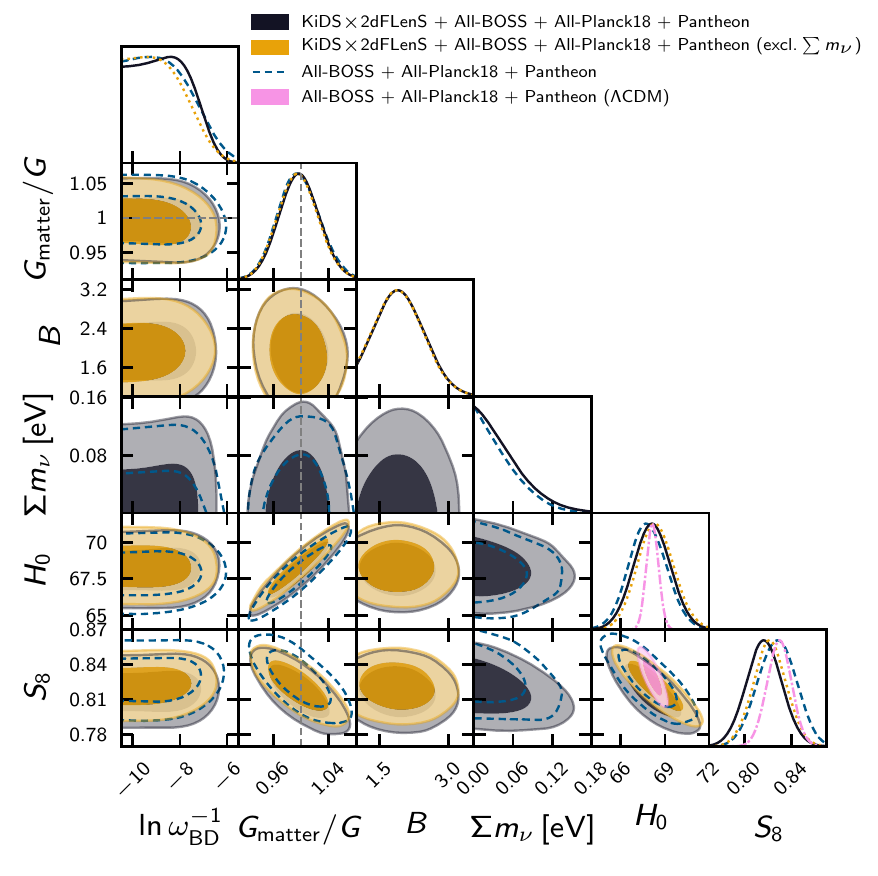}
\vspace{-3.79em}
\caption{\label{figall}
Marginalized posterior distributions (inner $68\%$~CL, outer $95\%$~CL) of the JBD parameter, $\smash{\ln \omega_{\rm BD}^{-1}}$, the present effective gravitational constant, $G_{\rm matter}/G$, the baryonic feedback amplitude, $B$, the sum of neutrino masses, $\sum m_{\nu}$, the Hubble constant, $H_0$ (in units of $\smash{{\rm km} \, {\rm s}^{-1} {\rm Mpc}^{-1}}$), and $\smash{S_8 = \sigma_8 \sqrt{\Omega_{\mathrm m}/0.3}}$ from different combinations of the KiDS, 2dFLenS, BOSS, Pantheon, and Planck datasets. All other standard cosmological and systematics parameters are simultaneously varied. For visual clarity, we have zoomed in on the $\smash{\ln \omega_{\rm BD}^{-1}}$ axis where the distributions flatten towards the GR limit at $-\infty$ (in practice towards the negative end of the prior range at $\smash{\ln \omega_{\rm BD}^{-1}  = -17}$).
}
\end{figure*}

\subsection{KiDS$\times$2dFLenS $+$ All-BOSS $+$ Pantheon}
\label{distsec2}

\subsubsection{Impact of including Pantheon distances}

In comparing the cosmological parameter constraints from KiDS$\times$2dFLenS $+$ All-BOSS $+$ Pantheon with the earlier constraints from KiDS$\times$2dFLenS $+$ All-BOSS (Sec.~\ref{wlgcsec2}), we find that the addition of Pantheon supernova distance measurements has a marginal impact. In $\Lambda$CDM, we constrain $B>3.1$ ($95\%$~CL), $A_{\rm IA} = 1.13^{+0.53}_{-0.46}$, $H_0 = 70.8^{+6.0}_{-3.7} ~ {\rm km} \, {\rm s}^{-1} {\rm Mpc}^{-1}$, and $S_8 = 0.746^{+0.028}_{-0.028}$. As compared to a setup that excludes the Pantheon dataset, this reflects a minor change in the upper bound of the baryonic feedback amplitude (by $\Delta B = -0.2$), along with marginal shifts in the posterior mean of the IA amplitude (by $\Delta A_{\rm IA} = 0.01$, with sub-percent level change in the uncertainty), the Hubble constant (by $\Delta H_0 = 0.2 ~ {\rm km} \, {\rm s}^{-1} {\rm Mpc}^{-1}$, with $5\%$ decrease in the uncertainty), and $S_8$ (by $10^{-3}$, with sub-percent level change in the uncertainty). 

We find similarly small changes in these parameters for an extended cosmology that includes massive neutrinos and unrestricted JBD gravity. The extended cosmological parameters are constrained to $\sum m_{\nu} < 1.3$~eV ($95\%$~CL), $\omega_{\rm BD} > 79$ ($95\%$~CL), and $G_{\rm matter} / G = 1.05^{+0.18}_{-0.20}$, such that the lower and upper bounds differ by $\Delta \omega_{\rm BD} = -4$ and $\Delta \sum m_{\nu} = -0.1$~eV, respectively, while the constraint on the effective gravitational constant agrees to within sub-percent level from that found when the Pantheon dataset is excluded. 

\subsubsection{Robustness in the cosmological constraints}

In comparing the KiDS$\times$2dFLenS $+$ All-BOSS $+$ Pantheon constraints in different JBD cosmologies with one another (restricted and unrestricted JBD cosmologies with and without neutrino mass), we find a robustness in the constraints to the extended cosmological degrees of freedom. In general, $\sum m_{\nu} \lesssim 1$~eV ($95\%$~CL), $\omega_{\rm BD} \gtrsim 10^2$ ($95\%$~CL), and $G_{\rm matter}/G \sim 1$ to within $68\%$~CL where the uncertainty is at the $\Delta G_{\rm matter}/G \simeq 0.2$ level (the inequalities here respectively denote ``less/greater than approximately'').

These constraints are substantially weaker than those found for All-Planck18 $+$ All-BOSS $+$ Pantheon (Sec.~\ref{distsec1}), such that the respective bounds on $\omega_{\rm BD}$ and $\sum m_{\nu}$ are approximately an order of magnitude weaker and the constraint on $G_{\rm matter}$ is a factor of six weaker. The weak correlations between the modified gravity parameters and the sum of neutrino masses (along with weak lensing systematics such as baryonic feedback) are shown in Fig.~\ref{figkidsextra}, and are consistent with the comparably weak correlations found for KiDS$\times$\{2dFLenS$+$BOSS\} and KiDS$\times$2dFLenS $+$ All-BOSS (Secs.~\ref{wlgcsec1} and \ref{wlgcsec2}).

\subsubsection{$H_0$ and $S_8$ constraints}

Turning to the constraints on the derived $H_0$ and $S_8$ parameters, correlations with $G_{\rm matter}/G$ continue to give rise to both a shift in the posterior mean and increase in the uncertainty. Concretely, for the unrestricted JBD model, $H_0 = 74.7^{+8.5}_{-9.3} ~ {\rm km} \, {\rm s}^{-1} {\rm Mpc}^{-1}$ (while for the restricted JBD model, where $G_{\rm matter}/G$ is fixed to unity, $H_0 = 71.3^{+5.9}_{-4.0} ~ {\rm km} \, {\rm s}^{-1} {\rm Mpc}^{-1}$). In comparison with $\Lambda$CDM, these constraints correspond to a positive shift in the posterior mean by $\Delta H_0 = 0.5 ~ {\rm km} \, {\rm s}^{-1} {\rm Mpc}^{-1}$ (with percent-level increase in the uncertainty) when allowing $\omega_{\rm BD}$ to vary (i.e.~for the restricted JBD model), and an additional increase in the posterior mean by $\Delta H_0 = 3.4 ~ {\rm km} \, {\rm s}^{-1} {\rm Mpc}^{-1}$ (with an $80\%$ increase in the uncertainty) when further allowing $G_{\rm matter}/G$ to vary (i.e.~for the unrestricted JBD model).

For the $S_8$ parameter, the constraints are effectively at the sub-percent level between the restricted JBD model and $\Lambda$CDM, while $S_8 = 0.719^{+0.062}_{-0.085}$ in the unrestricted JBD model which corresponds to a decrease in the posterior mean by $\Delta S_8 = 0.027$ and a factor of 2.6 increase in the uncertainty (similar to that found when the Pantheon dataset is excluded from the analysis in Sec.~\ref{wlgcsec2}). While the increased uncertainly on $S_8$ is robust to varying $\sum m_{\nu}$, we note that allowing for this additional degree of freedom gives rise to an increase in the posterior mean by $\Delta S_8 = 0.018$ (such that $S_8 = 0.737^{+0.063}_{-0.085}$ in an unrestricted JBD model with massive neutrinos; similar to that found for KiDS$\times$\{2dFLenS$+$BOSS\}). As in the case of $H_0$, this impact of the sum of neutrino masses on the $S_8$ posterior mean decreases as we fix $G_{\rm matter}/G$ to unity in the restricted JBD and $\Lambda$CDM models (shown in Table~\ref{subtab4}).

\begin{table*}
\vspace{-0.7em}
\caption{\label{subtab6} Marginalized posterior means and $68\%$ confidence intervals for a subset of the cosmological parameters. For the JBD parameter, $\omega_{\rm BD}$, and the sum of neutrino masses, $\sum m_{\nu}$, we quote the $95\%$ confidence lower and upper bounds, respectively. We also quote the $95\%$ confidence upper bound for the baryonic feedback amplitude, $B$, and the tension, $T(H_0)$, with the Riess et al.~(2019)~\cite{riess2019} direct measurement of the Hubble constant. The sum of neutrino masses is in units of eV and the Hubble constant, $H_0$, is in units of km\,s$^{-1}$\,Mpc$^{-1}$. See Table~\ref{subtab1} for further details. There is a minor improvement in the $H_0$ (and $A_{\rm IA}$) constraint when the sum of neutrino masses is varied in the Planck 2015 setup, as explained in Appendix~\ref{planckapp}. For visual clarity, we do not show the constraints on $S_8=\sigma_8\sqrt{\Omega_{\rm m}/0.3}$ in a separate column, but quote them here as $S_8 = \left[0.822^{+0.014}_{-0.014}, 0.817^{+0.015}_{-0.015}, 0.818^{+0.015}_{-0.015}, 0.809^{+0.016}_{-0.015}\right]$ from the first to the fourth row, respectively.
}
\renewcommand{\footnoterule}{} 
\begin{tabularx}{\textwidth}{l@{\extracolsep{\fill}}cccccc|c}
\hline
Probe setup & $\omega_{\rm BD}$ & $G_{\rm matter}/G$ & $B$ & $A_{\rm IA}$ & $\sum m_{\nu}$ & $H_0$ & $T(H_0)$\\
\hline
${\rm KiDS}\times{\rm 2dFLenS}+{\rm All}$-${\rm BOSS}+{\rm All}$-${\rm Planck18}+{\rm Pantheon} \left({\rm JBD}+G_{\rm matter}\right)$ & $2230$ & $0.996^{+0.029}_{-0.029}$ & $2.8$ & $1.54^{+0.38}_{-0.39}$ & $\cdots$ & $68.37^{+1.24}_{-1.24}$ & $3.0$\\
${\rm KiDS}\times{\rm 2dFLenS}+{\rm All}$-${\rm BOSS}+{\rm All}$-${\rm Planck18}+{\rm Pantheon} \left({\rm JBD}+G_{\rm matter}+\sum m_{\nu}\right)$ & $1540$ & $0.997^{+0.029}_{-0.029}$ & $2.8$ & $1.52^{+0.38}_{-0.38}$ & $0.12$ & $68.13^{+1.26}_{-1.25}$ & $3.1$\\
${\rm KiDS}\times{\rm 2dFLenS}+{\rm All}$-${\rm BOSS}+{\rm All}$-${\rm Planck15}+{\rm Pantheon} \left({\rm JBD}+G_{\rm matter}\right)$ & $2270$ & $1.010^{+0.030}_{-0.029}$ & $2.8$ & $1.49^{+0.38}_{-0.39}$ & $\cdots$ & $68.87^{+1.32}_{-1.32}$ & $2.7$\\
${\rm KiDS}\times{\rm 2dFLenS}+{\rm All}$-${\rm BOSS}+{\rm All}$-${\rm Planck15}+{\rm Pantheon} \left({\rm JBD}+G_{\rm matter}+\sum m_{\nu}\right)$ & $1640$ & $1.017^{+0.029}_{-0.030}$ & $2.8$ & $1.52^{+0.36}_{-0.39}$ & $0.21$ & $68.71^{+1.27}_{-1.26}$ & $2.8$\\
\hline
\end{tabularx}
\end{table*}

\subsubsection{$H_0$ and $S_8$ tensions}

Given the increase in the $S_8$ uncertainty in the unrestricted JBD model, the concordance with the Planck CMB increases, such that the respective $S_8$ constraints differ by $1.5\sigma$ as compared to $2.8\sigma$ in both $\Lambda$CDM and the restricted JBD model ($1.4\sigma$ and $2.5\sigma$, respectively, as the sum of neutrino masses is further varied). In $\Lambda$CDM, the posterior mean of $S_8$ differs from that obtained by All-Planck18 $+$ All-BOSS $+$ Pantheon by $\Delta S_8 = 0.084$. This difference in $S_8$ is maximal for an unrestricted JBD model with fixed neutrino masses, where $\Delta S_8 = 0.116$. However, given the increase in the uncertainty on $S_8$ (as noted, by a factor of $2.6$ for KiDS$\times$2dFLenS $+$ All-BOSS $+$ Pantheon and a factor of $1.5$ for All-Planck18 $+$ All-BOSS $+$ Pantheon, both relative to $\Lambda$CDM), this increase in $\Delta S_8$ indeed corresponds to a decrease in the statistical significance of the tension between the datasets, as illustrated in Fig.~\ref{figs8h0}.

While the constraint on the Hubble constant is consistent to within $68\%$~CL with both the direct measurement from Riess et al.~(2019)~\cite{riess2019} and that inferred by the Planck CMB, we note that the posterior mean is generally in greater agreement with the direct measurement (both in $\Lambda$CDM and the extended cosmologies). In other words, the Hubble constant constraint from KiDS$\times$2dFLenS $+$ All-BOSS $+$ Pantheon favors a $\Delta H_0 = 2.6 ~ {\rm km} \, {\rm s}^{-1} {\rm Mpc}^{-1}$ larger value as compared to All-Planck18 $+$ All-BOSS $+$ Pantheon in $\Lambda$CDM, and this difference increases to as much as $\Delta H_0 = 6.9 ~ {\rm km} \, {\rm s}^{-1} {\rm Mpc}^{-1}$ in an unrestricted JBD model with fixed neutrino masses (albeit with larger uncertainty). 
We show this difference in the $H_0$ constraints between KiDS$\times$2dFLenS $+$ All-BOSS $+$ Pantheon and All-Planck18 $+$ All-BOSS $+$ Pantheon in Fig.~\ref{figs8h0}.

\subsubsection{Model selection}

As shown in Appendix~\ref{extraparamsdicapp} (specifically Table~\ref{subtabdic}), despite a decrease in the tension between datasets, we find no model selection preference for the extended cosmological models relative to $\Lambda$CDM (given $\Delta {\rm DIC} \gtrsim 0$).

\section{Results: Fully combined data analysis of the JBD model}
\label{fullsec}

We now proceed to constraining the JBD model with the full combination of datasets. This includes the cosmic shear, galaxy-galaxy lensing, and overlapping redshift-space galaxy clustering (KiDS$\times$2dFLenS), BAO distances, AP distortions, and growth rates (All-BOSS), supernova distances (Pantheon), and the cosmic microwave background temperature, polarization, and lensing reconstruction (All-Planck). In Sec.~\ref{degsec} alone, we additionally include the small-scale CMB data of ACT.\footnote{With the exception of Sec.~\ref{degsec}, where we target degeneracies between parameters that modify the CMB damping tail, we do not include ACT DR4 in the final setup as its covariance with Planck 2018 is neglected in favor of further scale cuts, which has been validated to 5\% precision for certain single-parameter extensions in addition to $\Lambda$CDM~\cite{aiola20}.} The KiDS, 2dFLenS, and BOSS datasets are combined as in Sec.~\ref{wlgcsec2}, where we avoid double-counting information by restricting the $3\times2{\rm pt}$ dataset to KiDS$\times$2dFLenS instead of KiDS$\times$\{2dFLenS+BOSS\}, in order to separately utilize the full BOSS dataset (i.e.~All-BOSS).

\subsubsection{Assessing concordance as a requirement for combining datasets}

We have further taken care to only perform a combined~analysis of concordant datasets. As shown in Fig.~\ref{figs8}, the tension between KiDS$\times$\{2dFLenS+BOSS\} and All-Planck + All-BOSS + Pantheon in the $S_8 - \Omega_{\rm m}$ plane diminishes as we transition from $\Lambda$CDM to the unrestricted JBD cosmology. This holds to even greater extent as we consider the concordance between KiDS$\times$2dFLenS and All-Planck + All-BOSS + Pantheon (such that the $95\%$ confidence level contours fully overlap and there is agreeement in $S_8$ to within $1\sigma$ in the unrestricted JBD model). As observed in Fig.~\ref{figkidsextra}, in the unrestricted JBD model, Planck is also concordant in e.g.~the $S_8$--$H_0$ plane with KiDS$\times$2dFLenS + All-BOSS + Pantheon.

In assessing the concordance between KiDS$\times$2dFLenS $+$ All-BOSS $+$ Pantheon and All-Planck over the full parameter space, we find $\log \mathcal{I} = 0.93$ in in the unrestricted JBD model with massless neutrinos and $\log \mathcal{I} = 0.37$ in the unrestricted JBD model where the sum of neutrino masses is varied, corresponding to substantial and weak concordance between the datasets, respectively. This concordance decreases by $\Delta \log \mathcal{I} \approx -1$ when instead assessing the concordance between All-Planck $+$ All-BOSS $+$ Pantheon and KiDS$\times$2dFLenS. In other words, the ``order'' in which datasets are combined has an impact on the assessment of whether a fully combined analysis of the datasets is warranted. In our case, we take not only the concordance in the combination of All-Planck with the other datasets, but also the agreement in the parameters that lensing constrains most powerfully as motivation for performing a fully combined analysis of the datasets in the unrestricted JBD model. However, we do not quote constraints using the full combination of datasets in $\Lambda$CDM and the restricted JBD model where the datasets are evidently discordant.

\subsection{KiDS$\times$2dFLenS $+$ All-BOSS $+$ All-Planck $+$ Pantheon}
\label{fullsec2}

\subsubsection{JBD gravity and massive neutrinos}

We now consider ``KiDS$\times$2dFLenS $+$ All-BOSS $+$ All-Planck18 $+$ Pantheon'' in the unrestricted JBD model (see Fig.~\ref{figall} and Table~\ref{subtab6}). We obtain our strongest bounds on $\omega_{\rm BD} > 2230$~($95\%$~CL) and $G_{\rm matter}/G = 0.996^{+0.029}_{-0.029}$ as the sum of neutrino masses is fixed, and $\omega_{\rm BD} > 1540$~($95\%$~CL) and $G_{\rm matter}/G = 0.997^{+0.029}_{-0.029}$ in the unrestricted JBD model as the sum of neutrino masses is varied (where $\sum m_{\nu} < 0.12~{\rm eV}$ at $95\%$~CL). In both neutrino setups, the constraints are in agreement with the GR expectation. As compared to ``All-Planck18 $+$ All-BOSS $+$ Pantheon'', where KiDS$\times$2dFLenS is excluded, the $\omega_{\rm BD}$ bounds are strengthened by $\Delta \omega_{\rm BD} = \{190, 200\}$ in the \{massless, massive\} neutrino scenarios, respectively, while the uncertainties on $G_{\rm matter}/G$ are marginally narrowed (by $0.01$ in both neutrino scenarios). Meanwhile, the bound on the sum of neutrino masses is robust (at the $0.01$~eV level), as it is driven by the combination of Planck and BOSS.

\begin{figure*}
\vspace{-0.7em}
\includegraphics[width=1.0\hsize]{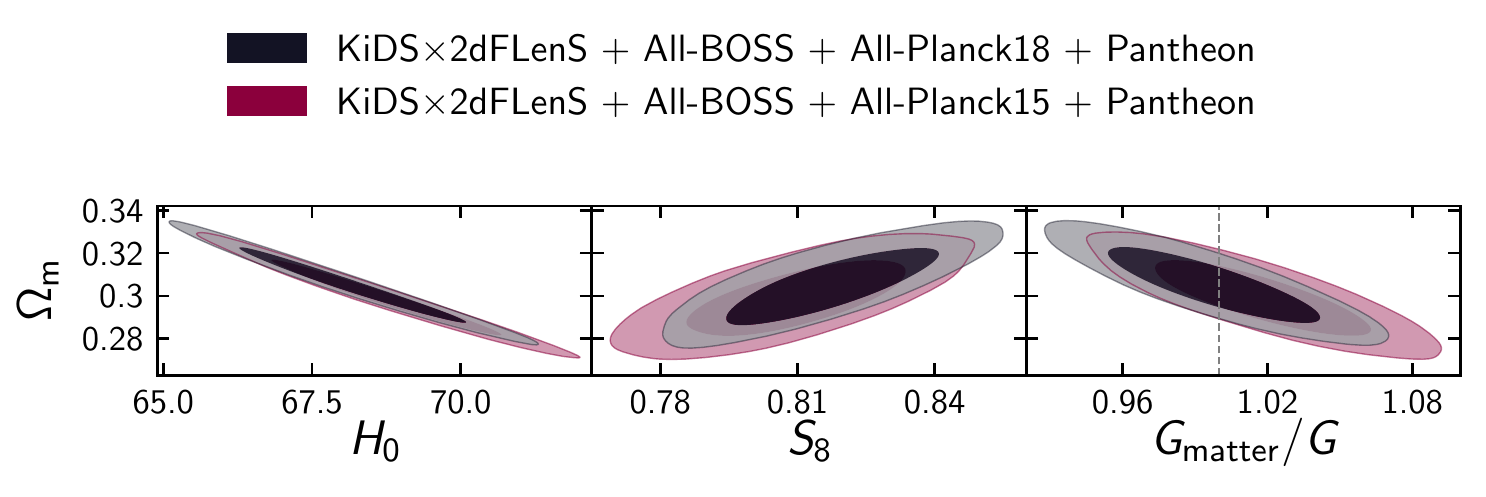}
\vspace{-2.6em}
\caption{\label{figmatter} 
Marginalized posterior distributions (inner $68\%$~CL, outer $95\%$~CL) of the matter density $\Omega_{\rm m}$ against the present effective gravitational constant $\smash{G_{\rm matter}/G}$, Hubble constant $H_0$ (in units of $\smash{{\rm km} \, {\rm s}^{-1} {\rm Mpc}^{-1}}$), and $\smash{S_8 = \sigma_8 \sqrt{\Omega_{\mathrm m}/0.3}}$ from the KiDS, 2dFLenS, BOSS, Pantheon, and Planck datasets. We consider an unrestricted JBD cosmology with massive neutrinos (JBD+$G_{\rm matter}$+$\sum m_{\nu}$), where all standard cosmological and systematics parameters are simultaneously varied. 
}
\end{figure*}

\subsubsection{Baryonic feedback and intrinsic alignments}

In accordance with earlier results using subsets of the data, there are weak correlations between $\{\omega_{\rm BD}, G_{\rm matter}, \sum m_{\nu}\}$ in this unrestricted JBD model. The parameters are also weakly correlated with the baryonic feedback amplitude, where $B < 2.8$ ($95\%$~CL) in both neutrino setups. This upper bound on the feedback amplitude is in agreement with the bounds from subsets of the data involving KiDS (Secs.~\ref{wlgcsec1}, \ref{wlgcsec2}, \ref{distsec2}) and differs from the ``no feedback'' (or ``dark matter only'') value of $B = 3.13$ at greater than $98\%$~CL.

The constraints on the IA amplitude are improved by more than $20\%$ relative to the equivalent data setup without Planck (i.e.~as compared to KiDS$\times$2dFLenS $+$ All-BOSS $+$ Pantheon). Concretely, we constrain $A_{\rm IA} = 1.52^{+0.038}_{-0.038}$ in the unrestricted JBD model with massive neutrinos, which remains practically identical as we fix the sum of neutrino masses. The constraints are mildly shifted toward larger values (by $\Delta A_{\rm IA} \sim 0.2$--$0.4$) with the addition of Planck, and are both positive at $4.0\sigma$. However, we note that the large posterior mean favored here is likely driven (at least partly) by the uncertainties in the photometric redshift distributions (see e.g.~Refs.~\cite{johnston19,wright2020,fortuna2020}).

\subsubsection{$H_0$ and $S_8$}

In addition to the primary parameters, we constrain $H_0 = 68.37^{+1.24}_{-1.24} ~ {\rm km} \, {\rm s}^{-1} {\rm Mpc}^{-1}$ and $S_8 = 0.822^{+0.014}_{-0.014}$ in the unrestricted JBD model with fixed neutrino masses, which are modified to $H_0 = 68.13^{+1.26}_{-1.25} ~ {\rm km} \, {\rm s}^{-1} {\rm Mpc}^{-1}$ and $S_8 = 0.817^{+0.015}_{-0.015}$ in the unrestricted JBD model where the sum of neutrino masses is additionally varied. The $H_0$ constraints are consistent at the $0.2\sigma$ level and are marginally stronger (by $\sim5\%$) than the respective constraints from All-Planck18 $+$ All-BOSS $+$ Pantheon. Given the positive shifts in the posterior mean as compared to All-Planck18 $+$ All-BOSS $+$ Pantheon, by $\Delta H_0 = \{0.57, 0.30\} ~ {\rm km} \, {\rm s}^{-1} {\rm Mpc}^{-1}$ in the \{fixed, varying\} neutrino mass scenarios, the discordance with the Riess et al.~(2019)~\cite{riess2019} measurement of the Hubble constant decreases marginally to $T(H_0) = \{3.0\sigma, 3.1\sigma\}$, respectively (i.e.~despite the narrower uncertainty).

By the inclusion of KiDS$\times$2dFLenS, the $S_8$ constraints from the full combination of datasets are shifted by $\Delta S_8 = \{-0.8\sigma, -0.7\sigma\}$ as compared to All-Planck18 $+$ All-BOSS $+$ Pantheon alone, and are consistent at the $0.3\sigma$ level between the two neutrino mass setups. However, the constraints are closer to the posterior mean favored by Planck, rather than KiDS$\times$2dFLenS, given its comparably higher constraining power.

\subsubsection{Correlations with the matter density}

In Fig.~\ref{figmatter}, given the full combination of datasets for both Planck 2018 and Planck 2015 (i.e.~``KiDS$\times$2dFLenS $+$ All-BOSS $+$ All-Planck18 $+$ Pantheon'' and ``KiDS$\times$2dFLenS $+$ All-BOSS $+$ All-Planck15 $+$ Pantheon''), we show a subset of the correlations with the matter density $(\Omega_{\rm m})$. As expected, the matter density is negatively correlated with $\{H_0, G_{\rm matter}\}$ and positively correlated with $S_8$. Here, the correlation for the novel pair $\{\Omega_{\rm m}, G_{\rm matter}\}$ is negative as the parameters are to first order multiplicative in the Poisson equation, such that a positive shift in one parameter is counteracted by an equally negative shift in the other parameter to obtain the same fit to the data;~a qualitatively similar relation holds for these two parameters in the Friedmann equation, where $H^2 \propto G_{\rm matter} \rho$ (as shown in Sec.~\ref{theorysec}). 

We constrain $\Omega_{\rm m} = 0.305^{+0.011}_{-0.012}$ in the unrestricted JBD model with massive neutrinos from the full combination of datasets, which is robust between the two Planck datasets (to within $0.5\sigma$, as $\Omega_{\rm m} = 0.298^{+0.011}_{-0.012}$ when Planck 2018 is replaced with Planck 2015), and is also robust to the exclusion of KiDS$\times$2dFLenS (to within $0.3\sigma$, as $\Omega_{\rm m} = 0.309^{+0.012}_{-0.012}$ for All-Planck18 $+$ All-BOSS $+$ Pantheon).

\begin{figure*}
\vspace{-1.3em}
\includegraphics[width=0.925\hsize]{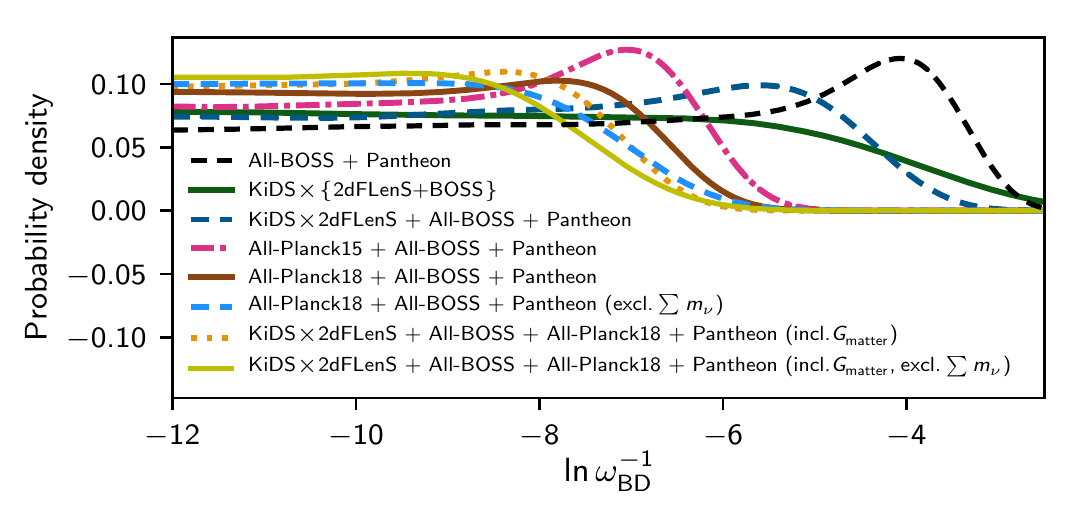}
\vspace{-2.1em}
\caption{\label{figwbdall} 
Marginalized posterior distributions for the JBD parameter $\smash{\ln \omega_{\rm BD}^{-1}}$. We simultaneously vary all standard cosmological and systematics parameters, along with the sum of neutrino masses $\smash{\sum m_{\nu}}$. We keep $\smash{G_{\rm matter}/G}$ fixed unless indicated otherwise. In the MCMC analysis, $\smash{\ln \omega_{\rm BD}^{-1}}$ is allowed to vary down to $-17$, zoomed in here for visual clarity (noting that the posteriors continue to stay flat below $\smash{\ln \omega_{\rm BD}^{-1} = -12}$). 
}
\end{figure*}

\subsubsection{Model selection}

Although the unrestricted JBD model is able to alleviate dataset discordances, we find no model selection preference for this model, as $\Delta{\rm DIC} = 2.0$ in our setup with fixed neutrino masses and $\Delta{\rm DIC} = 4.6$ when allowing the sum of neutrino masses to vary.

\subsubsection{Comparing against the Planck 2015 dataset}

We can further compare the differences in the parameter constraints from KiDS$\times$2dFLenS $+$ All-BOSS $+$ All-Planck $+$ Pantheon between the 2018 and 2015 datasets of Planck. As shown in Appendix~\ref{planckapp} (specifically Fig.~\ref{figplanck18unrest}), and similar to that found in the restricted JBD model (see Fig.~\ref{figplanck18rest}), 
there is strong consistency in the parameter constraints between the two setups. The sum of neutrino masses constitutes the main exception (along with the primordial scalar amplitude) due to its strong correlation with the optical depth, $\tau$, which is improved by 65\% in the data combination with All-Planck18 (as compared to the data combination with All-Planck15).\footnote{We note that for the data combination with All-Planck18, the $\tau$ constraint is effectively unchanged between the restricted and unrestricted JBD models, while it is degraded by $30\%$ for the data combination with All-Planck15 as we transition from the restricted to the unrestricted JBD model.} As a result, $\sum m_{\nu} < 0.12$~eV~($95\%$ CL) which reflects a $40\%$ improvement in the upper bound (relative to the bound from the data combination with All-Planck15). 

In combining KiDS$\times$2dFLenS, All-BOSS, Pantheon, and All-Planck15, we constrain $\omega_{\rm BD} > \{2270, 1640\}$ ($95\%$~CL) in the unrestricted JBD model with \{fixed, varying\} sum of neutrino masses, respectively (corresponding to differences of $|\Delta \omega_{\rm BD}| \lesssim 100$ with the equivalent bounds from the data combination with All-Planck18). The effective gravitational constant remains in agreement with the GR expectation, as $G_{\rm matter}/G = 1.017^{+0.029}_{-0.030}$, which corresponds to percent-level degradation in the uncertainty and shift of $|\Delta G_{\rm matter}/G| = 0.021$ in the posterior mean relative to that from the data combination with All-Planck18. 

Meanwhile, the constraint on the Hubble constant is positively shifted by $0.5\sigma$ (with effectively no change in the uncertainty), such that $H_0 = 68.71^{+1.27}_{-1.26} ~ {\rm km} \, {\rm s}^{-1} {\rm Mpc}^{-1}$ and $T(H_0) = 2.8$ (a decrease in the tension with Riess et al.~2019 by $0.3\sigma$). As expected, the baryonic feedback and IA amplitude constraints are effectively unchanged (as these are driven by KiDS$\times$2dFLenS);~however, given the weaker neutrino mass bounds in the data combination with All-Planck15 (relative to All-Planck18), the cosmological parameter constraints are generally less robust, in an absolute sense, to allowing $\sum m_{\nu}$ vary in the analysis (as shown in Table~\ref{subtabplanck15}).

\begin{figure*}
\vspace{-1.6em}
\includegraphics[width=0.9\hsize]{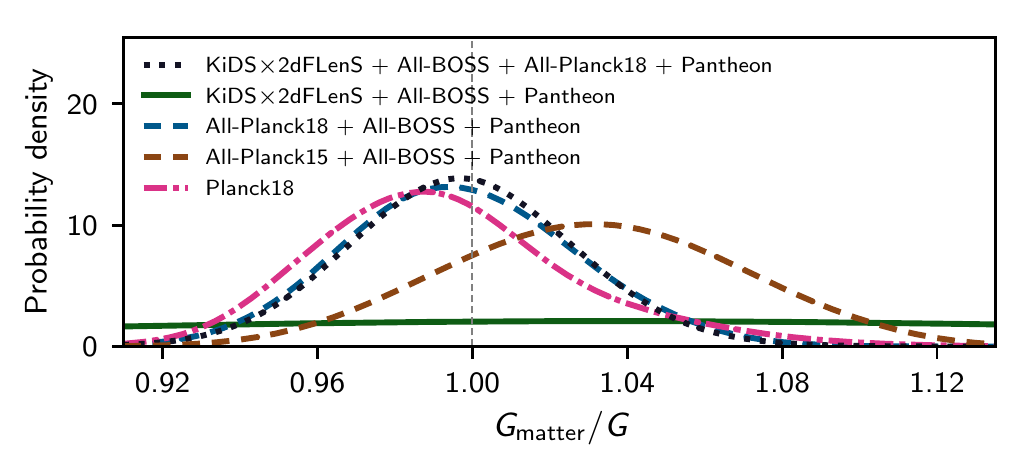}
\vspace{-2.1em}
\caption{\label{figgm} 
Marginalized posterior distributions for the present effective gravitational constant ($G_{\rm matter}/G$). We simultaneously vary all standard cosmological and systematics parameters, along with the JBD parameter $\ln \omega_{\rm BD}^{-1}$ and the sum of neutrino masses $\sum m_{\nu}$. In the MCMC analysis, $G_{\rm matter}/G$ is allowed to vary between $-1/2$ to $2$ (zoomed in here for visual clarity). The dashed grey vertical line indicates the GR expectation.
}
\end{figure*}

\subsection{Take-away of JBD constraints, parameter degeneracies, model selection, and dataset concordances}
\label{sumsec}

\subsubsection{JBD coupling constant and the effective gravitational constant}
\label{jbdta}

We summarize the constraints on the JBD coupling constant in Fig.~\ref{figwbdall}, considering different combinations of datasets (consisting of KiDS, 2dFLenS, BOSS, Planck, and Pantheon) and different cosmological models (with a focus on the possible inclusion of $\sum m_{\nu}$ and $G_{\rm matter}$). As shown in this figure (and discussed in Secs.~\ref{cmbresultssec}--\ref{fullsec2}), the strongest bounds on $\omega_{\rm BD}$ are from the Planck CMB, approximately an order of magnitude stronger than the bounds from KiDS, 2dFLenS, BOSS, and Pantheon (even when combined). The $\omega_{\rm BD}$ bound favors the modified gravity solution more strongly as massive neutrinos are simultaneously considered, while the transition from a restricted to unrestricted JBD model (i.e.~allowing for $G_{\rm matter}$ to vary) results in an $\omega_{\rm BD}$ bound that is in greater agreement with the GR expectation. 

We obtain the strongest bound on the JBD coupling constant in the unrestricted JBD model with fixed neutrino masses, where $\omega_{\rm BD} > 2230$~($95\%$~CL) from the combination of all datasets. As described in Sec.~\ref{fullsec2}, this bound weakens by approximately $\Delta \omega_{\rm BD} \simeq -700$ as we allow for the sum of neutrino masses to vary, and by an additional $\Delta \omega_{\rm BD} \simeq -600$ when we transition to a restricted JBD model (i.e.~fix $G_{\rm matter}/G=1$; noting that we remove KiDS$\times$2dFLenS to maintain dataset concordance in the restricted JBD model).

For the unrestricted JBD model, we summarize the constraints on the effective gravitational constant in Fig.~\ref{figgm}, where $G_{\rm matter}/G = \{0.996^{+0.029}_{-0.029},\,0.997^{+0.029}_{-0.029}\}$ as the sum of neutrino masses is \{fixed, varied\}, and which remains robust as we exclude any one of the datasets. The constraints on the JBD coupling constant and effective gravitational constant are also largely robust to whether the 2018 or 2015 dataset of Planck is considered (given the same observables). In particular, we note that the agreement with the GR expectation improves by $0.5\sigma$ with the 2018 dataset, which is driven by the improved measurement of the optical depth to reionization (as also seen in Fig.~\ref{figplanck18unrest}).

\subsubsection{Comparison of JBD constraints to earlier work}

We can compare our bounds on $\omega_{\rm BD}$ to those obtained in earlier work. As discussed in the forthcoming Sec.~\ref{modelingsec}, this is not straightforward given the different modeling choices. The most relevant comparison is to Avilez \& Skordis (2014)~\cite{Avilez:2013dxa} given the similar JBD modeling, where $\omega_{\rm BD} > \{1901,\,2441\}$ at $95\%$~CL in the \{restricted, unrestricted\} JBD model with Planck 2013 (temperature and lensing), respectively. In addition to the comparable bounds on $\omega_{\rm BD}$ to our analysis, these authors find a qualitatively similar difference in the $\omega_{\rm BD}$ bounds between the restricted and unrestricted JBD models.

However, while the $\omega_{\rm BD}$ bounds are comparable between the analyses, we note that these are one-sided parameter bounds and therefore not a reflection of the overall constraining power of the datasets considered. This can be seen by comparing the constraints on the effective gravitational constant in the unrestricted JBD model, where Avilez \& Skordis (2014)~\cite{Avilez:2013dxa} constrain this parameter to be in agreement with GR at the $\sigma\left({G_{\rm matter}/G}\right) \simeq 0.053$ level from Planck~2013 (temperature and lensing), ACT, and SPT, which we improve on by nearly a factor of two in our full analysis.

\subsubsection{Evolution of the effective gravitational constant and consistency with Big Bang Nucleosynthesis}
\label{bbnta}

As we constrain the coupling constant, $\omega_{\rm BD}$, to be larger than $\sim1\times10^3$ from the Planck CMB temperature and polarization alone and $\sim2\times10^3$ when Planck is combined with other probes (both at $95\%$~CL), the scalar field and thereby the effective gravitational constant is approximately constant with time, to within $0.5$--$1\%$ from the present to the BBN epoch (concretely from the present to matter-radiation equality, after which it freezes, as discussed in Sec.~\ref{theorysec}). Given the distinct $\omega_{\rm BD}$ and $G_{\rm matter}/G$ constraints in the unrestricted JBD model (Sec.~\ref{fullsec2}), this implies that we constrain the gravitational constant during BBN to $G_{\rm BBN}/G = 1.00 \pm 0.03$. This constraint holds for both fixed and varying sum of neutrino masses, and it additionally holds at the epoch of recombination as the effective gravitational constant evolves by a mere $0.05$--$0.1\%$ between recombination and BBN.

If we instead consider a restricted JBD model (Sec.~\ref{distsec1}), where the effective gravitational constant at present is not a free parameter but fixed to unity, we can place an upper bound on its value during recombination and BBN from our lower bound on the coupling constant. Here, we constrain $G_{\rm BBN}/G - 1 < 5.8\times10^{-3}$ and $G_{\rm recomb}/G - 1 < 5.0\times10^{-3}$ as the sum of neutrino masses is fixed, along with $G_{\rm BBN}/G - 1 < 8.7\times10^{-3}$ and $G_{\rm recomb}/G - 1 < 7.5\times10^{-3}$ as the sum of neutrino masses is varied (all at $95\%$~CL). For both the restricted and unrestricted JBD models, we can also compute the first and second order time-derivatives of the effective gravitational constant. In Appendix~\ref{timeapp}, we show that the \{first, second\} order derivatives are \{negative, positive\} since the onset of matter domination and that their magnitudes are presently less than approximately $10^{-13} \, {\rm year}^{-1}$ and $10^{-26} \, {\rm year}^{-2}$, respectively. In other words, as earlier discussed in Sec.~\ref{theorysec}, the effective gravitational constant is decreasing with time and it is doing so more slowly as time progresses.

The mild evolution of the gravitational constant and consistency with the standard model expectation is in agreement with the nucleosynthesis inference in Ref.~\cite{alvey20}, where the primordial helium and deuterium abundances are used to constrain $G_{\rm BBN}/G = 0.98 \pm 0.03$ (i.e.~similar precision to our constraint in the unrestricted JBD model, and evolution of the gravitational constant by $0.02$ to reach the standard model expectation at present).

\begin{figure*}
\vspace{-0.7em}
\includegraphics[width=1.02\hsize]{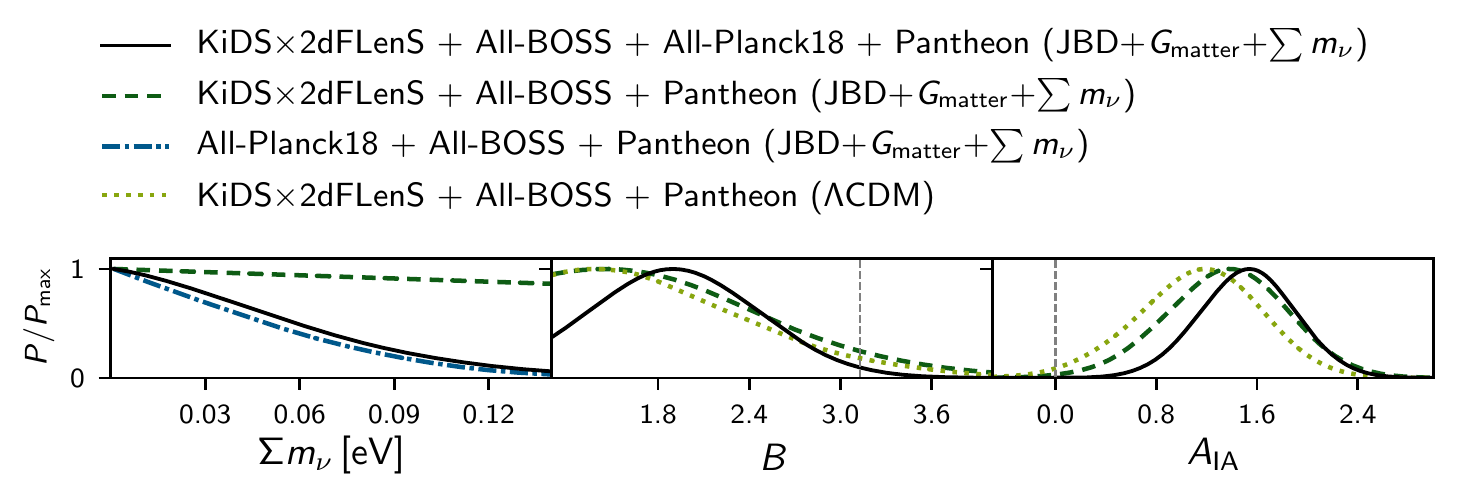}
\vspace{-3.0em}
\caption{\label{figextra}
Marginalized posterior distributions for the sum of neutrino masses, $\sum m_{\nu}$, the baryonic feedback amplitude, $B$, and the intrinsic alignment amplitude, $A_{\rm IA}$. We simultaneously vary all standard cosmological and systematics parameters in an unrestricted JBD model with massive neutrinos (along with $\Lambda$CDM for comparison). The grey vertical line at $B = 3.13$ corresponds to the ``no feedback'' scenario.
}
\end{figure*}

\subsubsection{Neutrino mass, baryonic feedback, and intrinsic alignments}

As summarized in Fig.~\ref{figextra}, our strongest bound on the sum of neutrino masses is $\sum m_{\nu} < 0.11~{\rm eV}$ ($95\%$~CL) from Planck, BOSS, and Pantheon, which is not particularly affected by the assumptions of the JBD modeling or by the inclusion of the KiDS$\times$2dFLenS dataset (at the $0.01$~eV level). The baryonic feedback amplitude is most strongly bounded in the unrestricted JBD model where all datasets are considered, such that $B < 2.8$ at $95\%$~CL (as compared to the ``no-feedback'' scenario of $B = 3.13$). This bound is only weakly sensitive to the cosmological model and specific datasets considered in addition to KiDS (such that it weakens by at most $\Delta B \simeq 0.6$ as we consider an exclusion of Planck and fixed neutrino masses).

In Fig.~\ref{figextra}, we also show that the IA amplitude is marginally larger in the JBD model relative to $\Lambda$CDM (by $\Delta A_{\rm IA} \approx 0.2$) and we improve the constraint on the amplitude by more than 20\% as Planck is considered alongside KiDS in the unrestricted JBD model. However, we emphasize that the large amplitude (positive by up to $4\sigma$) might be partly explained by the systematic uncertainties in the photometric redshift distributions (as noted in Sec.~\ref{fullsec2}). 

\subsubsection{$H_0$ and $S_8$}

\begin{figure*}
\vspace{-0.4em}
\minipage{\textwidth}
\includegraphics[width=0.95\hsize]{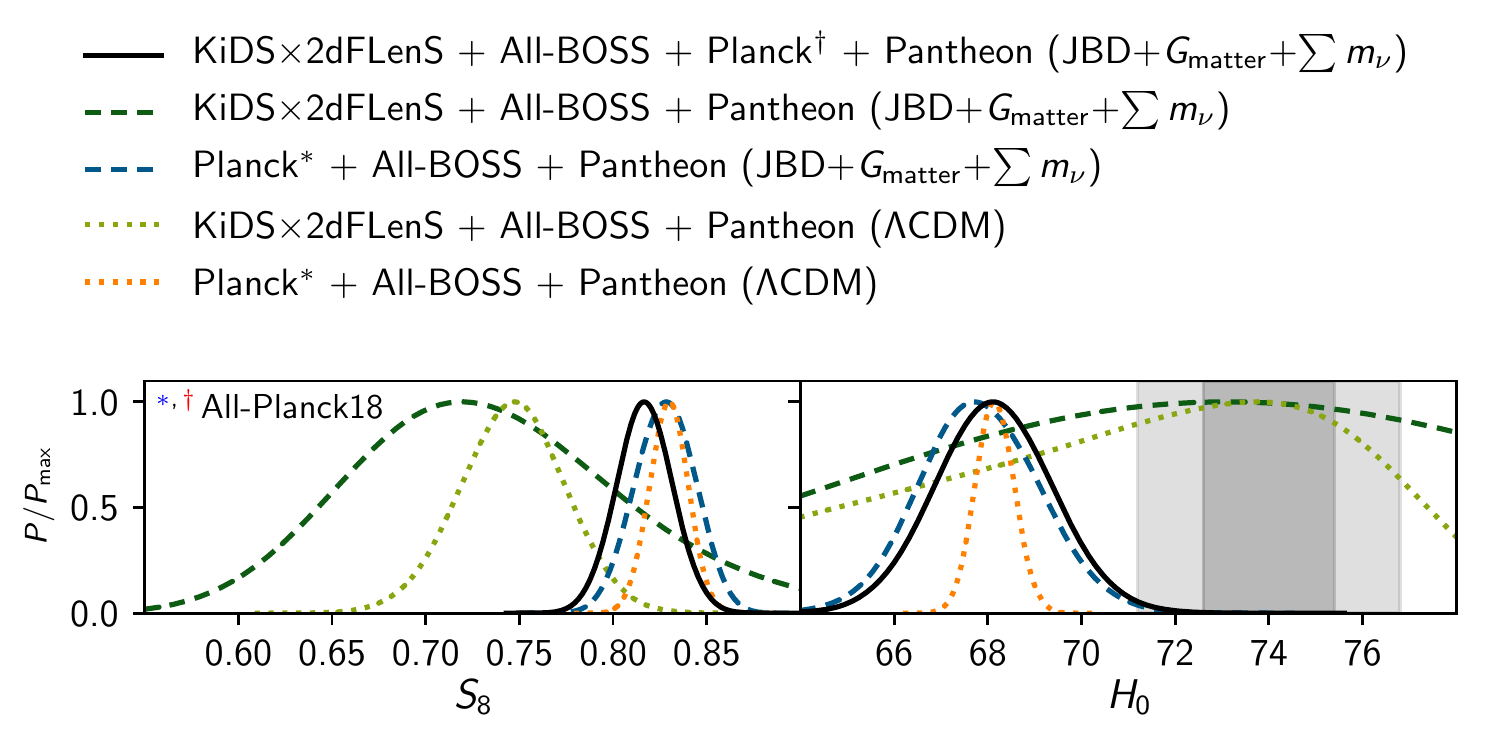}
\endminipage\hfill

\vspace{-4.4em}

\minipage{\textwidth}
\includegraphics[width=0.95\hsize]{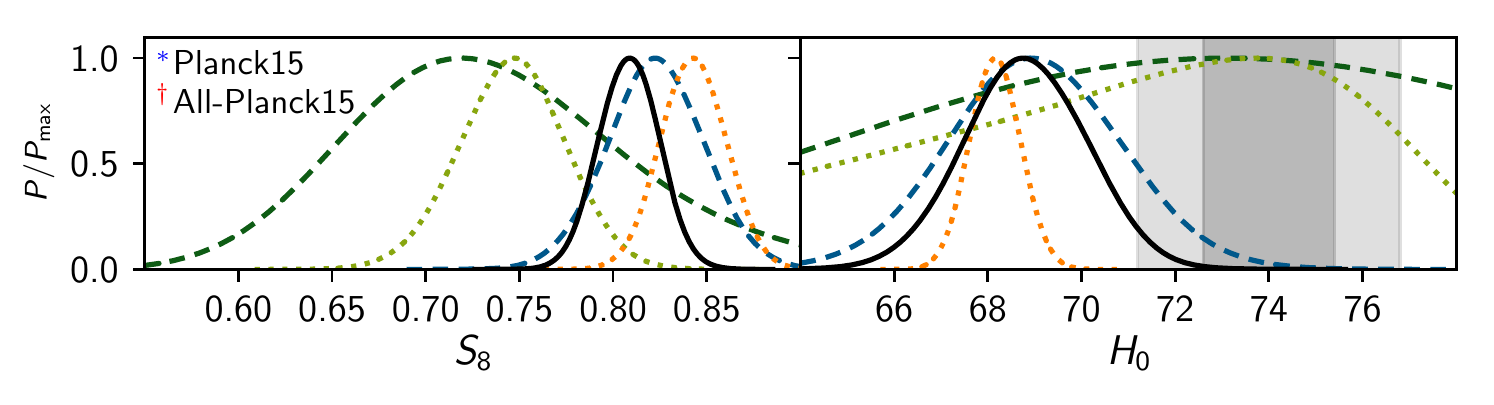}
\endminipage\hfill
\vspace{-1.6em}
\caption{\label{figs8h0} 
Marginalized posterior distributions for ${S_8 = \sigma_8 \sqrt{\Omega_{\mathrm m}/0.3}}$ ({\it left}) and the Hubble constant, $H_0$ ({\it right}), given in units of ${{\rm km} \, {\rm s}^{-1} {\rm Mpc}^{-1}}$. We simultaneously vary all standard cosmological and systematics parameters in an unrestricted JBD model with massive neutrinos (along with $\Lambda$CDM for comparison). The grey vertical bars show the $68\%$ CL (inner) and $95\%$ CL (outer) constraints on $H_0$ from Riess et al.~(2019)~\cite{riess2019}.
}
\end{figure*}

In Fig.~\ref{figs8h0}, we illustrate the $\{{H_0, S_8}\}$ discordances (for $S_8$ between the CMB and weak lensing, and $H_0$ between the CMB and the direct measurement of Riess et al.~2019~\cite{riess2019}). The BOSS and Pantheon datasets can be combined with either KiDS or Planck, such that the $S_8$ tension between KiDS and Planck becomes more significant.

We show that the $H_0$ and $S_8$ tensions are ameliorated by the widening of the posteriors as we consider an unrestricted JBD model (i.e.~a decrease in the $H_0$ tension down to approximately $3\sigma$, and in the $S_8$ tension to below $1\sigma$). We also show how the constraints improve in a combined analysis of all datasets in this extended model (which again increases the $H_0$ tension), and we contrast the results between the 2018 and 2015 datasets of Planck (notably finding that the $H_0$ tension decreases down to $2\sigma$ given the baseline Planck 2015 dataset which excludes the high-multipole polarization in particular). The fact that the $H_0$ and $S_8$ posteriors do not significantly shift, but are rather broadened, is further reflected in the agreement of $\omega_{\rm BD}$ and $G_{\rm matter}/G$ with the respective GR expectations. 

\subsubsection{$\Omega_{\rm m}$ versus $\Omega^{*}_{\rm m}$: impact on assessing the $S_8$ tension}

In assessing the $S_8$ tension, we have used $\Omega_{\rm m}$ for each dataset in the computation of the respective $S_8$ estimates, while in Sec.~\ref{theorysec} (specifically Eq.~\ref{densityparam}) we showed that $\Omega^{*}_{\rm m} = \Omega_{\rm m}/\phi$ is the density for which the sum of all densities add to unity. As a result, there is a freedom in whether we evaluate the $S_8$ tensions using the respective $\Omega_{\rm m}$ or $\Omega^{*}_{\rm m}$, which will increasingly differ as $\omega_{\rm BD}$ decreases and $G_{\rm matter}/G$ increasingly deviates from unity. 

As our constraints on the coupling constant are generally greater than $10^2$ (for which the difference in the matter density is at the percent level;~noting that $\Omega_{\rm m} > \Omega^{*}_{\rm m}$ as $\omega_{\rm BD}$ decreases), the dominant cause of the difference in $\Omega_{\rm m}$ and $\Omega^{*}_{\rm m}$ is due to the effective gravitational constant where the difference scales linearly (i.e.~$\Omega^{*}_{\rm m} \propto G_{\rm matter}/G$). The linear scaling of the density with the effective gravitational constant implies that $S^{*}_8 \propto \sqrt{G_{\rm matter}/G}$. In assessing the tension between KiDS and Planck, let us therefore consider a specific example. In Table~\ref{subtab4}, we find that $G_{\rm matter}/G$ is centered at $1.104$ for KiDS$\times$2dFLenS $+$ All-BOSS $+$ Pantheon (while $\omega_{\rm BD} \gg 1$), which implies that the $S_8$ tension of $1.5\sigma$ is lowered by $\gtrsim0.3\sigma$, such that the $S^{*}_8$ tension is $1.2\sigma$. However, as the constraints on JBD gravity tighten towards agreement with GR, as in the case of the fully combined datasets, where generally $\omega_{\rm BD} > 10^3$ and $|\Delta G_{\rm matter}/G| < 10^{-2}$, the difference between $S_8$ and $S^{*}_8$ becomes negligible.

\subsubsection{Model selection when including Riess et al.~(2019)}

Turning back to the Hubble constant, to assess whether the $H_0$ constraint from Planck (and data combinations with Planck) is more concordant with the direct measurement of Riess et al.~(2019)~\cite{riess2019} in an extended cosmological model, we have avoided a data analysis that includes both the Planck and Riess et al.~(2019) measurements (i.e.~our $H_0$ constraints have consistently excluded any ``external'' measurement from Riess et al.~2019). In this setup (which excludes Riess et al.~2019), both the restricted and unrestricted JBD models are at most weakly favored from a model selection standpoint relative to $\Lambda$CDM (see Table~\ref{subtabdic} in Appendix~\ref{extraparamsdicapp}).  

However, in assessing the model selection preference of the unrestricted JBD model, which does exhibit a smaller tension between Planck and Riess et al.~(2019)~\cite{riess2019}, we also consider MCMC inferences that include the Riess et al.~(2019) measurement. For a combined analysis of the 2018 dataset of Planck CMB temperature and polarization (i.e.~excluding lensing reconstruction) and Riess et al.~(2019), $\Delta \chi^2_{\rm eff} = -4.0$ and $\Delta {\rm DIC} = -4.8$, reflecting weak-to-moderate preference in favor of the extended model relative to $\Lambda$CDM. 

In a combined analysis of all datasets (${\rm KiDS}\times{\rm 2dFLenS}+{\rm All}$-${\rm BOSS}+{\rm All}$-${\rm Planck}+{\rm Pantheon}$, where Planck now further includes the 2018  lensing reconstruction) with Riess et al.~(2019), $\Delta \chi^2_{\rm eff} = -4.7$ and $\Delta {\rm DIC} = -2.7$, corresponding to weak preference in favor of the extended model. Hence, while the unrestricted JBD model is able to alleviate the $H_0$ and $S_8$ tensions, the extra parameters of the model are consistent with the GR expectation, and the extended model is only weakly favored in a model selection sense relative to $\Lambda$CDM. Given the current cosmological datasets, JBD gravity therefore does not simultaneously satisfy all of the conditions in Sec.~\ref{requirementssec} required to replace the standard model. 

\subsubsection{Constraining the effective field theory parameters}
\label{eftconstraints}

In Sec.~\ref{eftsec}, we discussed the relationship between the EFT and JBD parameters, noting that the tensor speed excess $\alpha_{\rm T} = 0$ (at all times) and that the other $\alpha_i = 0$ during radiation domination in JBD gravity. During the matter and cosmic accelerating epochs, the remaining EFT parameters can be obtained from the JBD coupling constant. In particular, as the scalar field unfreezes at the onset of matter domination, the Planck-mass run rate $\alpha_{\rm M} = (1+\omega_{\rm BD})^{-1}$, and as the Universe enters the epoch of cosmic acceleration, $\alpha_{\rm M} = 4(1 + 2\omega_{\rm BD})^{-1}$.

Given the lower bounds on the coupling constant in the unrestricted JBD model (Sec.~\ref{fullsec2}), we thereby constrain $\alpha_{\rm M} < \{0.44, 0.65\} \times 10^{-3}$ ($95\%$~CL) as the sum of neutrino masses is respectively \{fixed, varied\} during matter domination, and $\alpha_{\rm M} < \{0.90, 1.3\}\times 10^{-3}$ ($95\%$~CL) as the sum of neutrino masses is respectively \{fixed, varied\} during the epoch of cosmic acceleration. In the restricted JBD model (Sec.~\ref{distsec1}), we constrain $\alpha_{\rm M} < \{0.68,1.4\} \times 10^{-3}$ ($95\%$~CL) as the sum of neutrino masses is respectively \{fixed, varied\} during matter domination, and $\alpha_{\rm M} < \{1.4, 2.1\} \times 10^{-3}$ ($95\%$~CL) as the sum of neutrino masses is respectively \{fixed, varied\} during the epoch of cosmic acceleration.

These upper bounds in turn translate into lower bounds for $\alpha_{\rm B} = -\alpha_{\rm M}$. The bounds on $\alpha_{\rm M}$ moreover translate into upper bounds for $\alpha_{\rm K} = \omega_{\rm BD} \alpha_{\rm M}^2$, noting that for larger values of the coupling constant consistent with our constraints, $\alpha_{\rm K} \simeq \alpha_{\rm M}$ during matter domination and $\alpha_{\rm K} \simeq 2 \alpha_{\rm M}$ during the epoch of cosmic acceleration. 

\begin{figure}
\vspace{-0.7em}
\includegraphics[width=1.03\hsize]{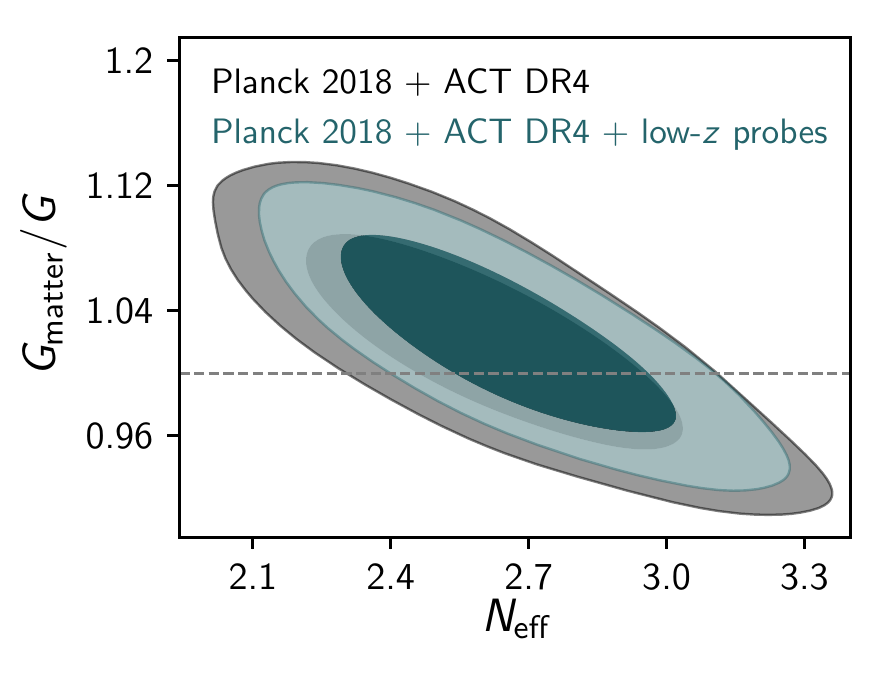}
\vspace{-2.9em}
\caption{\label{figneff} 
Marginalized posterior distributions (inner $68\%$~CL, outer $95\%$~CL) in the plane of the present effective gravitational constant, $\smash{G_{\rm matter}/G}$, and the effective number of neutrinos, $N_{\rm eff}$, from the Planck 2018 and ACT DR4 datasets. The two contours either include (dark cyan) or exclude (grey) additional datasets that probe the Universe at lower redshifts, which we take to be the Planck 2018 CMB lensing reconstruction along with ${\rm KiDS}\times{\rm 2dFLenS}+{\rm All}$-${\rm BOSS}+{\rm Pantheon}$. In this unrestricted JBD cosmology (\smash{JBD+$G_{\rm matter}$+$N_{\rm eff}$}), the sum of neutrino masses is kept fixed and the horizontal dashed line denotes the GR expectation ($\smash{G_{\rm matter}/G} = 1$).}
\end{figure}

\subsection{Degeneracies with small-scale physics in the CMB damping tail}
\label{degsec}

In Sec.~\ref{dampsec}, we showed how extended cosmological parameters targeted by small-scale CMB experiments, such as the effective number of neutrinos, primordial helium abundance, and the running of the spectral index $\smash{(N_{\rm eff}, Y_{\rm P}, {\rm d} n_{\rm s} / {\rm d} \ln k)}$ can be highly degenerate with the underlying gravitational theory in the CMB damping tail (which holds for both the temperature and polarization auto- and cross-power spectra). In the case of JBD gravity, this degeneracy is driven by the impact of the scalar field on the pre-recombination expansion rate (such that as $\omega_{\rm BD}$ is more tightly constrained, the impact of the scalar field on the expansion rate is increasingly dominated by $G_{\rm matter}/G$ through a constant rescaling of $(H/H_0)^2$).

In Fig.~\ref{figneff}, we concretely illustrate this degeneracy for the case of $N_{\rm eff}$ and $G_{\rm matter}/G$ with Planck 2018 and ACT DR4. As discussed in Sec.~\ref{cmbsec}, the covariance between the CMB datasets has not been included (which has been shown to be adequate in $\Lambda$CDM and certain single-parameter extensions to better than $5\%$~\cite{aiola20}) and we include ACT primarily to better understand the extent to which the parameter constraints are affected. The anti-correlation between $G_{\rm matter}/G$ and $N_{\rm eff}$ in Fig.~\ref{figneff} is consistent with the effect of these parameters on the CMB damping tail in Fig.~\ref{figclsdamping}, where $G_{\rm matter}/G > 1$ induces a suppression in the damping tail that can be counteracted by negative shifts in $N_{\rm eff}$. 

We constrain $G_{\rm matter}/G = 1.020^{+0.044}_{-0.045}$ and $N_{\rm eff} = 2.62^{+0.24}_{-0.29}$ from the CMB alone, which are $60$--$70$\% less stringent than the respective constraints in cosmologies where one of the parameters is fixed to the standard model expectation (noting that in fixing $G_{\rm matter}/G$ to unity, we also take the GR limit of $\omega_{\rm BD}$). When further combining the CMB datasets with the lower-redshift probes of ${\rm KiDS}\times{\rm 2dFLenS}$, ${\rm All}$-${\rm BOSS}$, ${\rm Pantheon}$, and the Planck CMB lensing reconstruction, we obtain a marginal ($\lesssim10\%$) improvement in $G_{\rm matter}/G = 1.023^{+0.042}_{-0.042}$ and $N_{\rm eff} = 2.65^{+0.22}_{-0.26}$. These constraints are consistent with, but $70$--$80$\% less stringent than, the scenario where one of the respective degrees of freedom is fixed. The corresponding shifts in the posterior means are $\Delta N_{\rm eff} \sim 0.1$ and $\Delta G_{\rm matter}/G \sim 0.02$ in all cases. While these are $\lesssim1\sigma$ shifts (relative to the fixed-case uncertainties), we note that they would need to decrease as the constraining power of the future datasets increase to avoid growing in statistical significance. 

\begin{figure*}
\vspace{-0.7em}
\includegraphics[width=1.02\hsize]{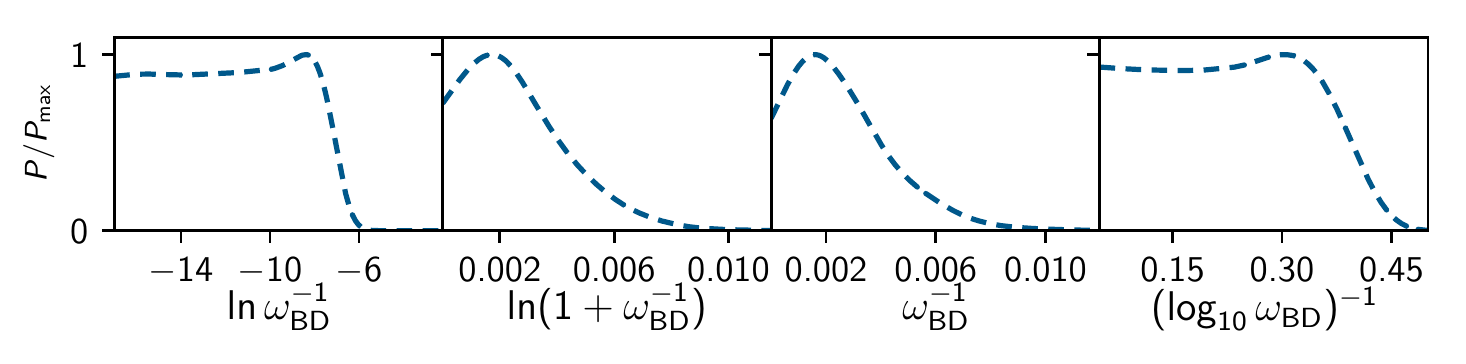}
\vspace{-2.6em}
\caption{\label{figparams} 
Marginalized posterior distributions for the JBD coupling constant from the full combination of datasets (${\rm KiDS}\times{\rm 2dFLenS}+{\rm All}$-${\rm BOSS}+{\rm All}$-${\rm Planck18}+{\rm Pantheon}$) considering four distinct parameterizations: $\ln \omega_{\rm BD}^{-1}$, $\ln(1+\omega_{\rm BD}^{-1})$, $\omega_{\rm BD}^{-1}$, and $(\log_{10} \omega_{\rm BD})^{-1}$. These are the effective primary parameters varied in the respective MCMC analysis (along with the other cosmological and systematics parameters).
}
\end{figure*}

The inclusion of the effective number of neutrinos as a free parameter moreover induces a widening of the $Y_{\rm P}$ posterior given the enforced BBN consistency relation (e.g.~Ref.~\cite{simha08}) between the two parameters and the baryon density $\Omega_{\rm b}h^2$ (i.e.~$Y_{\rm P}$ is a derived and not primary parameter in this scenario). Concretely, from the CMB alone, $\sigma(Y_{\rm P}) = 5.8 \times 10^{-5}$ in $\Lambda$CDM, $\sigma(Y_{\rm P}) = 2.4 \times 10^{-3}$ in $\Lambda$CDM with a varying $N_{\rm eff}$, and $\sigma(Y_{\rm P}) = 3.9 \times 10^{-3}$ in the unrestricted JBD model with a varying $N_{\rm eff}$. As compared to $\Lambda$CDM, we thereby find factors of 41 and 67 degradations in the constraining power, and note that the constraints improve by $\lesssim10\%$ as the CMB is considered together with the lower-redshift probes. The large widening of the uncertainty can further be contrasted with that found for the unrestricted JBD model with a fixed $N_{\rm eff}$. Here, $\sigma(Y_{\rm P}) = 8.2 \times 10^{-5}$ is widened by $40\%$ as compared to $\Lambda$CDM (and exhibits a degeneracy with $\Omega_{\rm b}h^2$ and $G_{\rm matter}/G$ as also noted in Ref.~\cite{zuma20}), but substantially less than when $N_{\rm eff}$ is varied.

As both $N_{\rm eff}$ and $G_{\rm matter}/G$ are varied in the analysis, their separate degeneracies with the Hubble constant are softened. 
In constraining $\smash{H_0 = 65.8^{+1.2}_{-1.5} ~ {\rm km} \, {\rm s}^{-1} {\rm Mpc}^{-1}}$ from the CMB alone and $\smash{H_0 = 67.0^{+1.1}_{-1.1} ~ {\rm km} \, {\rm s}^{-1} {\rm Mpc}^{-1}}$ from the CMB and lower-redshift probes in the unrestricted JBD model with a varying $N_{\rm eff}$, the constraints improve by $\lesssim10\%$ as either $N_{\rm eff}$ or $G_{\rm matter}/G$ is fixed. However, in the case of the $S_8$ parameter, it is largely uncorrelated with $N_{\rm eff}$, such that (for instance) $\smash{S_8 = 0.832^{+0.022}_{-0.025}}$ from the CMB alone in the unrestricted JBD model with a varying $N_{\rm eff}$, which improves to $\smash{S_8 = 0.840^{+0.015}_{-0.015}}$ in $\Lambda$CDM with a varying $N_{\rm eff}$, and the latter constraint persists as $N_{\rm eff}$ is additionally fixed. 

In summary, we have provided a concrete example that the degeneracies observed in Sec.~\ref{dampsec} have a substantial impact on the parameter constraints from Planck and ACT, and we have  found that these degeneracies are not broken by the current non-CMB datasets. We leave a more detailed investigation for future work, including a full quantification of the degeneracies between $G_{\rm matter}/G$, $N_{\rm eff}$, $Y_{\rm P}$, and ${\rm d} n_{\rm s} / {\rm d} \ln k$ for both current and next-generation CMB and large-scale structure surveys.

\subsection{Impact of JBD modeling choices}
\label{modelingsec}

We next highlight two interrelated caveats to the modified gravity constraints presented, in the form of the prior range and parameterization of the JBD coupling constant. As the coupling constant favors GR in the limit $\omega_{\rm BD} \rightarrow \infty$, we seek to parameterize it in a way that allows for both weak and strong levels of modified gravity to be well sampled. Our fiducial parameterization is taken to be $\mathcal{P}(\omega_{\rm BD}) = \ln \omega_{\rm BD}^{-1}$ in accordance with the choice in Avilez \& Skordis (2014)~\cite{Avilez:2013dxa},\footnote{We note that this parameterization is qualitatively similar to an earlier $\ln[(4\omega_{\rm BD})^{-1}]$ parameterization advocated by Acquaviva et al.~(2004)~\cite{Acquaviva:2004ti}.} and we have uniformly varied $\ln \omega_{\rm BD}^{-1}$ as a primary parameter in the range [-17, -2.3] in our fiducial MCMC analyses (as described in Sec.~\ref{priorsec}). 

\subsubsection{Impact of the JBD prior range on the $\omega_{\rm BD}$ bounds}

In Appendix~\ref{priorapp} (specifically Fig.~\ref{figprior}), we explore the impact of extending the lower prior bound such that $\ln \omega_{\rm BD}^{-1}$ is uniformly sampled in the range $\left[{-47, -2.3}\right]$ (we do not extend the upper prior bound as it is already ruled out by the data). As increasingly negative values of $\ln \omega_{\rm BD}^{-1}$ continue to be favored by the data, given its consistency with GR, we find that the $\ln \omega_{\rm BD}^{-1}$ constraint (and by extension the constraint on $\omega_{\rm BD}$) is sensitive to the prior range. In other words, the $\ln \omega_{\rm BD}^{-1}$ posterior has a relatively sharp transition and eventually flattens given the inability of the data to distinguish between lower values of $\ln \omega_{\rm BD}^{-1}$ (between e.g.~$\ln \omega_{\rm BD}^{-1}$ of $-10$ and $-15$). As a result, the $95\%$ confidence region of $\ln \omega_{\rm BD}^{-1}$ is pushed to increasingly negative values as the prior range is widened (and by extension the lower boundary of $\omega_{\rm BD}$ is increasingly positive). This is an inescapable feature of any logarithmic parameter that is uniformly sampled without a well-motivated or constrained finite boundary.

In the appendix, as an example, we have considered the data combination ${\rm KiDS}\times{\rm 2dFLenS}+{\rm All}$-${\rm BOSS}+{\rm All}$-${\rm Planck18}+{\rm Pantheon}$, where for our fiducial prior range of $\ln \omega_{\rm BD}^{-1}$ we obtain $\omega_{\rm BD} > 1540$ ($95\%$~CL), and for our extended prior range we obtain $\omega_{\rm BD} > 17600$ ($95\%$~CL). In other words, the $\omega_{\rm BD}$ bounds from cosmology need to be interpreted with caution. However, given the weak correlation of $\ln \omega_{\rm BD}^{-1}$ with the other cosmological and systematics parameters (including with $G_{\rm matter}/G$), as shown in the appendix, the constraints on these other parameters are robust. 

\subsubsection{Impact of the JBD parameterization on the $\omega_{\rm BD}$ bounds}

In Fig.~\ref{figparams}, we moreover explore how the constraint on the JBD coupling constant changes as a result of how we parameterize it. In addition to the fiducial parameterization ($\mathcal{P} = \ln \omega_{\rm BD}^{-1}$), designed to allow for a wide range of JBD gravity to be well sampled \cite{Avilez:2013dxa}, we consider the parameterization of Wu \& Chen (2009)~\cite{Wu:2009zb}, where $\mathcal{P} = \ln(1+\omega_{\rm BD}^{-1})$, along with a simple $\mathcal{P} = \omega_{\rm BD}^{-1}$ parameterization. We expect these last two parameterizations to be effectively identical in the limit of large $\omega_{\rm BD}$ given the Mercator series (where to first order $\ln(1+\omega_{\rm BD}^{-1}) \approx \omega_{\rm BD}^{-1}$ for $|\omega_{\rm BD}^{-1}| \leq 1$ and $\omega_{\rm BD}^{-1} \neq -1$). A benefit of these two parameterizations is that the GR limit is obtained at zero instead of infinity (negative infinity in the fiducial parameterization). However, in an MCMC analysis with a uniform prior on $\mathcal{P}$, they tend to penalize weaker JBD gravity scenarios by not adequately sampling large $\omega_{\rm BD}$.\footnote{As a concrete example, translating the prior range for $\ln \omega_{\rm BD}^{-1}$ to $\ln(1+\omega_{\rm BD}^{-1})$, the latter is sampled in the range $[4.1 \times 10^{-8}, 0.095]$. Hence, $10\leq\omega_{\rm BD}<100$ covers 90\% of the prior space of $\ln(1+\omega_{\rm BD}^{-1})$, while $\omega_{\rm BD}>100$ is restricted to 10\% of the space, $\omega_{\rm BD}>1000$ is restricted to 1\% of the space, and so on. 
This hinders the ability of the MCMC to adequately sample parts of the $\ln(1+\omega_{\rm BD}^{-1})$ space that corresponds to large $\omega_{\rm BD}$.}

\begin{figure}
\vspace{-0.7em}
\includegraphics[width=1.03\hsize]{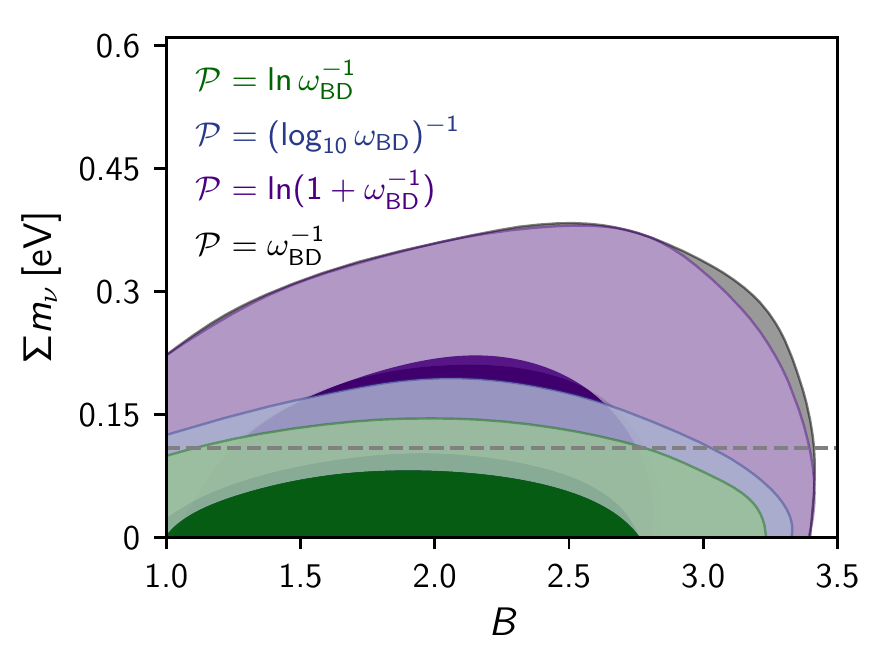}
\vspace{-2.9em}
\caption{\label{figparams2d} 
Marginalized posterior distributions (inner $68\%$~CL, outer $95\%$~CL) in the plane of the sum of neutrino masses, $\sum m_{\nu}$, and the baryonic feedback amplitude, $B$, from the full combination of datasets (${\rm KiDS}\times{\rm 2dFLenS}+{\rm All}$-${\rm BOSS}+{\rm All}$-${\rm Planck18}+{\rm Pantheon}$). The different contours correspond to four distinct parameterizations $\mathcal{P}$ of the JBD coupling constant $\left({\ln \omega_{\rm BD}^{-1}, (\log_{10} \omega_{\rm BD})^{-1}, \ln(1+\omega_{\rm BD}^{-1}), \omega_{\rm BD}^{-1}}\right)$. For comparison, the horizontal dashed line shows the $95\%$ upper bound on $\sum m_{\nu}$ from ${\rm All}$-${\rm Planck18}+{\rm All}$-${\rm BOSS}+ {\rm Pantheon}$ in $\Lambda$CDM.
}
\end{figure}

We moreover consider a parameterization of the form $\mathcal{P} = (\log_{10} \omega_{\rm BD})^{-1}$, created to allow for a wide range of $\omega_{\rm BD}$ to be sampled more uniformly $($as compared to for instance $\mathcal{P} = \omega_{\rm BD}^{-1}$ or $\mathcal{P} = \ln(1+\omega_{\rm BD}^{-1}))$, while at the same time having the GR-limit at $\mathcal{P} = 0$ (thus avoiding a GR-limit at infinity). Given an example prior range of $\mathcal{P} \in [0.05, 0.5]$, $\omega_{\rm BD}$ is sampled in the range $[10^2, 10^{20}]$, where a non-negligible region of $\mathcal{P}$ probes each decade in $\omega_{\rm BD}$. Here, concrete values of $\omega_{\rm BD} = \{10^2, 10^4, 10^{10}, 10^{20}\}$ correspond to $\mathcal{P} = \{0.5, 0.25, 0.1, 0.05\}$, while in the case of the $\omega_{\rm BD}^{-1}$ or $\ln(1+\omega_{\rm BD}^{-1})$ parameterizations $\mathcal{P} \simeq \{10^{-2}, 10^{-4}, 10^{-10}, 10^{-20}\}$. Hence, with the new parameterization, we expect to adequately sample a wider range of JBD gravity strengths.

For each parameterization, we can translate the constraints on $\mathcal{P}$ (posteriors shown in Fig.~\ref{figparams}) into constraints on $\omega_{\rm BD}$. As a concrete example, for the same dataset combination of ${\rm KiDS}\times{\rm 2dFLenS}+{\rm All}$-${\rm BOSS}+{\rm All}$-${\rm Planck}+{\rm Pantheon}$, in the case of the four parameterizations $\mathcal{P} = \{\ln \omega_{\rm BD}^{-1}, \ln(1+\omega_{\rm BD}^{-1}), \omega_{\rm BD}^{-1}, (\log_{10} \omega_{\rm BD})^{-1}\}$, the coupling constant is respectively constrained to $\omega_{\rm BD} > \{1540, 160, 160, 350\}$ ($95\%$~CL). As expected, the JBD constraints are approximately the same for the $\ln(1+\omega_{\rm BD}^{-1})$ and $\omega_{\rm BD}^{-1}$ parameterizations. We also note that the constraints using the latter three parameterizations do not suffer from the same dependence on the prior interval of $\omega_{\rm BD}$ as the fiducial parameterization, given their GR limit at $\mathcal{P} \rightarrow 0$ (for completeness, however, we have maintained the same prior interval as for the fiducial parameterization in the case of the $\ln(1+\omega_{\rm BD}^{-1})$ and $\omega_{\rm BD}^{-1}$ parameterizations). 

\subsubsection{Impact of the JBD parameterization on the effective gravitational constant, neutrino mass, and baryonic feedback}

The constraint on the effective gravitational constant is $\smash{G_{\rm matter}/G = 0.997^{+0.029}_{-0.029}}$ in the fiducial parameterization, and deviates by at most to $\smash{G_{\rm matter}/G = 0.970^{+0.033}_{-0.033}}$ for both the $\ln(1+\omega_{\rm BD}^{-1})$ and $\omega_{\rm BD}^{-1}$ parameterizations, which corresponds to a $15\%$ increase in the uncertainty and $0.9\sigma$ shift away from the GR expectation. While the other cosmological and systematics parameters are largely robust to the choice of JBD parameterization, in Fig.~\ref{figparams2d} we highlight an exception to this for the constraints in the plane of the sum of neutrino masses and the baryonic feedback amplitude. These constraints demonstrate the interplay of modified gravity, neutrino mass, and baryonic feedback, where the contours expectedly expand in the $\sum m_{\nu}$--$B$ plane for the parameterizations that allow for stronger JBD gravity. 

Considering the four different JBD parameterizations, we find $95\%$ upper bounds on $B$ in the range $2.8$ to $3.1$ and on $\sum m_{\nu}$ in the range $0.12$~eV to $0.32$~eV. In other words, notably, the bound on the sum of neutrino masses degrades by up to a factor of three as we consider a JBD parameterization that favors a stronger coupling constant. This exploration of the effects of the prior and parameterization of the JBD coupling constant on the parameter constraints underscores the need for clarity of the assumptions that enter the cosmological analysis, and illustrates the more complete inference that is possible from a broader consideration of these assumptions. 

\section{Conclusions}
\label{discussionsec}

As the precision and accuracy of cosmological datasets continue to improve, we will increasingly be able to test extensions to the standard cosmological model. One such extension consists of a modification of the gravitational theory, General Relativity, underpinning the expansion and the growth of structure in the Universe. To this end, we have performed a robust exploration of ``modified gravity'' with current cosmological data, capturing its impact on nonlinear scales with numerical simulations, and focusing on possible degeneracies with other cosmological and astrophysical degrees of freedom, such as the sum of neutrino masses and baryonic feedback, which we simultaneously constrain. 

We specifically consider the scalar-tensor theory of Jordan-Brans-Dicke (JBD)~\cite{Brans:1961sx}, where Newton's constant is promoted to a dynamical field as the scalar curvature becomes coupled to a hypothesized scalar field. We consider this model as a testbed for cosmological analyses of modified gravity (and extended cosmologies more broadly), given its rich history and the role it plays in some of the fundamentally motivated extensions to the standard model of particle physics: in particular in string theory, extra-dimensional theories, and the decoupling limit of theories with higher spin fields~\cite{Clifton:2011jh}. While JBD gravity can be considered the simplest modified gravity theory~\cite{Clifton:2011jh,amendola20}, it approximates a wider range of scalar-tensor theories (within Horndeski) on cosmological scales where gradients are suppressed~\cite{Avilez:2013dxa}. JBD gravity is also one of the remaining viable theories after the LIGO-Virgo measurement of the speed of gravitational waves \cite{ligovirgo,Baker:2017hug,Creminelli:2017sry,Ezquiaga:2017ekz}.

We provide an analytical and numerical description of JBD gravity in the linear regime, detailing its impact on the background evolution and linear perturbations (through modifications of the \eftcamb Einstein-Boltzmann solver). We extend this modeling of JBD gravity to nonlinear scales by performing a hybrid suite of $N$-body simulations to calibrate the \hmcode fitting function for the matter power spectrum to within 5--10\% precision. As \hmcode is further calibrated to simulations that separately include baryonic feedback and massive neutrinos~\cite{Mead15,Mead16}, we use this single fitting function to describe the nonlinear matter power spectrum in a Universe with cold dark matter, baryons, massive neutrinos, and modified gravity (neglecting the sub-dominant differences with an approach where all are simultaneously included in a single simulation suite~\cite{puchwein13, al19, ha20, mummery17, mccarthy18, baldi14, giocoli18, wkwz19}).

We methodically constrain the JBD model (via MCMC computations using our extended \cosmolss analysis package), mainly considering the CMB temperature, polarization, and lensing reconstruction from Planck 2018~\cite{planck2018}, the ``$3 \times 2{\rm pt}$'' combined dataset of cosmic shear, galaxy-galaxy lensing, and overlapping redshift-space galaxy clustering from KiDS$\times$2dFLenS (restricted to 450 deg$^2$)~\cite{Joudaki:2017zdt}, the Pantheon supernova distances~\cite{pantheon18}, along with the BOSS DR12 measurements of BAO distances, Alcock-Paczynski effect, and the growth rate~\cite{alam17}. We consider both a restricted JBD model with the coupling constant, $\omega_{\rm BD}$, as a new degree of freedom, along with an unrestricted JBD model where the effective gravitational constant at present, $G_{\rm matter}/G$, is additionally varied. In GR, $\omega_{\rm BD} \rightarrow \infty$ and $G_{\rm matter}/G = 1$. For both types of JBD gravity (and GR), we consider setups where the sum of neutrino masses, $\sum m_{\nu}$, is either fixed or allowed to vary. The baryonic feedback amplitude, $B$, moreover varies when we probe nonlinear scales. 

In the restricted JBD model, the Planck CMB temperature and polarization anisotropies constrain $\omega_{\rm BD} > 1150$, which degrades to $\omega_{\rm BD} > 430$ as the CMB polarization is excluded, and improves to $\omega_{\rm BD} > 1380$ when combined with the anisotropy measurements of the Atacama Cosmology Telescope (ACT DR4;~both at $95\%$~CL). For both datasets, the Hubble constant, $H_0$, rises as the strength of JBD gravity increases, with a ``hook shape'' in the $\omega_{\rm BD}$--$H_0$ plane. However, the uncertainties in our marginalized $H_0$ and $\smash{S_8 = \sigma_8 \sqrt{\Omega_{\rm m}/0.3}}$ increase by $\lesssim10\%$ and the parameters shift by $\lesssim0.3\sigma$ due to modified gravity alone, and so the discordances in these parameters persist; the former at the level of $\sim4\sigma$ with the direct measurement of the Hubble constant from Riess et al.~(2019)~\cite{riess2019}, and the latter at $\sim2.5\sigma$ with weak lensing (e.g.~CFHTLenS~\cite{Joudaki:2016mvz}, KiDS~\cite{hildebrandt19,asgari20b}, DES~\cite{desy1shear}, HSC~\cite{hikage19}, and combinations thereof~\cite{joudaki20,asgari20};~however also see~\cite{jee16,hamana20}).

The full Planck dataset (temperature, polarization, lensing reconstruction) in combination with the lower redshift datasets of BOSS and Pantheon constrain $\omega_{\rm BD} > 1460$~($95\%$~CL) as the sum of neutrino masses is kept fixed, and $\omega_{\rm BD} > 970$~($95\%$~CL) as the sum of neutrino masses is additionally varied (where $\sum m_{\nu} < 0.11$~eV at $95\%$~CL). We further combine the joint datasets of Planck, BOSS, and Pantheon with the $3\times2{\rm pt}$ dataset of KiDS$\times$2dFLenS in the unrestricted JBD cosmology, where the $2.3\sigma$ $S_8$ discrepancy in $\Lambda$CDM is reduced to $0.7\sigma$ ($0.6\sigma$ as the sum of neutrino masses is varied), and where there is weak-to-substantial concordance between the datasets over the full parameter space as estimated by the $\log\mathcal{I}$ statistic (which we connected to other known tension statistics). 

The Planck, BOSS, and Pantheon joint constraint on $H_0$ is also weakened in the unrestricted JBD cosmology, such that the tension with Riess et al.~(2019) decreases to $3.0\sigma$ for the full combination of datasets ($3.1\sigma$ as the sum of neutrino masses is varied). In this model, we constrain $\omega_{\rm BD} > 2230$~($95\%$~CL) and $G_{\rm matter}/G = 0.996^{+0.029}_{-0.029}$ given fixed neutrino masses, along with $\omega_{\rm BD} > 1540$~($95\%$~CL) and $G_{\rm matter}/G = 0.997^{+0.029}_{-0.029}$ as the sum of neutrino masses is simultaneously varied, both in excellent agreement with GR. These constraints on the coupling constant and present effective gravitational constant are driven by the Planck CMB and are the strongest to date, where in particular the $G_{\rm matter}/G$ constraints are improved by nearly a factor of two relative to comparable past analyses (e.g.~Ref.~\cite{Avilez:2013dxa}). The $3\%$ constraint on $G_{\rm matter}/G$ is also comparable to the precision of the BBN constraint in Ref.~\cite{alvey20}. 

Given our bounds on the JBD coupling constant, we constrained the EFT parameters $\alpha_{\rm M} \lesssim 10^{-3}$, $\alpha_{\rm B} \gtrsim -10^{-3}$, and $\alpha_{\rm K} \lesssim 2\times10^{-3}$ since the onset of matter domination (in addition to $\alpha_{\rm T} = 0$ in JBD theory). As the evolution of the effective gravitational constant is uniquely determined by the coupling constant, we moreover showed that $|{\rm d} \ln G_{\rm matter} / {\rm d}t|_{a=1} \lesssim 10^{-13} \, {\rm year}^{-1}$ ($95\%$~CL) in JBD gravity.

In the unrestricted JBD model, the neutrino mass bound is marginally weakened to $\sum m_{\nu} < 0.12$~eV~($95\%$~CL). The combined datasets improve the constraint on the intrinsic alignment amplitude by $40\%$ relative to KiDS$\times$2dFLenS alone (by $20\%$ relative to KiDS$\times$\{2dFLenS+BOSS\}), such that $A_{\rm IA} = 1.52^{+0.38}_{-0.38}$ is positive at $4.0\sigma$. However, we note that this large amplitude is not found favored in direct measurements of $A_{\rm IA}$~\cite{johnston19} and is likely driven by the photometric redshift uncertainties~(e.g.~\cite{,wright2020,fortuna2020}). We also constrain the baryonic feedback amplitude to $B<2.8$ ($95\%$~CL), which corresponds to a shift of $\Delta B = -0.7$ in the upper bound relative to the $\Lambda$CDM constraint of KiDS$\times$2dFLenS ($\Delta B = -0.4$ relative to KiDS$\times$\{2dFLenS+BOSS\}), and mildly favors a deviation from the ``no feedback'' scenario of $B = 3.13$. 

Employing the deviance information criterion, we find no meaningful model selection preference for JBD gravity (relative to $\Lambda$CDM) for the different dataset combinations and specific cosmologies considered. Given the alleviation of the $\{H_0, S_8\}$ discordances in the unrestricted JBD model, we additionally performed a model selection assessment with the Riess et al.~(2019) measurement of the Hubble constant included in the analysis. For Planck 2018 combined with Riess et al.~(2019), we find $\Delta{\rm DIC} = -4.8$, which corresponds to~weak-to-moderate preference in favor of the extended model. However, for the full combination of datasets (which here includes Riess et al.~2019), this decreases to $\Delta{\rm DIC} = -2.7$, corresponding to weak preference in favor of the extended model.

In addition to our fiducial cosmological constraints, we have examined their sensitivity to the modeling choices. In particular, we have illustrated how the choice of parameterization of the coupling constant can affect the constraints on $G_{\rm matter}/G$, $B$, and $\sum m_{\nu}$, in the latter case degrading the upper bound by up to a factor of three. We have further highlighted the possible degeneracy of the effective gravitational constant with other physics that affect the CMB damping tail. As the effective gravitational constant gives rise to a response in the CMB temperature and polarization power spectra that coherently strengthens towards smaller scales, it is correlated with physics such as the primordial helium abundance, the effective number of neutrinos, and the running of the spectral index, targeted by CMB surveys such as AdvACT~\cite{advact}, SPT-3G~\cite{spt3g}, and the Simons Observatory~\cite{simonsobs}. Given the CMB datasets of Planck and ACT (along with the lower-redshift datasets of KiDS, 2dFLenS, BOSS, Pantheon) we provided an illustration of an up to $80\%$ degradation in the constraints on the effective gravitational constant and the effective number of neutrinos when analyzed simultaneously. 

We note that our bounds on the JBD coupling constant are more than an order of magnitude weaker than the bounds from astrophysical probes (i.e.~$\omega_{\rm BD} \gtrsim 10^3$ from cosmology as compared to $\omega_{\rm BD} \gtrsim 10^4$--$10^5$ from astrophysics). Concretely, our strongest bound on the coupling constant is a factor of $18$ weaker than that obtained from Shapiro time delay measurements by the Cassini satellite~\cite{Bertotti:2003rm}, a factor of $5$ weaker than the lower bound from the analysis of the pulsar--white dwarf binary PSR J1738+0333~\cite{2012MNRAS.423.3328F}, and a factor 63 weaker than the stellar triple system PSR J0337+1715~\cite{archibald18,voisin20}. However, we emphasize the usefulness of constraining modified gravity through multiple pathways. In the case of JBD theory, as it can be considered an approximation to a wider class of Horndeski scalar-tensor theories on cosmological scales~\cite{Avilez:2013dxa}, which may be endowed with screening mechanisms on astrophysical scales, these stronger astrophysical bounds might not be representative of the true strength of its corrections on the cosmological observables.

We expect approximately an order of magnitude improvement in the constraints on JBD gravity with next-generation (Stage-IV) surveys of the CMB, the large-scale structure, and the radio sky~\cite{Alonso:2016suf}, which can be further improved in a combined analysis with a future set of gravitational wave standard siren events (as future electromagnetic and gravitational wave surveys have been shown to yield comparable improvements in the constraint on the EFT parameter $\alpha_{\rm M}$~\cite{bh20}). This will allow for cosmological constraints on JBD gravity that are at a similar precision to the most powerful astrophysical probes. 

For the expected improvements in cosmological inferences to be realized, a series of conditions need to be met, in particular:~continued progress in the nonlinear modeling of the modified gravity, the disentanglement of possible degeneracies with other cosmological and systematics degrees of freedom, and concordance between distinct cosmological datasets (required for multi-probe analyses; noting that the concordance might only emerge in the extended cosmology, as we have shown). The impact of modeling choices, for instance concerning the choice of modified gravity parameterization and prior ranges, also needs to be highlighted in a robust cosmological analysis. We have illustrated these different ingredients for JBD gravity, and anticipate the advent of more powerful datasets which will either signpost deviations to or confirm the standard cosmological model to ever higher precision.

We publicly release the MCMC chains together with the likelihood code and data products needed to reproduce them in \url{https://github.com/sjoudaki/CosmoJBD}.
 
\section*{Acknowledgements}
{We thank S. Alam, D. Alonso, A. Avilez, D. Bartlett, E. Bellini, E. Calabrese, H. Desmond, B. Hu, M. Kamionkowski, K. Kuijken, E. M. Mueller, T. Nanayakkara, V. Niro, D. Parkinson, M. Raveri, A. Silvestri, C. Skordis, and M. Zumalacarregui for useful discussions. We thank Shadab Alam for providing the BOSS DR12 final consensus likelihood, and Nora Elisa Chisari for performing an internal KiDS Collaboration review of the manuscript. We also thank Jonathan Patterson and Jon Wakelin for HPC support. We acknowledge the use of the \camb~\cite{LCL,hu14eft}, \cosmomc~\cite{Lewis:2002ah}, and GetDist~\cite{getdist19} packages. SJ and PGF acknowledge support from the Beecroft Trust, the Science and Technology Facilities Council (STFC), and the European Research Council under grant agreement No.~693024. We acknowledge the use of the Oxford computing cluster Glamdring. Part of this work was performed using the DiRAC Data Intensive service at Leicester operated by the University of Leicester IT Services, and DiRAC@Durham managed by the Institute for Computational Cosmology, which form part of the STFC DiRAC HPC Facility acknowledging BEIS and STFC grants STK0003731, STR0023631, STR0010141, STP0022931, STR0023711, STR0008321.}


\bibliographystyle{apsrev4-2}
\bibliography{JBD}

\appendix

\begin{figure*}
\includegraphics[width=\hsize]{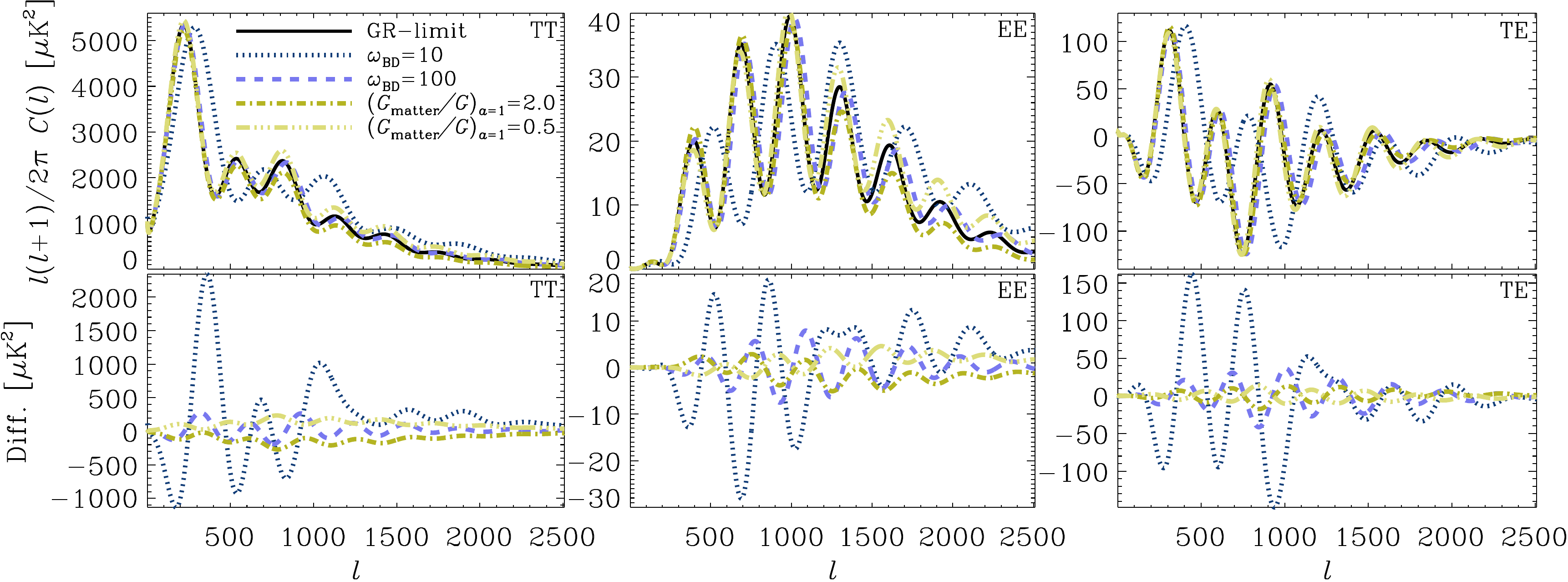}
\vspace{-2.1em}
\caption{CMB temperature ($T$) and polarization ($E$) power spectra in a cosmology with JBD gravity along with their respective differences relative to GR, defined as $A_{\rm JBD} - A_{\rm GR}$, where $A \in \{TT, EE, TE\}$ (noting that $TT$ here is the same as in Fig.~\ref{figcls} and shown for comparison with $EE$ and $TE$). For our GR limit, we have effectively imposed $\omega_{\rm BD} \rightarrow \infty$ and $G_{\rm matter}/G = 1$. For the JBD model, we show the four cases $\omega_{\rm BD} = 10$, $\omega_{\rm BD} = 100$, $G_{\rm matter}/G = 0.5$, and $G_{\rm matter}/G = 2.0$ (such that $\omega_{\rm BD} \rightarrow \infty$ when $G_{\rm matter}/G \neq 1$, and $G_{\rm matter}/G = 1$ when $\omega_{\rm BD} \neq \infty$). We emphasize that our use of ``$G_{\rm matter}/G$'' here is shorthand for $(G_{\rm matter}/G)|_{a=1}$ (as defined in Eq.~\ref{jbdpsi}). 
}
\label{figclspol}
\end{figure*}

\begin{figure*}
\includegraphics[width=\hsize]{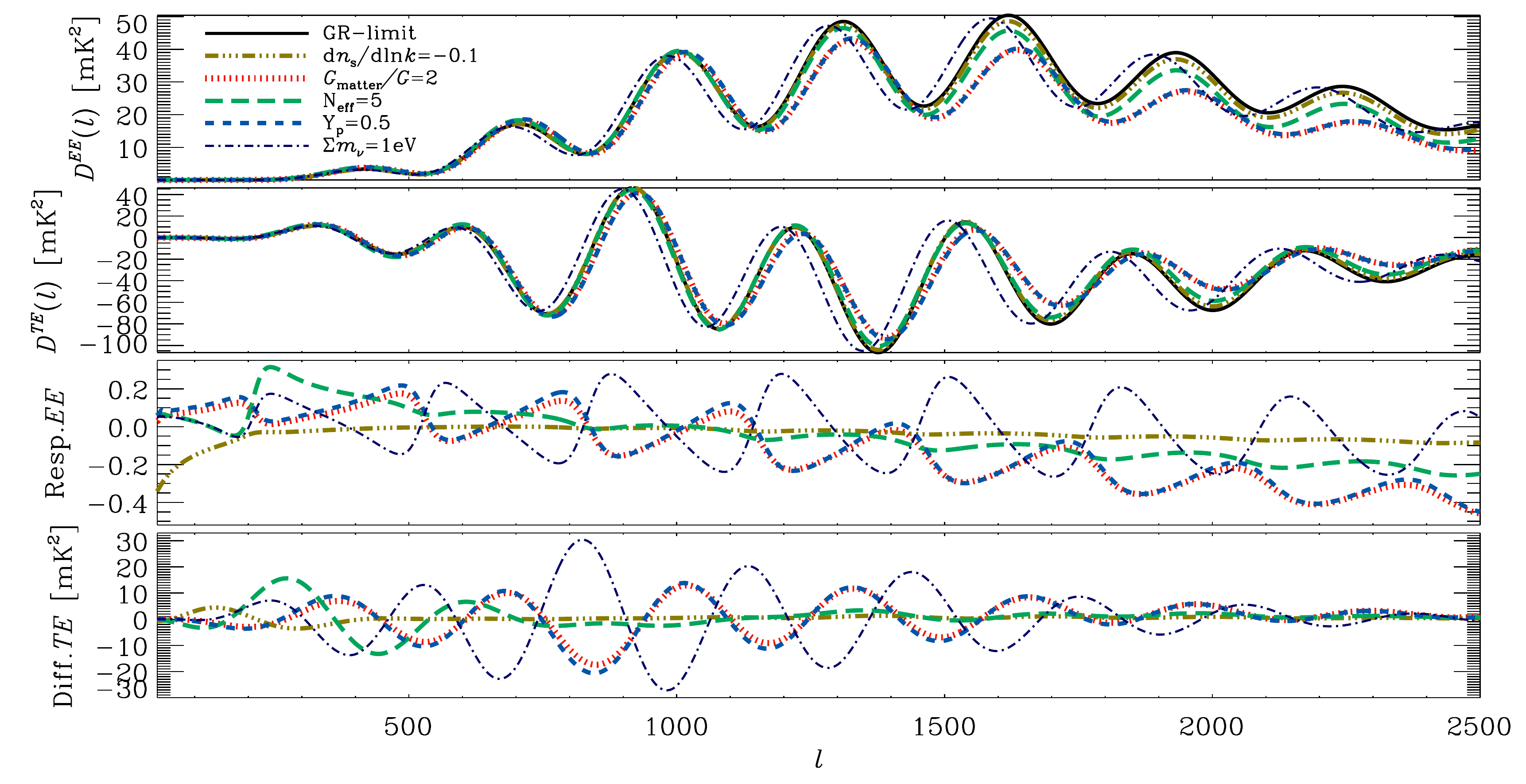}
\vspace{-2.3em}
\caption{CMB polarization power spectra ($EE$) and temperature-polarization cross-spectra ($TE$) in extended cosmological parameter spaces, where $D(\ell) = \ell^3 (\ell+1)/{2\pi}~C(\ell)$, along with the $EE$ responses, defined as $C^{\rm extended}(\ell)/C^{\Lambda{\rm CDM}}(\ell) - 1$, and the $TE$ differences, defined as $C^{\rm extended}(\ell) - C^{\Lambda{\rm CDM}}(\ell)$. We consider deviations in the running of the spectral index, ${\rm d} n_{\rm s} / {\rm d} \ln k$, the effective number of neutrinos, $N_{\rm eff}$, the sum of neutrino masses, $\sum m_{\nu}$, the primordial helium abundance, $Y_{\rm P}$, and the present effective gravitational constant, $G_{\rm matter}/G$.  
}
\label{figclsdampingpol}
\end{figure*}

\begin{figure*}
\hspace{-0.05cm}
\includegraphics[width=0.92\hsize]{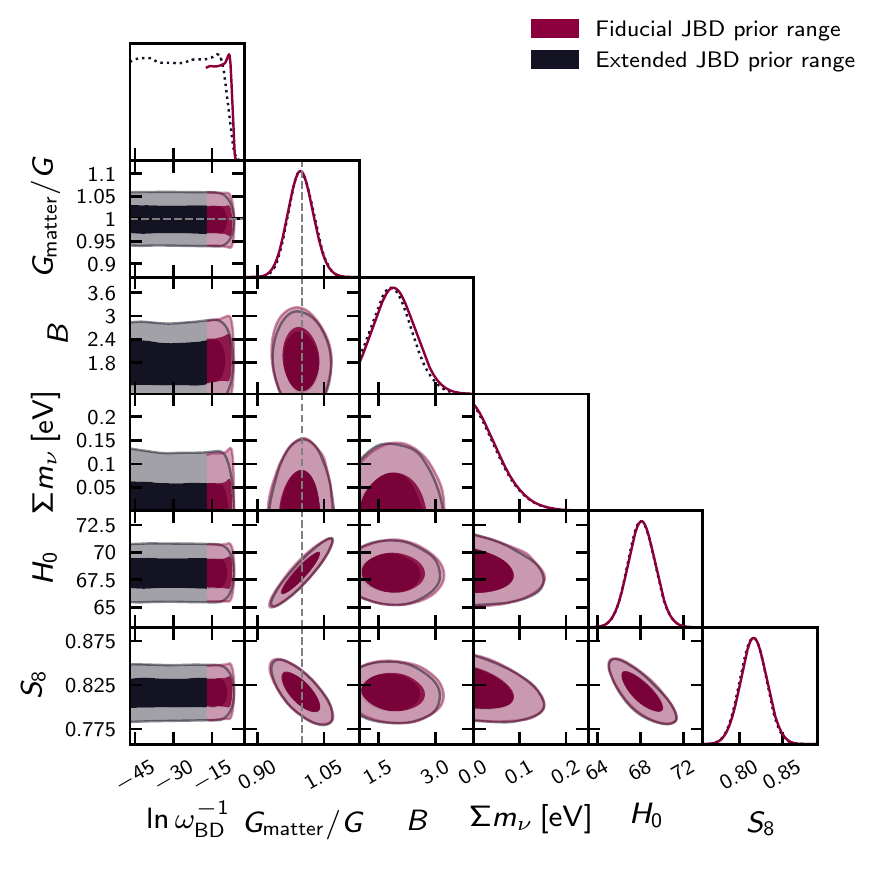}
\vspace{-2.5em}
\caption{\label{figprior}
Marginalized posterior distributions (inner $68\%$~CL, outer $95\%$~CL) of the JBD parameter, $\ln \omega_{\rm BD}^{-1}$, the present effective gravitational constant, $G_{\rm matter}/G$, the baryonic feedback amplitude, $B$, the sum of neutrino masses, $\sum m_{\nu}$, the Hubble constant, $H_0$ (in units of ${\rm km} \, {\rm s}^{-1} {\rm Mpc}^{-1}$), and $S_8 = \sigma_8 \sqrt{\Omega_{\mathrm m}/0.3}$, considering two different choices of the prior range of the primary parameter $\ln \omega_{\rm BD}^{-1}$. In the fiducial setup, 
$-17 \leq \ln \omega_{\rm BD}^{-1} \leq -2.3$, while in the extended setup, $-47 \leq \ln \omega_{\rm BD}^{-1} \leq -2.3$. All other standard cosmological and systematics parameters are simultaneously~varied.
}
\end{figure*}

\begin{table*}
\vspace{-0.8em}
\caption{\label{subtabdic} The changes in the best-fit $\smash{\chi^2_{\rm eff}}$ and deviance information criterion (DIC) relative to $\Lambda$CDM for different data combinations and parameter extensions. A negative change indicates preference in favor of the extended model. For completeness, the absolute numbers for the best-fit $\smash{\chi^2_{\rm eff}}$ and DIC of the two most central runs are: ${\rm All}$-${\rm Planck18}+{\rm All}$-${\rm BOSS}+{\rm Pantheon} \left({{\rm JBD}+\sum m_{\nu}}\right)$, where $\smash{\chi^2_{\rm eff} = 3820.4}$ and ${\rm DIC} = 3865.1$, 
and ${\rm KiDS}\times{\rm 2dFLenS}+{\rm All}$-${\rm BOSS}+{\rm All}$-${\rm Planck18}+{\rm Pantheon} \left({{\rm JBD+G_{\rm matter}+\sum m_{\nu}}}\right)$, where $\smash{\chi^2_{\rm eff} = 4019.7}$ and ${\rm DIC} = 4072.7$. 
}
\renewcommand{\footnoterule}{}
\begin{tabularx}{\textwidth}{l@{\extracolsep{\fill}}rr}
\hline
Probe setup & $\Delta \chi^2_{\rm eff}$ & $\Delta {\rm DIC}$\\
\hline
${\rm Planck18} \, \left({\Lambda {\rm CDM}+\sum m_{\nu}}\right)$ & $1.1$ & $3.9$ \\
${\rm Planck18} \, \left({{\rm JBD}}\right)$ & $-0.41$ & $1.0$ \\
${\rm Planck18} \, \left({{\rm JBD}+\sum m_{\nu}}\right)$ & $1.6$ & $3.3$ \\
${\rm Planck18} \, \left({{\rm JBD}+G_{\rm matter}+\sum m_{\nu}}\right)$ & $2.7$ & $4.0$ \\
${\rm ACT~DR4} \, \left({{\rm JBD}}\right)$ & $-0.10$ & $0.22$ \\
${\rm Planck18}+{\rm ACT~DR4} \, \left({{\rm JBD}}\right)$ & $-0.04$ & $0.34$ \\
${\rm All}$-${\rm Planck18} \, \left({\Lambda {\rm CDM}+\sum m_{\nu}}\right)$ & $0.83$ & $3.0$ \\
${\rm All}$-${\rm Planck18} \, \left({{\rm JBD}}\right)$ & $-0.84$ & $1.3$ \\
${\rm KiDS}\times{\rm 2dFLenS} \left({\Lambda{\rm CDM}+\sum m_{\nu}}\right)$ & $0.097$ & $0.030$ \\
${\rm KiDS}\times{\rm 2dFLenS} \left({{\rm JBD}}\right)$ & $0.15$ & $-0.41$ \\
${\rm KiDS}\times{\rm 2dFLenS} \left({{\rm JBD}+\sum m_{\nu}}\right)$ & $-0.32$ & $0.62$ \\
${\rm KiDS}\times{\rm 2dFLenS} \left({{\rm JBD}+G_{\rm matter}}\right)$ & $0.12$ & $1.2$ \\
${\rm KiDS}\times{\rm 2dFLenS} \left({{\rm JBD}+G_{\rm matter}+\sum m_{\nu}}\right)$ & $-0.41$ & $2.3$ \\
${\rm All}$-${\rm BOSS}+{\rm Pantheon} \left({\Lambda {\rm CDM}}+\sum m_{\nu}\right)$ & $0.0044$ & $-0.072$ \\
${\rm All}$-${\rm BOSS}+{\rm Pantheon} \left({\rm JBD}\right)$ & $-0.0038$ & $-0.033$ \\
${\rm All}$-${\rm BOSS}+{\rm Pantheon} \left({\rm JBD}+\sum m_{\nu}\right)$ & $0.0012$ & $-0.10$ \\
${\rm All}$-${\rm BOSS}+{\rm Pantheon} \left({\rm JBD}+G_{\rm matter}\right)$ & $-0.061$ & $-0.060$ \\
${\rm All}$-${\rm BOSS}+{\rm Pantheon} \left({\rm JBD}+G_{\rm matter}+\sum m_{\nu}\right)$ & $-1.4$ & $1.3$ \\
${\rm KiDS}\times\{{\rm 2dFLenS}+{\rm BOSS}\} \left({\Lambda{\rm CDM}+\sum m_{\nu}}\right)$ & $0.34$ & $0.55$ \\
${\rm KiDS}\times\{{\rm 2dFLenS}+{\rm BOSS}\} \left({{\rm JBD}}\right)$ & $0.46$ & $-0.098$ \\
${\rm KiDS}\times\{{\rm 2dFLenS}+{\rm BOSS}\} \left({{\rm JBD,~no~feedback}}\right)$ & $3.2$ & $1.4$ \\
${\rm KiDS}\times\{{\rm 2dFLenS}+{\rm BOSS}\} \left({{\rm JBD}+\sum m_{\nu}}\right)$ & $-1.3$ & $2.1$ \\
${\rm KiDS}\times\{{\rm 2dFLenS}+{\rm BOSS}\} \left({{\rm JBD}+\sum m_{\nu},~{\rm no~feedback}}\right)$ & $3.9$ & $2.8$ \\
${\rm KiDS}\times\{{\rm 2dFLenS}+{\rm BOSS}\} \left({{\rm JBD}+G_{\rm matter}}\right)$ & $3.3$ & $1.4$ \\
${\rm KiDS}\times\{{\rm 2dFLenS}+{\rm BOSS}\} \left({{\rm JBD}+G_{\rm matter}+\sum m_{\nu}}\right)$ & $0.62$ & $2.2$ \\
${\rm KiDS}\times{\rm 2dFLenS}+{\rm All}$-${\rm BOSS}+{\rm Pantheon} \left({\Lambda{\rm CDM}+\sum m_{\nu}}\right)$ & $-1.2$ & $-0.48$ \\
${\rm KiDS}\times{\rm 2dFLenS}+{\rm All}$-${\rm BOSS}+{\rm Pantheon} \left({{\rm JBD}}\right)$ & $0.43$ & $-0.60$ \\
${\rm KiDS}\times{\rm 2dFLenS}+{\rm All}$-${\rm BOSS}+{\rm Pantheon} \left({{\rm JBD}+\sum m_{\nu}}\right)$ & $-1.8$ & $0.0055$ \\
${\rm KiDS}\times{\rm 2dFLenS}+{\rm All}$-${\rm BOSS}+{\rm Pantheon} \left({{\rm JBD}+G_{\rm matter}}\right)$ & $0.50$ & $1.5$ \\
${\rm KiDS}\times{\rm 2dFLenS}+{\rm All}$-${\rm BOSS}+{\rm Pantheon} \left({{\rm JBD}+G_{\rm matter}+\sum m_{\nu}}\right)$ & $-1.8$ & $1.9$ \\
${\rm All}$-${\rm Planck18}+{\rm All}$-${\rm BOSS}+{\rm Pantheon} \left({{\Lambda{\rm CDM}}+\sum m_{\nu}}\right)$ & $2.1$ & $0.53$ \\
${\rm All}$-${\rm Planck18}+{\rm All}$-${\rm BOSS}+{\rm Pantheon} \left({{\rm JBD}}\right)$ & $0.14$ & $-0.22$ \\
${\rm All}$-${\rm Planck18}+{\rm All}$-${\rm BOSS}+{\rm Pantheon} \left({{\rm JBD}+\sum m_{\nu}}\right)$ & $0.076$ & $2.4$ \\
${\rm All}$-${\rm Planck18}+{\rm All}$-${\rm BOSS}+{\rm Pantheon} \left({{\rm JBD}+G_{\rm matter}}\right)$ & $1.6$ & $3.3$ \\
${\rm All}$-${\rm Planck18}+{\rm All}$-${\rm BOSS}+{\rm Pantheon} \left({{\rm JBD}+G_{\rm matter}+\sum m_{\nu}}\right)$ & $1.5$ & $4.1$ \\
${\rm KiDS}\times{\rm 2dFLenS}+{\rm All}$-${\rm BOSS}+{\rm All}$-${\rm Planck18}+{\rm Pantheon} \left({{\rm JBD}+G_{\rm matter}}\right)$ & $0.45$ & $1.7$\\
${\rm KiDS}\times{\rm 2dFLenS}+{\rm All}$-${\rm BOSS}+{\rm All}$-${\rm Planck18}+{\rm Pantheon} \left({{\rm JBD}+G_{\rm matter}+\sum m_{\nu}}\right)$ & $-2.0$ & $4.6$\\
${\rm Planck18}+{\rm Riess~2019}\, \left({{\rm JBD}+G_{\rm matter}+\sum m_{\nu}}\right)$ & $-4.0$ & $-4.8$\\
${\rm KiDS}\times{\rm 2dFLenS}+{\rm All}$-${\rm BOSS}+{\rm All}$-${\rm Planck18}+{\rm Pantheon}+{\rm Riess~2019}\, \left({{\rm JBD}+G_{\rm matter}+\sum m_{\nu}}\right)$ & $-4.7$ & $-2.7$\\
\hline
\end{tabularx}
\end{table*}

\begin{table*}
\vspace{-0.8em}
\caption{\label{subtabappall} Marginalized posterior means and $68\%$ confidence intervals for the Hubble constant, $H_0$, in units of km\,s$^{-1}$\,Mpc$^{-1}$, and $\smash{S_8=\sigma_8\sqrt{\Omega_{\rm m}/0.3}}$. The symbol ``$\diamond$'' implies that the parameter is effectively unconstrained by the data, and the symbol ``$\circ$'' implies that the tension $T$ is not meaningful to quote (i.e.~$T \sim 0$).
}
\renewcommand{\footnoterule}{} 
\begin{tabularx}{\textwidth}{l@{\extracolsep{\fill}}cc|cc}
\hline
Probe setup & $H_0$ & $S_8$ & $T(H_0)_{\rm Riess19}$ & $T(S_8)_{\rm Planck18}$\\
${\rm KiDS}\times{\rm 2dFLenS} \left({\Lambda{\rm CDM}}\right)$ & $\diamond$ & $0.736^{+0.039}_{-0.038}$ & $\circ$ & $2.4$\\
${\rm KiDS}\times{\rm 2dFLenS} \left({\Lambda{\rm CDM}+\sum m_{\nu}}\right)$ & $\diamond$  & $0.723^{+0.037}_{-0.037} $  & $\circ$ & $2.6$ \\
${\rm KiDS}\times{\rm 2dFLenS} \left({{\rm JBD}}\right)$ & $\diamond$  & $0.738^{+0.039}_{-0.039} $  & $\circ$ & $2.4$ \\
${\rm KiDS}\times{\rm 2dFLenS} \left({{\rm JBD}+\sum m_{\nu}}\right)$ & $\diamond$  & $0.725^{+0.035}_{-0.035} $  & $\circ$ & $2.5$ \\
${\rm KiDS}\times{\rm 2dFLenS} \left({{\rm JBD}+G_{\rm matter}}\right)$ & $\diamond$  & $0.768^{+0.083}_{-0.105}$ & $\circ$ & $0.8$ \\
${\rm KiDS}\times{\rm 2dFLenS} \left({{\rm JBD}+G_{\rm matter}+\sum m_{\nu}}\right)$ & $\diamond$  & $0.772^{+0.080}_{-0.095}$  & $\circ$ & 0.7 \\
\end{tabularx}
\renewcommand{\footnoterule}{} 
\begin{tabularx}{\textwidth}{l@{\extracolsep{\fill}}cc|ccc}
\hline
 & $H_0$ & $S_8$ & $T(H_0)_{\rm Riess19}$ & $T(S_8)_{\rm Planck18}$ & $T(S_8)_{{\rm KiDS}\times{\rm 2dFLenS}}$\\
${\rm All}$-${\rm BOSS}+{\rm Pantheon} \left({\Lambda {\rm CDM}}\right)$ & $71.1^{+6.3}_{-4.3}$  & $0.805^{+0.051}_{-0.052} $  & $0.54$ & $0.6$ & $1.1$\\
${\rm All}$-${\rm BOSS}+{\rm Pantheon} \left({\Lambda {\rm CDM}}+\sum m_{\nu}\right)$ & $69.7^{+6.2}_{-4.3}$  & $0.812^{+0.051}_{-0.051}$  & $0.79$ & $0.3$ & $1.4$ \\
${\rm All}$-${\rm BOSS}+{\rm Pantheon} \left({\rm JBD}\right)$ & $73.7^{+4.9}_{-7.9}$  & $0.802^{+0.051}_{-0.051}$  & $0.04$ & $0.7$ & $1.0$ \\
${\rm All}$-${\rm BOSS}+{\rm Pantheon} \left({\rm JBD}+\sum m_{\nu}\right)$ & $72.3^{+5.2}_{-7.5}$  &  $0.810^{+0.051}_{-0.051}$  & $0.26$ & $0.4$ & $1.3$ \\
${\rm All}$-${\rm BOSS}+{\rm Pantheon} \left({\rm JBD}+G_{\rm matter}\right)$ & $\diamond$  & $\diamond$  & $\circ$ & $\circ$ & $\circ$ \\
${\rm All}$-${\rm BOSS}+{\rm Pantheon} \left({\rm JBD}+G_{\rm matter}+\sum m_{\nu}\right)$ & $\diamond$ & $\diamond$ & $\circ$ & $\circ$ & $\circ$ \\
\hline
\end{tabularx}
\end{table*}

\begin{table*}
\footnotesize
\vspace{-0.1em}
\caption{\label{subtabplanck15} Marginalized posterior means and $68\%$ confidence intervals for a subset of the cosmological parameters when analyzing the Planck 2015 dataset (instead of Planck 2018). For the JBD parameter, $\omega_{\rm BD}$, and the sum of neutrino masses, $\sum m_{\nu}$, we quote the $95\%$ confidence lower and upper bounds, respectively. The sum of neutrino masses, $\sum m_{\nu}$, is in units of eV, the Hubble constant, $H_0$, is in km\,s$^{-1}$\,Mpc$^{-1}$, and $\smash{S_8=\sigma_8\sqrt{\Omega_{\rm m}/0.3}}$. A table element with ``$\cdots$'' implies that the parameter is not varied in the analysis. There is a minor improvement in the $H_0$ (and $A_{\rm IA}$) constraint as we allow the sum of neutrino masses to vary in ${\rm KiDS}\times{\rm 2dFLenS}+{\rm All}$-${\rm BOSS}+{\rm All}$-${\rm Planck15}+{\rm Pantheon}$, as the fiducial $\sum m_{\nu}$ is located at the boundary of the posterior when varied, such that the widest range in $H_0$ is favored at this boundary, and due to the weak correlation between the parameters (see Fig.~\ref{figplanck18unrest}). The tensions $T(H_0)$ and $T(S_8)$ are against Riess et al.~2019~\cite{riess2019} and KiDS$\times$\{2dFLenS+BOSS\}, respectively (in the latter case only against KiDS$\times$2dFLenS when Planck is combined with BOSS). See the caption of Table~\ref{subtab1} for further details.}
\renewcommand{\footnoterule}{} 
\begin{tabularx}{\textwidth}{l@{\extracolsep{\fill}}ccccc|cc}
\hline
Probe setup & $\omega_{\rm BD}$ & $G_{\rm matter}/G$ & $\sum m_{\nu}$ & $H_0$ & $S_8$ & $T(H_0)$ & $T(S_8)$\\
\hline
${\rm Planck15} \, \left({\Lambda {\rm CDM}}\right)$ & $\cdots$ & $\cdots$ & $\cdots$  & $67.94^{+1.00}_{-0.98} $ & $0.853^{+0.025}_{-0.025} $  & $3.5$ & $2.5$\\
${\rm Planck15} \, \left({\Lambda {\rm CDM}+\sum m_{\nu}}\right)$ & $\cdots$ & $\cdots$ & $0.64$  & $65.59^{+1.63}_{-2.43} $ & $0.843^{+0.026}_{-0.026} $  & $3.4$ & $2.5$\\
${\rm Planck15} \, \left({{\rm JBD}}\right)$ & $530$ & $\cdots$ & $\cdots$ & $68.27^{+0.93}_{-1.29} $  & $ 0.850^{+0.025}_{-0.025} $  & $3.2$ & $2.4$\\
${\rm Planck15} \, \left({{\rm JBD}+\sum m_{\nu}}\right)$ & $860$ & $\cdots$ & $0.62$  & $ 65.95^{+2.49}_{-1.92} $  & $ 0.842^{+0.026}_{-0.025} $  & $3.3$ & $2.4$\\
${\rm Planck15} \, \left({{\rm JBD}+G_{\rm matter}}\right)$ & $850$ & $1.024^{+0.046}_{-0.053}$ & $\cdots$ & $69.86^{+2.57}_{-2.88} $  & $ 0.842^{+0.034}_{-0.034} $  & $1.6$ & $1.6$\\
${\rm Planck15}+{\rm ACT~DR3} \, \left({\Lambda {\rm CDM}}\right)$ & $\cdots$ & $\cdots$ & $\cdots$ & $67.76^{+0.94}_{-0.95} $  & $0.853^{+0.024}_{-0.024} $  & $3.7$ & $2.5$\\
${\rm Planck15}+{\rm ACT~DR3} \, \left({{\rm JBD}}\right)$ & $900$ & $\cdots$ & $\cdots$ & $67.97^{+0.92}_{-1.10} $  & $0.851^{+0.023}_{-0.024} $  & $3.5$ & $2.5$\\
${\rm All}$-${\rm Planck15} \, \left({\Lambda{\rm CDM}}\right)$ & $\cdots$ & $\cdots$ & $\cdots$ & $68.02^{+0.64}_{-0.64}$ & $0.834^{+0.012}_{-0.013}$ & $3.9$ & $2.4$ \\
${\rm All}$-${\rm Planck15} \, \left({{\rm JBD}}\right)$ & $1110$ & $\cdots$ & $\cdots$ & $68.18^{+0.65}_{-0.74}$ & $0.835^{+0.013}_{-0.013}$ & $3.7$ & $2.3$ \\
${\rm Planck15}+{\rm All}$-${\rm BOSS}+{\rm Pantheon} \left({\Lambda{\rm CDM}}\right)$ & $\cdots$ & $\cdots$ & $\cdots$ & $68.16^{+0.54}_{-0.54} $  & $ 0.843^{+0.018}_{-0.017} $  & $3.9$ & $2.5$ \\
${\rm Planck15}+{\rm All}$-${\rm BOSS}+{\rm Pantheon} \left({\Lambda{\rm CDM}+\sum m_{\nu}}\right)$ & $\cdots$ & $\cdots$ & $0.20$ & $67.85^{+0.59}_{-0.60}$  & $ 0.832^{+0.020}_{-0.020} $  & $4.0$ & $2.5$ \\
${\rm Planck15}+{\rm All}$-${\rm BOSS}+{\rm Pantheon} \left({{\rm JBD}}\right)$ & $840$ & $\cdots$ & $\cdots$ & $68.26^{+0.53}_{-0.63}$  & $0.844^{+0.018}_{-0.018}$ & $3.8$ & $2.5$ \\
${\rm Planck15}+{\rm All}$-${\rm BOSS}+{\rm Pantheon} \left({{\rm JBD}+\sum m_{\nu}}\right)$ & $480$ & $\cdots$ & $0.22$ & $68.02^{+0.60}_{-0.71}$  & $0.832^{+0.020}_{-0.020}$ & $3.8$ & $2.5$ \\
${\rm Planck15}+{\rm All}$-${\rm BOSS}+{\rm Pantheon} \left({{\rm JBD}+G_{\rm matter}}\right)$ & $1460$ & $1.014^{+0.045}_{-0.045}$ &  $\cdots$ & $68.79^{+1.89}_{-1.90} $  & $ 0.839^{+0.023}_{-0.025} $  & $2.2$ & $0.7$ \\
${\rm Planck15}+{\rm All}$-${\rm BOSS}+{\rm Pantheon} \left({{\rm JBD}+G_{\rm matter}+\sum m_{\nu}}\right)$ & $1140$ & $1.029^{+0.045}_{-0.044}$ & $0.22$ & $69.03^{+1.80}_{-1.83}$ & $0.822^{+0.026}_{-0.026}$ & $2.2$ & $0.6$ \\
${\rm All}$-${\rm Planck15}+{\rm All}$-${\rm BOSS}+{\rm Pantheon} \left({\Lambda{\rm CDM}}\right)$ & $\cdots$ & $\cdots$ & $\cdots$ & $68.11^{+0.45}_{-0.45}$  & $0.832^{+0.011}_{-0.011}$ & $4.0$ & $2.4$\\
${\rm All}$-${\rm Planck15}+{\rm All}$-${\rm BOSS}+{\rm Pantheon} \left({{\rm JBD}}\right)$ & $1050$ & $\cdots$ & $\cdots$ & $68.22^{+0.46}_{-0.52}$  & $0.833^{+0.011}_{-0.011}$ & $3.9$ & $2.3$\\
${\rm All}$-${\rm Planck15}+{\rm All}$-${\rm BOSS}+{\rm Pantheon} \left({{\rm JBD}+\sum m_{\nu}}\right)$ & $590$ & $\cdots$ & $0.19$ & $67.88^{+0.59}_{-0.61}$  & $0.828^{+0.012}_{-0.012}$ & $4.0$ & $2.6$\\
\hline
\end{tabularx}
\renewcommand{\footnoterule}{} 
\begin{tabularx}{\textwidth}{l@{\extracolsep{\fill}}ccccccc|cc}
& $\omega_{\rm BD}$ & $G_{\rm matter}/G$ & $B$ & $A_{\rm IA}$ & $\sum m_{\nu}$ & $H_0$ & $S_8$ & $T(H_0)$\\
${\rm KiDS}\times{\rm 2dFLenS}+{\rm All}$-${\rm BOSS}+{\rm Planck15} \left({{\rm JBD}+G_{\rm matter}+\sum m_{\nu}}\right)$ & $840$ & $1.034^{+0.042}_{-0.046}$ & $3.0$ & $1.50^{+0.41}_{-0.41}$ & $0.30$ & $69.51^{+1.80}_{-1.79}$ & $0.790^{+0.025}_{-0.025}$ & $2.0$\\
${\rm KiDS}\times{\rm 2dFLenS}+{\rm All}$-${\rm BOSS}+{\rm All}$-${\rm Planck15}+{\rm Pantheon} \left({\rm JBD}+G_{\rm matter}\right)$ & $2270$ & $1.010^{+0.030}_{-0.029}$ & $2.8$ & $1.49^{+0.38}_{-0.39}$ & $\cdots$ & $68.87^{+1.32}_{-1.32}$ & $0.818^{+0.015}_{-0.015}$ & $2.7$\\
${\rm KiDS}\times{\rm 2dFLenS}+{\rm All}$-${\rm BOSS}+{\rm All}$-${\rm Planck15}+{\rm Pantheon} \left({\rm JBD}+G_{\rm matter}+\sum m_{\nu}\right)$ & $1640$ & $1.017^{+0.029}_{-0.030}$ & $2.8$ & $1.52^{+0.36}_{-0.39}$ & $0.21$ & $68.71^{+1.27}_{-1.26}$ & $0.809^{+0.016}_{-0.015}$ & $2.8$\\
\hline
\end{tabularx}
\end{table*}

\begin{figure*}
\hspace{-0.05cm}
\includegraphics[width=0.93\hsize]{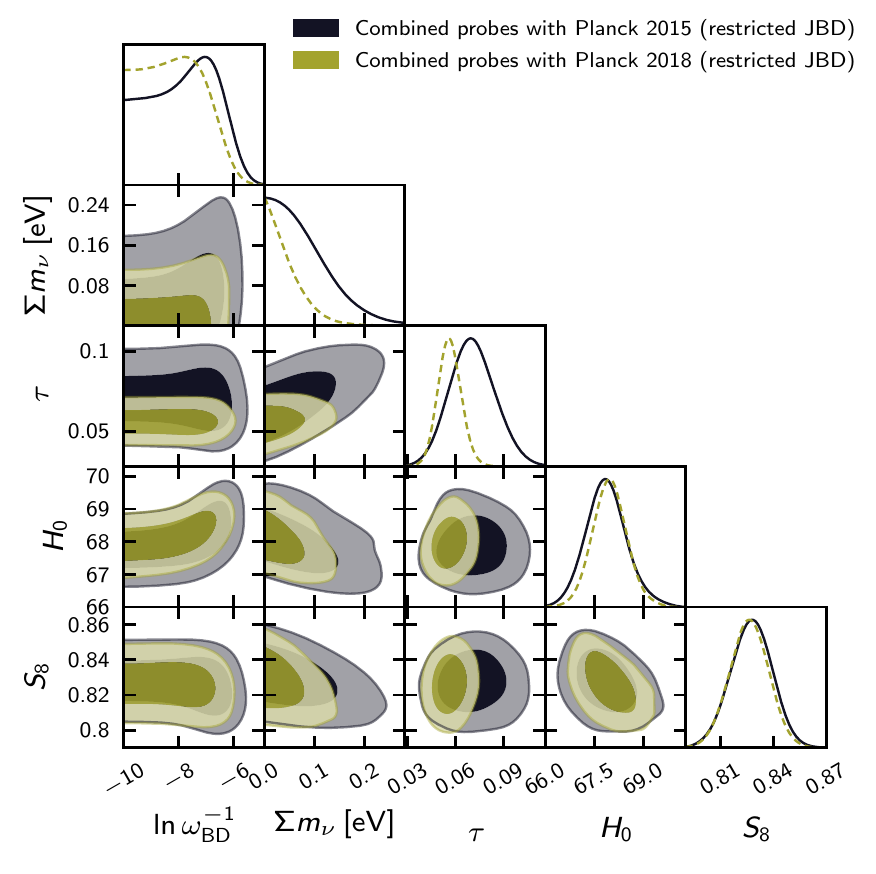}
\vspace{-2.5em}
\caption{\label{figplanck18rest} 
Marginalized posterior distributions (inner $68\%$~CL, outer $95\%$~CL) of the JBD parameter, $\ln \omega_{\rm BD}^{-1}$, the sum of neutrino masses, $\smash{\sum m_{\nu}}$, the optical depth, $\tau$, the Hubble constant, $H_0$ (in units of $\smash{{\rm km} \, {\rm s}^{-1} {\rm Mpc}^{-1}}$), and $\smash{S_8 = \sigma_8 \sqrt{\Omega_{\mathrm m}/0.3}}$ from All-Planck $+$ All-BOSS $+$ Pantheon, where for ``All-Planck'' we consider either Planck 2015 (TT+TE+EE+lowTEB+lensing) or Planck 2018 (TT+TE+EE+lowl+lowE+lensing). All other standard cosmological and systematics parameters are simultaneously varied in this restricted JBD model. For visual clarity, we have zoomed in on the $\smash{\ln \omega_{\rm BD}^{-1}}$ axis where the distributions flatten towards the GR limit at $-\infty$ (in practice towards the end of the prior range at $\smash{\ln \omega_{\rm BD}^{-1}  = -17}$).
}
\end{figure*}

\begin{figure*}
\hspace{-0.05cm}
\includegraphics[width=0.972\hsize]{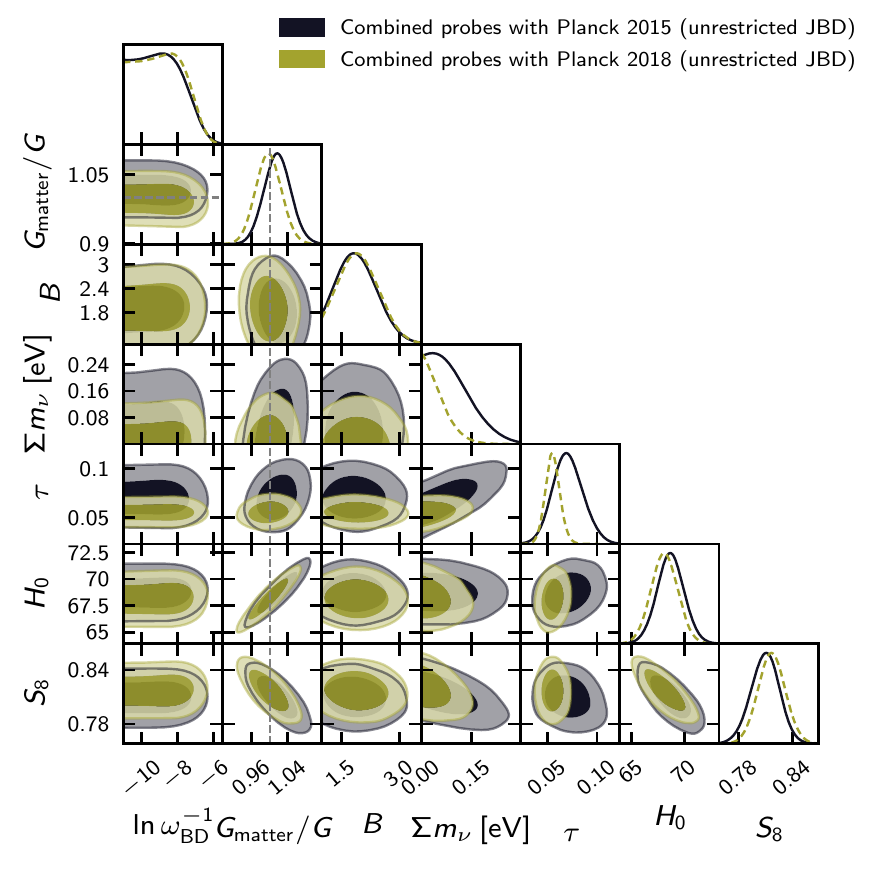}
\vspace{-2.5em}
\caption{\label{figplanck18unrest} 
Marginalized posterior distributions (inner $68\%$~CL, outer $95\%$~CL) of the JBD parameter, $\ln \omega_{\rm BD}^{-1}$, the present effective gravitational constant, $\smash{G_{\rm matter}/G}$, the baryonic feedback amplitude, $B$, the sum of neutrino masses, $\smash{\sum m_{\nu}}$, the optical depth, $\tau$, the Hubble constant, $H_0$ (in units of $\smash{{\rm km} \, {\rm s}^{-1} {\rm Mpc}^{-1}}$), and $\smash{S_8 = \sigma_8 \sqrt{\Omega_{\mathrm m}/0.3}}$ from the full dataset combination KiDS$\times$2dFLenS $+$ All-BOSS $+$ All-Planck $+$ Pantheon, where for ``All-Planck'' we consider either Planck 2015 (TT+TE+EE+lowTEB+lensing) or Planck 2018 (TT+TE+EE+lowl+lowE+lensing). All other standard cosmological and systematics parameters are simultaneously varied. For visual clarity, we have zoomed in on the $\smash{\ln \omega_{\rm BD}^{-1}}$ axis where the distributions flatten towards the GR limit at $-\infty$ (in practice towards the negative end of the prior range at $\smash{\ln \omega_{\rm BD}^{-1}  = -17}$).
}
\end{figure*}

\section{Evolution of the effective gravitational constant}
\label{timeapp}

We now consider the time-variation of the effective gravitational constant ($G_{\rm matter}$) in JBD theory. Given the scalar field ($\phi$) and its relation to the effective gravitational constant in Sec.~\ref{theorysec}, it is straightforward to express
\begin{equation}
\label{geffeqn}
G_{\rm matter}(a) = G_{\rm matter,0} a^{-f(\omega_{\rm BD})},
\end{equation}
where $f(\omega_{\rm BD}) = -\ln(G_{\rm matter}/G_{\rm matter,0}) / \ln(a)$ is a constant of time and is given by
\begin{equation}
\label{gefftime}
f(\omega_{\rm BD})=\left\{\begin{array}{cc}
                0 & {\rm radiation}\\
                1/(1+\omega_{\rm BD}) & {\rm matter}\\
                4/(1+2\omega_{\rm BD}) & {\rm acceleration},\\
               \end{array}\right.
\end{equation}
during the radiation, matter, and cosmic accelerating epochs. Here, $G_{\rm matter,0}$ is the value of the effective gravitational constant at the onset of each epoch.

We can subsequently express the first time-derivative as
\begin{equation}
\label{geffderiveqn}
{\rm d} \ln G_{\rm matter} / {\rm d}t = -f(\omega_{\rm BD}) H(a),
\end{equation}
and the second time-derivative as
\begin{equation}
\label{geff2deriveqn}
{\rm d}^2 \ln G_{\rm matter} / {\rm d}t^2 = -f(\omega_{\rm BD}) \, {\rm d}H(a) / {\rm d}t.
\end{equation}
The second time-derivative can also be expressed in the form
\begin{align}
\label{geff2deriveqn}
\frac{{\rm d}^2 G_{\rm matter} / {\rm d}t^2}{G_{\rm matter}} &= \,
{\rm d} \ln G_{\rm matter} / {\rm d}t \nonumber\\
&\times \, \left[{\rm d} \ln G_{\rm matter} / {\rm d}t \, + \, {\rm d} \ln H(a) / {\rm d}t \right], 
\end{align}
which we can approximate as $16(1+2\omega_{\rm BD})^{-2}H^2(a)$ during the cosmic accelerating epoch and $\left((5+3\omega_{\rm BD})/2\right) (1+\omega_{\rm BD})^{-2}H^2(a)$ during matter domination (along with a zero second derivative during radiation domination). Hence, while the first derivative is negative, the second derivative is positive during the cosmic accelerating and matter-dominated epochs given our positivity prior on $\omega_{\rm BD}$ (except in the GR limit of $\omega_{\rm BD} \rightarrow \infty$ where both derivatives vanish).

We now consider the full combination of datasets to evaluate these derivatives at the present time. In the unrestricted JBD model, this includes the KiDS, 2dFLenS, BOSS, Planck, and Pantheon datasets, and in the restricted JBD model, this includes the BOSS, Planck, and Pantheon datasets (as in Sec.~\ref{fullsec}). In the unrestricted JBD model, we thereby constrain $|{\rm d} \ln G_{\rm matter} / {\rm d}t|_{a=1} < 6.5 \times 10^{-14} \, {\rm year}^{-1}$ as the sum of neutrino masses is fixed and $|{\rm d} \ln G_{\rm matter} / {\rm d}t|_{a=1} < 9.4 \times 10^{-14} \, {\rm year}^{-1}$ as the sum of neutrino masses is varied (both at $95\%$~CL). In the restricted JBD model, we constrain $|{\rm d} \ln G_{\rm matter} / {\rm d}t|_{a=1} < 9.7 \times 10^{-14} \, {\rm year}^{-1}$ as the sum of neutrino masses is fixed and $|{\rm d} \ln G_{\rm matter} / {\rm d}t|_{a=1} < 1.4 \times 10^{-13} \, {\rm year}^{-1}$ as the sum of neutrino masses is varied (both at $95\%$~CL).

For the second derivative, in the unrestricted JBD model, we provide the upper bound $\left(({\rm d}^2 G_{\rm matter} / {\rm d}t^2) / G_{\rm matter}\right)|_{a=1} < 4.2 \times 10^{-27} \, {\rm year}^{-2}$ as the sum of neutrino masses is fixed and $\left(({\rm d}^2 G_{\rm matter} / {\rm d}t^2) / G_{\rm matter}\right)|_{a=1} < 8.8 \times 10^{-27} \, {\rm year}^{-2}$ as the sum of neutrino masses is varied (both at $95\%$~CL). In the restricted JBD model, we constrain $\left(({\rm d}^2 G_{\rm matter} / {\rm d}t^2) / G_{\rm matter}\right)|_{a=1} < 9.4 \times 10^{-27} \, {\rm year}^{-2}$ as the sum of neutrino masses is fixed and $\left(({\rm d}^2 G_{\rm matter} / {\rm d}t^2) / G_{\rm matter}\right)|_{a=1} < 2.1 \times 10^{-26} \, {\rm year}^{-2}$ as the sum of neutrino masses is varied (both at $95\%$~CL). Here, given its stronger bounds on the coupling constant, the constraints on the time evolution of the effective gravitational constant are stronger in the unrestricted JBD model as compared to the restricted JBD model. 

\section{Impact of JBD gravity on the CMB polarization}
\label{cmbpolapp}

In addition to showing the impact of JBD gravity on the CMB temperature power spectrum in Figs.~\ref{figcls} and~\ref{figclsdamping}, we illustrate the impact of JBD gravity on the CMB polarization power spectrum and polarization-temperature cross-spectrum in Fig.~\ref{figclspol} (along with the CMB temperature power spectrum again for comparison). Similar to the CMB temperature power spectrum, the peaks of the polarization power spectrum are primarily shifted by the coupling constant, $\omega_{\rm BD}$, while the damping tail is coherently suppressed (enhanced) as the present effective gravitational constant $(G_{\rm matter}/G)|_{a=1}$ is greater (smaller) than unity. However, while a positive shift in the effective gravitational constant induces a suppression in the temperature power spectrum across scales, the polarization power spectrum primarily exhibits an enhancement for $\ell \lesssim 10^3$ (as also pointed out in Ref.~\cite{zz03}). For the CMB temperature-polarization cross-spectrum, the peaks oscillate about zero, such that a positive $\omega_{\rm BD}$ and $(G_{\rm matter}/G)|_{a=1} < 1$ counteract one another. The different signatures in the CMB temperature and polarization power spectra can thereby be used to place stringent constraints on JBD gravity, as shown in Sec.~\ref{cmbresultssec}.

While we are showing the differences between the JBD and GR (technically GR-limit) power spectra in Fig.~\ref{figclspol}, the absolute value of the response $|\mathcal{R}_\ell| = |C_\ell^{\rm JBD}/C_\ell^{\rm GR}-1|$ coherently increases with $\ell$ in the damping tail for both the temperature and polarization auto spectra as $(G_{\rm matter}/G)|_{a=1}$ deviates from unity. This is shown explicitly for the polarization in Fig.~\ref{figclsdampingpol} (and for the temperature in Figs.~\ref{figcls} and~\ref{figclsdamping}), and implies that the impact of the present effective gravitational constant will be correlated in both the temperature and polarization power spectra with that of other physics affecting the small-scale CMB, such as the primordial helium abundance in particular (discussed in Sec.~\ref{dampsec}), but also the effective number of neutrinos, mass of the neutrinos, and running of the spectral index (targeted by surveys such as AdvACT~\cite{advact}, SPT-3G~\cite{spt3g}, and Simons Observatory~\cite{simonsobs}). 

\section{Impact of the JBD prior range}
\label{priorapp}

In Fig.~\ref{figprior}, we show the impact of the uniform prior range of the primary parameter $\ln \omega_{\rm BD}^{-1}$, where $\omega_{\rm BD}$ is the JBD coupling constant, on the cosmological parameter constraints in an unrestricted JBD model (considering the KiDS, 2dFLenS, BOSS, Planck, and Pantheon datasets combined). While the constraints are shown for only a subset of the parameters (where $H_0$ and $S_8$ are derived parameters), the agreement in the constraints between the fiducial and extended prior cases is strong for the other cosmological and systematics parameters simultaneously varied in the analysis (see Table~\ref{tabpriors} for a list of additional primary parameters). We discuss these results in Sec.~\ref{modelingsec}.

\section{Model selection and additional parameter constraints}
\label{extraparamsdicapp}

In Table~\ref{subtabappall}, we provide parameter constraints and assess dataset tensions for a subset of additional combinations of probes (namely KiDS$\times$2dFLenS alone, and All-BOSS $+$ Pantheon alone). In Table~\ref{subtabdic}, we provide the changes in the best-fit $\chi^2_{\rm eff}$ and deviance information criterion (DIC) relative to $\Lambda$CDM for a range of different combinations of probes and parameter extensions. The datasets consist of KiDS, 2dFLenS, Planck, BOSS,  Pantheon, and ACT, and the parameter extensions consist of $\Lambda {\rm CDM}+\sum m_{\nu}$, ${\rm JBD}$, ${\rm JBD}+\sum m_{\nu}$, ${\rm JBD}+G_{\rm matter}$, and ${\rm JBD}+G_{\rm matter}+\sum m_{\nu}$. Here, a negative DIC (and $\chi^2_{\rm eff}$) implies a model selection preference in favor of the extended model (the significance of which is described in Sec.~\ref{modelselec}). We find that none of the cases favor an extended model beyond $\Delta {\rm DIC} \simeq -5$ which is the threshold of moderate preference in favor of the extended model.

\section{Multi-probe parameter constraints including either Planck 2015 or Planck 2018}
\label{planckapp}
\vspace{-1em}

For completeness, we consider a comparison of the cosmological constraints for dataset combinations that respectively include the Planck 2015 and Planck 2018 CMB datasets. The datasets that we combine are BOSS (growth rates, AP distortions, and BAO distances) and Pantheon (SN distances) together with Planck which includes the CMB temperature, polarization, and lensing reconstruction from either the 2015~\cite{planck2015} or 2018~\cite{planck2018} dataset (i.e.~``TT,TE,EE+lowP+lensing'' and ``TT,TE,EE+lowE+lensing'' for Planck 2015 and 2018, respectively). In constraining the unrestricted JBD model, we further consider the KiDS$\times$2dFLenS 3$\times$2pt dataset of cosmic shear, galaxy-galaxy lensing, and redshift-space galaxy clustering (all datasets as described in Sec.~\ref{datasec}). 

We show a comparison of the constraints on a subset of the parameter space $\{\ln \omega_{\rm BD}^{-1}, \sum m_{\nu}, \tau, H_0, S_8\}$ for the restricted JBD model in Fig.~\ref{figplanck18rest} and an expanded subset $\{\ln \omega_{\rm BD}^{-1}, G_{\rm matter}/G, \sum m_{\nu}, \tau, B, H_0, S_8\}$ for the unrestricted JBD model in Fig.~\ref{figplanck18unrest}. A range of Planck 2015 constraints are also summarized in Table~\ref{subtabplanck15}. As expected, we find substantial improvements in the constraints on the optical depth (and scalar amplitude) and thereby the sum of neutrino masses when the Planck 2018 dataset is considered (instead of Planck 2015). However, the constraints on the other parts of the subspace, in particular the constraints on JBD gravity, baryonic feedback, the Hubble constant, and $S_8$ seem largely robust between the two dataset combinations. This also seems to hold for the correlations and degeneracies between the parameters over the full subspace.

For the dataset combination KiDS$\times$2dFLenS $+$ All-BOSS $+$ All-Planck15 $+$ Pantheon, we note that there is a minor unexpected improvement in the $H_0$ (and $A_{\rm IA}$) constraint when the sum of neutrino masses is varied in the unrestricted JBD model, as the fiducial $\sum m_{\nu}$ is located at the boundary of the posterior when varied, where the widest range in $H_0$ is favored, and due to the weak correlation between the parameters (quoted in Table~\ref{subtabplanck15} and shown in Fig.~\ref{figplanck18unrest}). We obtain the expected change in the uncertainty on the Hubble constant with the inclusion of massive neutrinos as Planck 2018 is considered instead of Planck 2015. For both the Planck 2015 and 2018 dataset combinations, we also observe that the anti-correlation between the sum of neutrino masses and the Hubble constant diminishes as we transition from the restricted to the unrestricted JBD model, due to the positive correlation of both parameters with the effective gravitational constant.

In Table~\ref{subtabplanck15}, we can also compare the parameter constraints from KiDS$\times$2dFLenS $+$ All-BOSS $+$ All-Planck15 $+$ Pantheon to those from the data combination KiDS$\times$2dFLenS $+$ All-BOSS $+$ Planck15 in the unrestricted JBD model. As expected, we find weaker parameter constraints for the latter data combination, as $\omega_{\rm BD} > 840$ ($95\%$~CL) and $G_{\rm matter}/G = 1.034^{+0.042}_{-0.046}$ in the unrestricted JBD model with massive neutrinos (where $\sum m_{\nu} < 0.30~{\rm eV}$ at $95\%$~CL). These weaker constraints (notably by 50\% for $G_{\rm matter}$) are driven by the absence of the CMB polarization and lensing reconstruction (i.e.~Planck instead of All-Planck, rather than the Pantheon SNe). The constraints on $H_0$ and $S_8$ are also weakened (by $40\%$ and $60\%$, respectively), as $H_0 = 69.5^{+1.8}_{-1.8} ~ {\rm km} \, {\rm s}^{-1} {\rm Mpc}^{-1}$ and $S_8 = 0.790^{+0.025}_{-0.025}$. This highlights the significance of the additional CMB observables in both strengthening the parameter constraints in the unrestricted JBD model and increasing the agreement with the GR expectation.

\section{JBD theory in \eftcamb}
\label{theoryapp}

Since the scalar field is expected to remain frozen during radiation domination at early times due to the Hubble friction term in its equation of motion (Eq.~\ref{sfeq}), we set its initial velocity, $\dot{\phi}_{i}$, to zero and find its initial value, $\phi_{i}$, by means of a binary search enforcing $\phi(a=1)$ to be the desired value. The background evolution is commenced at $a_i = 10^{-10}$, and we ensure that the flatness condition is respected and verified up to a tolerance of $\sim10^{-4}$, which is important for the overall soundness and stability of the model. In the language of \eftcamb~\cite{hu14eft}, the effective field theory functions that describe the background dynamics and linear perturbations of JBD gravity are given by (e.g.~\cite{Bellini:2017avd})
\begin{eqnarray}{\label{eftfunc}}
&&\Omega^{\rm EFT}(t) = \phi - 1 \nonumber \\
&&\gamma_{i\in\{1,2,3\}}^{\rm EFT} = 0 \nonumber \\
&&\Lambda^{\rm EFT}(t) = \frac{1}{2} \frac{\omega_{\rm{BD}}}{\phi} \dot{\phi}^{2} - V(\phi) \nonumber \\
&&{\rm{c^{\rm EFT}}}(t) = \frac{1}{2} \frac{\omega_{\rm{BD}}}{\phi} \dot{\phi}^{2},
\end{eqnarray}
where the potential $V(\phi)$ is fixed to the cosmological constant as discussed in Sec.~\ref{theorysec}, and we have added the superscripts ``EFT'' to avoid confusion with 
$\Omega$, $\gamma$, $\Lambda$, and $c$ defined elsewhere. We note that $\Lambda^{\rm EFT}(t)$ and $c^{\rm EFT}(t)$ are not independent functions, but can be expressed in terms of $\Omega^{\rm EFT}(t)$ (encapsulating the scalar field coupling to gravity and matter in the Jordan and Einstein frames, respectively), the Hubble parameter, and the matter density and pressure (e.g.~\cite{gubitosi13,hu14eft}), and that a direct one-to-one correspondence can be established between the EFT functions and the $\alpha_i$ parameterization (e.g.~\cite{planckmg15,Alonso:2016suf,zuma17,Bellini:2017avd}).

\end{document}